\documentclass[aps,prd,superscriptaddress,floatfix,nofootinbib,preprintnumbers,eqsecnum,twocolumn]{revtex4-1}
\pdfoutput=1  
\usepackage[utf8]{inputenc}
\usepackage[dvips]{graphicx}
\usepackage[dvipsnames]{xcolor}

\usepackage{amsmath,float,graphicx,placeins,amssymb,multirow,pgfplots,pgfplotstable}
\usepackage{bm}
\usepackage{empheq}
\usepackage{tikz}
\usepackage{comment}
\usepackage{enumerate,slashed,cancel,comment,bbm,rotating,placeins,mathtools,multirow,hhline}
\usepackage[normalem]{ulem}
\usepackage{array}

\usepackage{epstopdf}
\usepackage{hyperref}
\usepackage{mathrsfs}
\usepackage{amssymb}
\usepackage{physics}
\usepackage{booktabs}

\usepackage{color}
\usepackage{relsize}
\usepackage{graphics}
\usepackage{epstopdf}
\usepackage{hyperref}
\usepackage{mathrsfs}
\usepackage{amssymb}
\usepackage{physics}
\usepackage{booktabs}
\renewcommand{\figurename}{Fig.}
\usepackage[normalem]{ ulem }
\usepackage{amsthm}
\usepackage{amsmath}
\usepackage{cancel}
\usepackage{float}
\usepackage{fontawesome} 
\usepackage[caption=false]{subfig}
\usepackage{tabularx,ragged2e} 
 \newcolumntype{L}{>{\RaggedRight\arraybackslash}X}

\colorlet{mylinkcolor}{ForestGreen}

\colorlet{mycitecolor}{Red}

\colorlet{myurlcolor}{violet}

\usepackage[dvipsnames]{xcolor}
\definecolor{darkgreen}{RGB}{0,100,0}

\usepackage{lipsum}
\usepackage{diagbox}
\usetikzlibrary{shapes.geometric,arrows}
\graphicspath{}

\def\beq{\begin{eqnarray}}
\def\eeq{\end{eqnarray}}
\def\beqn{\begin{eqnarray}}
\def\eeqn{\end{eqnarray}}
\def\be{\begin{eqnarray}} 
\def\ee{\end{eqnarray}}

\newcommand{\expt}[1]{\left\langle #1 \right\rangle}

\hypersetup{
 linkcolor = {blue!80!black},
 citecolor = {blue!70!black},
 urlcolor = {blue!80!black},
 colorlinks = true
}
\def\barOmega{{\overline{\Omega}}}

\newcommand{\Hc}{\mathcal{H}}

\newcommand{\kk}{\bm{k}}
\newcommand{\pp}{\bm{p}}
\newcommand{\xx}{\bm{x}}

\def\ie{{\it i.e.}\/}
\def\eg{{\it e.g.}\/}
\def\etc{{\it etc}.\/}

\def\IR{\relax{\rm I\kern-.18em R}}
 \font\cmss=cmss10 \font\cmsss=cmss10 at 7pt
\def\IQ{\relax{\rm I\kern-.18em Q}}
\def\IZ{\relax\ifmmode\mathchoice
 {\hbox{\cmss Z\kern-.4em Z}}{\hbox{\cmss Z\kern-.4em Z}}
 {\lower.9pt\hbox{\cmsss Z\kern-.4em Z}}
 {\lower1.2pt\hbox{\cmsss Z\kern-.4em Z}}\else{\cmss Z\kern-.4em Z}\fi}

\def\M0{M^{(0)}}
\def\f0{f^{(0)}}
\def\t0{t^{(0)}}
\def\X0{X^{(0)}}
\def\a0{a^{(0)}}
\def\H0{H^{(0)}}

\def\nn{\nonumber}

\def\wbar{\overline{w}}
\def\barOmega{\overline{\Omega}}

\pgfplotsset{compat=1.17}
\begin{document}

\title{Gravitational Waves as a Probe~of  Multi-Fluid Cosmologies:\\   General Formalism and the Case of Cosmological Stasis}

\def\andname{\hspace*{-0.5em}} 

\author{Keith R. Dienes}
 \email[Email address: ]{dienes@arizona.edu}
 \affiliation{Department of Physics, University of Arizona, Tucson, AZ 85721 USA}
 \affiliation{Department of Physics, University of Maryland, College Park, MD 20742 USA}
 
 \author{Fazlollah Hajkarim}
 \email[Email address: ]{fazlollah.hajkarim@ou.edu}
 \affiliation{Homer L. Dodge Department of Physics and Astronomy, University of Oklahoma, Norman, OK  73019  USA}
 
\author{Lucien Heurtier}
\email[Email address: ]{lucien.heurtier@kcl.ac.uk}
\affiliation{Theoretical Particle Physics and Cosmology, King’s College London,\\
Strand, London WC2R 2LS, United Kingdom}

\author{Fei Huang}
\email[Email address: ]{huangfei@aust.edu.cn}
 \affiliation{Center for Fundamental Physics, School of Mechanics and Optoelectronic Physics, Anhui University of Science and Technology, Huainan, Anhui 232001, China}

\author{\\ Kuver Sinha}
 \email[Email address: ]{kuver.sinha@ou.edu}
 \affiliation{Homer L. Dodge Department of Physics and Astronomy, University of Oklahoma, Norman, OK  73019  USA}
 
 \author{Brooks Thomas}
  \email[Email address: ]{thomasbd@lafayette.edu}
 \affiliation{Department of Physics, Lafayette College, Easton, PA  18042 USA}

\author{Tao Xu}
 \email[Email address: ]{tao.xu@ou.edu}
 \affiliation{Jockey Club Institute for Advanced Study, Hong Kong University of Science and Technology, Clear Water Bay, Kowloon, Hong Kong}
\affiliation{Homer L. Dodge Department of Physics and Astronomy, University of Oklahoma, Norman, OK  73019  USA}

\begin{abstract}
Scalar-induced gravitational waves (SIGWs) provide a powerful probe of the 
expansion history of the universe prior to the Big Bang.  Indeed, the production 
of such gravity waves depends on the evolution of first-order scalar perturbations 
within the early universe, while their subsequent propagation depends on the 
background equation of state.  In this paper we extend previous discussions by 
studying SIGWs in cosmologies containing multiple fluids.  As we shall demonstrate, 
the gravitational-wave signal within such multi-fluid cosmologies is modified in 
two distinct ways: not only are there new velocity effects which have no 
single-fluid analogues, but there are also significant alterations to the subsequent 
redshifting of modes that enter the horizon before the multi-fluid era ends.  
Thus SIGWs provide a probe of not only the overall expansion rate within such 
cosmologies, but also their internal multi-fluid composition.  As an example, we 
then consider the case of gravitational waves arising during an epoch of cosmological 
stasis.  During such an epoch, the abundances of energy components with different 
equations of state remain constant, thereby giving rise to a universe with a constant 
effective equation of state which nevertheless remains genuinely multi-fluid 
despite cosmological expansion.  Despite their constant equations of state, we find 
that such stasis cosmologies also give rise to tensor power spectra and present-day 
gravitational-wave abundances which are significantly different from those expected 
within a standard radiation-dominated universe.
\end{abstract}

\maketitle

\tableofcontents


\section{Introduction\label{sec:intro}}


The  observation of gravitational waves (GWs)  from binary black-hole mergers  and, subsequently,  
binary neutron-star mergers, has ushered in a new era in 
astronomy~\cite{Abbott:2016nmj,Abbott:2016blz,TheLIGOScientific:2017qsa}. 
While current experiments are able to detect GWs from astrophysical events
and have obtained preliminary evidence~\cite{NANOGrav:2023gor} for the 
existence of a stochastic GW background, future facilities are expected to be 
sensitive to phenomena in the very early universe --- phenomena such as phase transitions, 
cosmic strings, inflationary dynamics, or primordial black-hole formation --- making GWs potentially 
transformative for particle physics and cosmology~\cite{Audley:2017drz, Sathyaprakash:2012jk,
Seto:2001qf,Sato:2017dkf,Crowder:2005nr,Janssen:2014dka,Arzoumanian:2018saf,Lentati:2015qwp,Caldwell:2022qsj}.
As such, the fact that GWs are sensitive to such phenomena makes them excellent probes of cosmological history.    

One particularly interesting open question along these lines concerns the nature of the 
cosmological history between inflation and the epoch of Big Bang Nucleosynthesis (BBN).~
While standard cosmological histories assume a universe dominated by radiation between 
inflation and the BBN epoch, alternatives to this standard picture have increasingly 
become a topic of central importance (for reviews, see, {\it e.g.}\/, Refs.~\cite{Batell:2024dsi,Allahverdi:2020bys,Kane:2015jia}). 
One such possibility which has received particular attention in recent years involves 
an era of early matter domination before BBN.~  The possible existence of such matter-dominated 
eras is motivated by the ubiquity of scalar fields in string compactifications, including 
not only moduli~\cite{Allahverdi:2013noa,Giudice:2000ex} but also towers of 
Kaluza-Klein (KK) states arising from spacetime compactification.  Early matter-dominated eras 
can leave a host of imprints on the cosmological properties of  dark 
matter~\cite{Erickcek:2015jza,Dutta:2009uf}, on large-scale 
structure~\cite{Erickcek:2011us,Fan:2014zua}, and on the spectrum of GWs sourced by 
early-universe phenomena~\cite{Guo:2020grp, Bernal:2020ywq, Hajkarim:2019nbx}.

In contrast to alternative scenarios such as early matter domination in which the universe 
is dominated by a single non-traditional fluid, it is also possible to consider 
alternative cosmological histories in which {\it two or more}\/ fluids dominate 
simultaneously, essentially retaining comparable or even fixed abundances across an 
extended period.  Of course, there exist many cosmological situations that simultaneously 
feature multiple fluids --- examples include the intervals when the universe is transitioning 
from domination by one fluid to domination by another.  However, such situations are
 typically confined to a short transition period and affect only a narrow range of 
 scales~\cite{Kumar:2024hsi, Inomata:2025wiv}. 

Our interest, by contrast, concerns entire extended {\it epochs}\/ across which multiple 
fluids persist and retain abundances of sizes sufficient to affect the resulting cosmology.   
At first glance, it may seem  that fluids with different equations of state cannot retain 
comparable abundances over an extended period --- such fluids necessarily have energy 
densities which scale differently with the cosmological scale factor, a feature which 
ultimately causes the energy density of one fluid to quickly dominate over the other(s).  
However, it has been shown~\cite{Dienes:2021woi,Dienes:2023ziv} that this expectation is not 
correct, and that there can exist periods of {\it cosmological stasis}~\cite{Dienes:2021woi,
Dienes:2023ziv,Barrow:1991dn,Dienes:2022zgd, Dienes:2024wnu,Halverson:2024oir,Barber:2024vui,
Barber:2024izt,Huang:2025odd,Dienes:2025tox,Long:2025wjw,Barenboim:2026zgj,Barenboim:2026txj,
Barber:2026bse,Barenboim:2026ive} during which the abundances of the different constituent 
fluids remain constant despite cosmological expansion.  Indeed, such stasis epochs can 
persist across many $e$-folds of cosmic time.  Moreover, such stasis solutions emerge 
naturally (and potentially even unavoidably) in many cosmological scenarios built on 
beyond-the-Standard-Model (BSM) physics.  Finally, such stasis solutions are not the results 
of fine tunings --- rather, they are {\it dynamical attractors}\/ of the cosmological 
dynamics within the BSM scenarios in which they arise.  Therefore, in a manner which is 
largely independent of the initial conditions, such BSM cosmologies will naturally enter 
into and remain in a multi-fluid stasis configuration with a constant effective equation 
of state persisting across many $e$-folds of cosmic expansion.

Despite its apparent calm, much is happening ``under the hood'' during an epoch of stasis.  
For example, although the abundances of the different fluids remain constant, such stasis 
epochs involve a continual transfer of energy between the fluids involved.  Indeed, it is 
this transfer of energy which counteracts the effects of cosmological expansion and allows 
the different fluid abundances to each remain constant.

The goal of this paper is to understand the properties of the GWs that arise 
during multi-fluid cosmological epochs.  In this way, we also hope to determine the extent 
to which future GW detectors will be able to probe cosmological histories containing such 
epochs.  Although much of our analysis will be completely general, epochs of cosmological 
stasis will often serve as our primary example and motivation.

Throughout this paper, our primary interest will  focus on {\it scalar-induced}\/  
gravitational waves (SIGWs) --- \ie,  GWs which are sourced by features associated 
with the multi-fluid cosmological epoch itself.  In particular, as we shall later 
discuss, such GWs are unavoidably induced at second order in General Relativity by the scalar 
perturbations which arise at first order in such 
cosmologies~\cite{Acquaviva:2002ud, Ananda:2006af,Domenech:2021ztg,Baumann:2007zm,Kohri:2018awv}.  
As a result, such gravity waves exhibit a spectrum that is sensitive not only to the background 
cosmological scale factor, but also to the multi-fluid composition of the universe during such epochs. 
Note that such gravity waves are to be distinguished from the {\it primordial}\/  gravity waves 
that already existed prior to the stasis epoch, and which can also be used as probes of the stasis
epoch~\cite{Barenboim:2026zgj, Barenboim:2026txj}.  We shall discuss the relation between 
these different phenomena later in this paper.

This paper is organized as follows. 
In Sect.~\ref{singlefluidreview} we review the origins of scalar-induced gravitational waves 
within a {\it single}\/-field cosmology, focusing on the form of the tensor source for 
such waves.  This discussion will also serve to establish our notational conventions.  
Then, in  Sect.~\ref{multifluus}, we derive the corresponding source tensor for GWs 
within a {\it two}\/-fluid cosmology, emphasizing the decomposition of this source into 
geometric, bulk-velocity, slip-velocity, and anisotropic-stress contributions.  
In Sect.~\ref{powerspectrumenergydens}, we then calculate the corresponding two-fluid induced tensor power spectrum 
and the resulting GW energy density.
In Sect.~\ref{stasissec}, we proceed to apply these results to the case of a tower-based 
matter/radiation stasis of the sort discussed in Refs.~\cite{Dienes:2021woi, Dienes:2023ziv}
and derive the scalar and velocity transfer functions needed for the tensor calculation.  
 Finally, in Sect.~\ref{results}, we present numerical 
results for a benchmark stasis cosmology and compare these results with those expected 
within a standard radiation-dominated cosmology.  We conclude in Sect.~\ref{discussions} 
with a discussion of our results and directions for further research, while 
further details concerning various analytical estimates and numerical approximations 
made in the course of this work are provided in an Appendix.


\section{The source tensor in single-fluid cosmologies}\label{singlefluidreview}


In this section, we briefly review the main features of SIGWs generated in single-fluid 
cosmologies. There are many standard references; we follow Refs.~\cite{Baumann:2007zm,Hajkarim:2019nbx}. 

\subsection{Source tensor $S_{ij}$}

We work in conformal time $\eta$ with scale factor $a(\eta)$ and conformal Hubble 
parameter
\begin{equation}
  \Hc ~\equiv~ \frac{a'}{a}~, 
\end{equation}
where a prime denotes a derivative with respect to $\eta$.  
Spatial indices are raised 
and lowered with the Kronecker delta $\delta_{ij}$, and 
$\nabla^2 \equiv \delta^{ij}\partial_i\partial_j$. 
We begin with the perturbed Friedmann-Robertson-Walker metric in the Newtonian gauge,
\begin{eqnarray}
  d s^2 &=& a^2(\eta)\Big\{
      -(1+2\Phi^{(1)}+2\Phi^{(2)})\,d\eta^2
      +2V_i^{(2)}\,d\eta\,d x^i \nonumber \\ 
    && \!\!\!\! +\Big[(1-2\Psi^{(1)}-2\Psi^{(2)})\delta_{ij}
      +\tfrac12 h_{ij}\Big]d x^id x^j\Big\}\,,
  \label{eq:metric}
\end{eqnarray}
where the superscripts $(1)$ and $(2)$ denote first- and second-order 
quantities, respectively.  Since we are interested in SIGWs, we shall 
assume that the first-order perturbation to the transverse, traceless 
tensor perturbation $h_{ij}$ vanishes, and thus that 
$h_{ij} \approx h^{(2)}_{ij}$.
 
The spatial Einstein tensor at second order has the schematic form
\begin{eqnarray}
&&  \!\!\!\!\!\!\! G^{(2)i}{}_j ~=~ a^{-2}\bigg[
      \frac14\big(  h^i{}_{j}{}{''}  
      +2\Hc h^i{}_{j}{}'-\nabla^2 h^i{}_j\big)
      +Q^i{}_j ~~~~~~~~\nonumber \\
&&    + \left(\Phi^{(2)},\Psi^{(2)},V_i^{(2)}\text{ terms}\right)
      +\left(\text{diagonal part}\right)\delta^i_j
      \bigg]\,,\nn\\~~~
\end{eqnarray}
where the explicit non-diagonal quadratic scalar piece is given by
\begin{eqnarray}
\label{eq:Qij}
  Q^i{}_j &=&\ 2\Phi\,\partial^i\partial_j\Phi
     -2\Psi\,\partial^i\partial_j\Phi
     +4\Psi\,\partial^i\partial_j\Psi
     +\partial^i\Phi\,\partial_j\Phi \nonumber \\
    &&~~ -\partial^i\Phi\,\partial_j\Psi
     -\partial^i\Psi\,\partial_j\Phi
    +3\,\partial^i\Psi\,\partial_j\Psi~.
  \label{eq:QijOriginal}
\end{eqnarray} 
Note that we have used the notation $\Phi\equiv\Phi^{(1)}$ and $\Psi\equiv\Psi^{(1)}$. 

The second order stress-energy tensor for a single fluid is given by 
\begin{eqnarray} 
  T^{(2)i}{}_j &=& \Big[
    (\rho^{(0)}+P^{(0)})\,v^{(1)i}v^{(1)}_{j} \nonumber \\
  &  & + P^{(0)}\Pi^{(2)i}{}_j + P^{(1)}\Pi^{(1)i}{}_j
    + P^{(2)}\delta^i_j \Big]~,~~~~
  \label{eq:T2ijsingle}    
\end{eqnarray}
where $\rho$, $P$, $v$, and $\Pi$ denote the  energy density, pressure, velocity, 
and anisotropic stress of the fluid, respectively, and where a superscript $(0)$ 
denotes a zeroth-order --- \ie, background --- quantity.

At this point, it is customary to introduce the transverse traceless (TT) projector 
$\widehat{\mathcal{T}}^{\,\ell m}{}_{ij}$, which extracts the transverse, traceless part of
any tensor on which it acts.  Using this on both sides of the spatial Einstein equation, 
we obtain
\begin{equation}
\widehat{\mathcal{T}}^{\,\ell m}{}_{ij}\,G^{(2)}_{\ell m} ~=~ 
  \kappa^2\,\widehat{\mathcal{T}}^{\,\ell m}{}_{ij}\,T^{(2)}_{\ell m}~,
  \label{eq:projEin}
\end{equation}
where we have defined $\kappa^2 \equiv 8\pi G = M^{-2}_P$ with 
$M_P$ denoting the reduced Planck mass
and where $T^{(2)}$ denotes the second-order stress-energy tensor.  Note that after the 
TT projector is applied, only the tensor operator acting on $h_{ij}$ and the TT parts of 
$Q_{ij}$ and $T^{(2)}_{ij}$ survive. We shall keep only those contributions in $T^{(2)}$ 
which are quadratic in first-order quantities.

With the above definitions, the equation of motion for the tensor perturbation $h^{ij}$ becomes
\begin{equation}
  h''_{ij}+2\Hc h'_{ij}-\nabla^2 h_{ij} ~=~
    -4\,\widehat{\mathcal{T}}^{\,\ell m}{}_{ij}\,S_{\ell m}~,
  \label{eq:tensorEOM}
\end{equation}
where $S_{\ell m}$ is the source term that collects all of the contributions from the 
$Q$ and $T^{(2)}$ terms.  Note that $S_{ij}$ is always a function of $\eta$.  However,
for notational simplicity we will often suppress the $\eta$ dependence of $S_{ij}$ and
its Fourier transform in what follows.

Our goal is now to determine $S_{ij}$ explicitly.  The fluid-dependent quantities can be 
eliminated in favor of $\Phi$ and $\Psi$ by making use of the standard first-order relations
\begin{eqnarray}
  P^{(1)} &=& c_s^2\, \rho^{(1)} \nonumber\\
 \rho^{(1)} &=& -\frac{2}{\kappa^2 a^2} \left[ 3\mathcal{H} 
    \left( \mathcal{H}\Phi - \Psi^{\prime} \right) + \nabla^2 \Psi \right] 
    \nonumber\\
  v_i^{(1)} &=& -\frac{2}{\kappa^2 a^2 (\rho^{(0)} + P^{(0)})} \, 
    \partial_i \left( \Psi^{\prime} + \mathcal{H}\Phi \right)
    \nonumber\\
  \Pi^{(1)\,i}_{\;\;\;j} &=& -\frac{1}{\kappa^2 a^2 P^{(0)}} 
    \left( \partial^i \partial_j - \frac{1}{3} \delta^i_j \nabla^2 \right)
    \left( \Phi - \Psi \right)\,,~~~~~
\label{singlefluidfirstorder1}
\end{eqnarray}
where $c_s^2$ is the adiabatic sound speed of the fluid.
The source term $S_{ij}$ can then be expressed in terms of first-order scalars and 
fluid variables as
\begin{eqnarray} 
  && \!\!\!\! 
  S_{ij}  ~\equiv~ Q_{ij}
      - \frac{4}{3(1+w)\mathcal{H}^2}
      \partial_i \left( \Psi' + \mathcal{H}\Phi \right)
      \partial_j \left( \Psi' + \mathcal{H}\Phi \right)~~ 
      \nonumber \\
    && 
    - \frac{2 c_s^2}{3 w \mathcal{H}^2}
      \left[ 3\mathcal{H} \left( \mathcal{H}\Phi - \Psi' \right) 
      + \nabla^2 \Psi \right]
      \partial_i \partial_j (\Phi - \Psi)\, ,
  \label{source1}
\end{eqnarray}
where $w$ is the equation-of-state parameter for the single fluid which is assumed 
to dominate the energy density of the universe in this scenario.
Conceptually, it is convenient to decompose the total contribution to the source 
tensor in Eq.~\eqref{source1} to be a sum of two pieces.  The first of these pieces,
which consists simply of $Q_{ij}$, depends only on the potentials $\Phi$ and $\Psi$.
We can therefore consider this piece of $S_{ij}$ to be a purely ``geometric'' piece.
By contrast, the second piece, which comprises all of the remaining terms in Eq.~(\ref{source1}), 
originates from the $v^{(1)}$ dependent terms  in the stress-energy tensor of 
Eq.~\eqref{eq:T2ijsingle}.  We therefore refer to this piece of $S_{ij}$ as the 
``bulk-velocity piece.''  While we retain the third term in Eq.~(\ref{source1})
at this stage for completeness, we note that this term generically
vanishes in the special case wherein the anisotropic stress likewise vanishes.

\subsection{Source tensor in Fourier space}\label{sourcefourier}

We now perform a Fourier decomposition of $S_{ij}$.  In what follows, we define 
the Fourier transform $X(\kk)$ of any position-space quantity $X(\xx)$ such that   
\begin{eqnarray}
  X(\xx)&=&\int\frac{d^3k}{(2\pi)^{3/2}}\,e^{i\kk\cdot\xx}\,X(\kk)
    \nonumber \\ 
  X(\kk)&=&\int\frac{d^3x}{(2\pi)^{3/2}}\,e^{-i\kk\cdot\xx}\,X(\xx)~.
\end{eqnarray}
Note that we shall refer to both members of a Fourier-transform pair using the same 
symbol $X$, as above,  allowing their stated arguments to disambiguate them.
Derivatives with respect to coordinates are therefore given in Fourier space by
\begin{equation}
  \partial_i X(\xx) \leftrightarrow i k_i X(\kk)\, ,
    \quad
    \partial_i\partial_j X(\xx) \leftrightarrow -k_i k_j X(\kk)~,~~
\end{equation}
while the convolution theorem tells us that the Fourier transforms of products of 
position-space quantities are given by
\begin{equation}
  \big[X\,Y\big](\kk) ~=~ 
    \int\frac{d^3p}{(2\pi)^{3/2}}\,X(\pp)\,Y(\kk-\pp)~.
  \label{eq:square_bracket_relation}
\end{equation}
In this paper we shall therefore define
\beq
\bm{q}~\equiv~ \kk-\pp~.
\label{qdef}
\eeq

We begin with the Fourier expansion of $h_{ij}$, which may be expressed as  
\begin{equation}\label{eq:hFourier}
h_{ij}(\xx,\eta)
\,=  \int\!\frac{d^3k}{(2\pi)^{3/2}}\,e^{i\kk\cdot\xx}
\Big[h_{\kk}(\eta)\,e_{ij}(\kk)+\overline h_{\kk}(\eta)\,\overline e_{ij}(\kk)\Big]\,.
\end{equation}
In accord with standard practice, we choose our orthonormal basis vectors 
$\bm{e}(\kk)$ and $\overline{\bm{e}}(\kk)$ to be perpendicular to $\kk$ in order 
to obtain the polarization tensors
\begin{eqnarray} \label{eq:eij}
  e_{ij}(\kk)&=& \frac{1}{\sqrt{2}}\left[e_i(\kk)e_j(\kk)
      -\overline e_i(\kk)\overline e_j(\kk)\right]
      \nonumber \\
    \overline e_{ij}(\kk)&=&\frac{1}{\sqrt{2}}\left[e_i(\kk)\overline 
      e_j(\kk)+\overline e_i(\kk)e_j(\kk)\right]~.~~
\end{eqnarray}
These polarization tensors satisfy
\begin{equation}
  k^i e_{ij}=0\,,\quad e^i{}_i=0\,, \quad
  k^i \overline e_{ij}=0\,,\quad \overline e^i{}_i=0~.
\end{equation}
Moreover, for $\bm{q}=\kk-\pp$, we have 
\begin{equation}
\label{eq:transversalityIdentity}
  e^{ij}(\kk)p_i q_j \,=\, e^{ij}p_i(k_j-p_j)
    \,=\, -e^{ij}(\kk)\,p_i p_j~,
\end{equation}
where we have recognized that $e^{ij}(\kk)k_j=0$.

The Fourier expansion of the TT-projected source can be expressed in terms of the 
polarization tensors and $S_{ij}$ as
\begin{eqnarray} 
  &&\widehat{\mathcal{T}}^{\,\ell m}{}_{ij}\,S_{\ell m} ~=~
    \int\frac{d^3k}{(2\pi)^{3/2}}\,e^{i\kk\cdot\xx} \nonumber \\
  && ~~~~~~~\times
  \left[e_{ij}(\kk)e^{\ell m}(\kk)+\overline e_{ij}(\kk)\overline 
    e^{\ell m}(\kk)\right]S_{\ell m}(\kk)~. \nonumber \\
\label{eq:TTprojS}
\end{eqnarray}
Given these expansions for $h_{ij}(\xx,\eta)$ and the TT-projected 
source term, 
we find that the Fourier-space tensor perturbations satisfy the equation of motion
\begin{equation}
  h_{\kk}''+2\Hc h_{\kk}'+k^2 h_{\kk} ~=~ S(\kk,\eta)~,
\label{eq:modeEq}
\end{equation}
where for the $e_{ij}$ polarization we have
\begin{equation}
  S(\kk,\eta)~=~-4\,e^{\ell m}(\kk)\,S_{\ell m}(\kk,\eta)\,\,.
  \label{eq:SkDef}
\end{equation}
Indeed, the equation of motion in Eq.~(\ref{eq:modeEq}) is essentially the Fourier 
transform of the equation of motion in Eq.~\eqref{eq:tensorEOM}.

Using Eq.~\eqref{source1}, we can then show that at second order $S(\kk,\eta)$ is a 
convolution of first-order scalar perturbations at different 
wavenumbers~\cite{Baumann:2007zm}:
\begin{widetext}
\begin{align}
  S(\bm{k},\eta)
  ~=~ 4 \int \frac{d^{3}\tilde{\bm{k}}}{(2\pi)^{3/2}}\,
      e^{\ell m}(\bm{k})\,\tilde{k}_{\ell}\tilde{k}_{m}
      &\Bigg\{
        \left[
          \frac{7+3w}{3(1+w)} - \frac{2c_{s}^{2}}{w}
        \right]
        \Phi_{\tilde{\bm{k}}}(\eta)\,\Phi_{\bm{k}-\tilde{\bm{k}}}(\eta)
        +\left(
           1 - \frac{2c_{s}^{2}\tilde{k}^{2}}{3w\mathcal{H}^{2}}
         \right)
         \Psi_{\tilde{\bm{k}}}(\eta)\,\Psi_{\bm{k}-\tilde{\bm{k}}}(\eta)
        \notag\\
      &\quad
      +\frac{2c_{s}^{2}}{w}
        \left(1+\frac{\tilde{k}^{2}}{3\mathcal{H}^{2}}\right)
          \Phi_{\tilde{\bm{k}}}(\eta)\,\Psi_{\bm{k}-\tilde{\bm{k}}}(\eta)
          +\left[
           \frac{8}{3(1+w)} + \frac{2c_{s}^{2}}{w}
         \right]
         \frac{1}{\mathcal{H}}\,
         \Phi_{\tilde{\bm{k}}}(\eta)\,\Psi'_{\bm{k}-\tilde{\bm{k}}}(\eta)
        \notag\\
      &\quad
      -\frac{2c_{s}^{2}}{w\mathcal{H}}\,
         \Psi_{\tilde{\bm{k}}}(\eta)\,\Psi'_{\bm{k}-\tilde{\bm{k}}}(\eta)
        +\frac{4}{3(1+w)\mathcal{H}^{2}}\,
         \Psi'_{\tilde{\bm{k}}}(\eta)\,\Psi'_{\bm{k}-\tilde{\bm{k}}}(\eta)
      \Bigg\}~.
  \label{singlefluidsignal}
\end{align}
\end{widetext}

Having obtained the source tensor for a given single-fluid cosmology, we can now proceed 
to obtain the tensor power spectrum using standard methods.  However, given a particular 
form for $S_{ij}$, we shall see that the procedure for calculating the tensor power spectrum 
is identical for single-fluid and multi-fluid cosmologies.  We shall therefore defer our 
description of this calculation until Sect.~\ref{powerspectrumenergydens} and instead now 
turn to discuss the structure of $S_{ij}$ which emerges in multi-fluid cosmologies --- a 
structure which differs from that which emerges in single-fluid cosmologies in several 
crucial ways.

   
\section{The source tensor in multi-fluid cosmologies}\label{multifluus}


In this section, we derive an expression for the source tensor $S_{ij}$ in 
multi-fluid cosmologies.  We begin in Sects.~\ref{sect3a} and \ref{sect3b}
by expressing the second-order stress-energy tensor in terms of the relevant 
thermodynamic quantities.  In Sect.~\ref{source2fluids} we then use this form of 
$T^{(2)i}{}_j$ in order to derive a general expression for $S_{ij}$.  
In Sect.~\ref{subsec:simplified_kernels}, we then consider the simpler form to which 
this general expression reduces in the special case in which the anisotropic 
stress vanishes.

Even at first glance, it is apparent that the case of multiple fluids will 
not be a trivial generalization of the single-fluid case.  For example, 
the relations in Eq.~(\ref{singlefluidfirstorder1})
allow us to eliminate $\rho^{(1)}, v_i^{(1)}$, and $\Pi^{(1)\,i}_{\;\;\;j}$ in favor of 
$\Phi$, $\Psi$, and their derivatives.  The source in 
Eq.~(\ref{source1}) can therefore be expressed purely in terms of $\Phi$ and $\Psi$. 
For a multi-fluid cosmology with $N$ fluids, on the other hand, we have $N$ velocity 
vectors, which can be decomposed into one center-of-mass degree of freedom and $N-1$ 
relative velocities.  For $N>1$, we therefore cannot simply eliminate the velocities in favor 
of the potentials.  For similar reasons, we also cannot eliminate the anisotropic-stress 
degrees of freedom.

\subsection{Two-fluid stress-energy tensor and thermodynamic quantities\label{sect3a}}

We consider two fluids, labeled by the index $A\in\{M,\gamma\}$.
We use the labels $M$ and $\gamma$ to denote these two fluids in 
anticipation that we shall later apply the formalism we develop in this section
to the case of a matter/radiation stasis and use these labels to refer to 
the matter and radiation components involved in the stasis dynamics, respectively.
However, we emphasize that our treatment in this section is completely general and 
does not depend on the identities of these two components.

Given our two fluids, the total stress-energy tensor is given by
\begin{equation}
  T^\mu{}_\nu ~=~ \sum_A(T_A)^\mu{}_{\nu}~.
\label{eq:TmunuTwoFluids}
\end{equation}
The total background energy, total pressure, and effective equation-of-state 
parameter for the universe as a whole are respectively given by
\begin{equation}
  \rho^{(0)} = \sum_A \rho_{A}^{(0)}\,,~~
  P^{(0)} =\sum_A  P_{A}^{(0)}\, ,~~
  \expt{w} \equiv \frac{P^{(0)}}{\rho^{(0)}}\,,
  \label{eq:rhoNoughtPNoughtBracketw}
\end{equation}
whereas the equation-of-state parameters for the individual fluids are given by
\begin{equation}
  w_A ~\equiv~ \frac{P_{A}^{(0)}}{\rho_{A}^{(0)}}~.
\end{equation}

We shall assume that the universe has already entered a multi-fluid stage.  
At this point we  shall remain agnostic about the duration of the multi-fluid era.
Moreover, we shall approximate the transition from this epoch to the standard 
radiation-dominated epoch that follows it as instantaneous.  As such, the $w_A$ will be 
assumed constant in our analysis, although our treatment in this section does 
not depend on their actual values.  For later use, it will be convenient to introduce 
the inertial weights
\begin{eqnarray}
  W_A &\equiv& \rho_A^{(0)}+P_A^{(0)} \nonumber \\
  W &\equiv& \rho^{(0)}+P^{(0)} ~=~  \sum_A W_A~.
  \label{eq:W_and_WA_defs}
\end{eqnarray}

At first order, the energy density and pressure are given by
\begin{equation}
  \rho^{(1)} = \sum_A
\rho_{A}^{(1)}\,,
~~~~
  P^{(1)} =
  \sum_A P_{A}^{(1)}\, , 
\end{equation}
while the sound speeds $c_{s,A}^2$ for the individual fluid components are defined via 
the relation
\begin{equation}
  P_A^{(1)}~=~c_{s,A}^2\,\rho_A^{(1)}~.
\end{equation}

The effective adiabatic sound speed for the multi-fluid system as a whole is given by
\begin{equation}
  c_s^2~\equiv~ \frac{{P^{(0)}}'}{{\rho^{(0)}}'}~,
\end{equation}
while  the non-adiabatic pressure perturbation is given by
\begin{eqnarray}
  P_{\rm nad}^{(1)} &\equiv& 
  P^{(1)}-c_s^2\,\rho^{(1)} \nonumber \\
    &=& \!\sum_A(c_{s,A}^2-c_s^2)\,\rho_A^{(1)}~.
  \label{eq:Pnad}
\end{eqnarray}

\subsection{First-order Einstein constraints\label{sect3b}}

Given that gravity couples to the total stress-energy of our multi-fluid system, 
the Einstein equations constrain the total energy and pressure.  We can then write down 
the relevant constraints for the total first-order energy density, pressure, 
and stress, as well as the bulk velocity of the two-fluid system.  

The expression for the perturbation to the total energy-density which follows 
from the Einstein equations at first order is
\begin{equation}\label{eq:rho1}
  \rho^{(1)} ~=~-\frac{2}{\kappa^2 a^2}
    \Big[3\Hc(\Hc\Phi-\Psi')+\nabla^2\Psi\Big]~.
\end{equation}
For the corresponding constraint on the velocity, it will be useful to distinguish 
the total or bulk velocity $v$ from the relative or slip velocity $\Delta v$ between 
the fluids (equivalently, we could have worked in terms of $v_M$ and 
$v_\gamma$).  We shall deal with $v$ first and postpone our discussion of $\Delta v$ 
for later.

The bulk velocity can be expressed in terms of the momentum-weighted sum
\begin{equation}
  v_i ~\equiv~ \frac{1}{W}\sum_A W_A\,v_{A i}~.
  \label{eq:bulkVdef}
\end{equation}
We may use the expression for the total momentum which follows from the 
first-order Einstein equations in order to express the sum in this 
expression in terms of $\Phi$ and $\Psi$, yielding 
\begin{equation}
  v_i ~=~ -\frac{2}{\kappa^2 a^2 W}\,\partial_i(\Psi'+\Hc\Phi)~.
  \label{eq:viConstraint}
\end{equation}
We shall find it convenient in what follows to define the total velocity 
divergence $\theta$ and the velocity divergence $\theta_A$ for each individual 
cosmological component:
\begin{equation}
  \theta ~\equiv~ \partial_j v^j\,,\qquad
  \theta_A ~\equiv~ \partial_j v_A^j\,.
\end{equation}
Indeed, taking the divergence of Eq.~(\ref{eq:viConstraint}) yields
\begin{equation}\label{eq:thetaConstraintReal}
  \theta ~=~ -\frac{2}{\kappa^2 a^2 W}\,\nabla^2(\Psi'+\Hc\Phi)~.
\end{equation}
In Fourier space, we then obtain 
\begin{equation}
  \theta_{\kk} ~=~ \frac{2k^2}{\kappa^2 a^2 W}\,(\Psi'+\Hc\Phi)_{\kk}~.
  \label{eq:thetaConstraintK}
\end{equation}

Assuming that the first-order velocity perturbation is purely scalar 
(and hence irrotational),
we can 
relate $\theta_{\kk}$ to $v_{\kk}$ via
the longitudinal relation
\begin{equation}
  v_{\kk}^{\,j} ~=~ -i\,\frac{k^j}{k^2}\,\theta_{\kk}~.
  \label{eq:vtheta}
\end{equation}
Indeed, this relation is consistent with the relation $\theta_{\kk} = i k_j v_{\kk}^j$. 

Finally, in a similar way we can derive a first-order expression for the 
total anisotropic stress $\Pi^{(1)i}{}_j$.  This total stress which can be expressed 
in terms of a pressure-weighted sum of the first-order stresses $\Pi_A^{(1)i}{}_j$ for 
the individual fluids via the relation
\begin{equation}
  P^{(0)}\,\Pi^{(1)i}{}_j ~\equiv~ \sum_A
    P_M^{(0)}\Pi_A^{(1)i}{}_j~.
\label{eq:PiTotalDef}
\end{equation}
Thus, once again making use of the first-order Einstein equations in order to
express the sum in terms of $\Phi$ and $\Psi$, we find 
\begin{equation}
  \Pi^{(1)i}{}_j ~=~
    -\frac{1}{\kappa^2 a^2 P^{(0)}}
    \left(\partial^i\partial_j
    -\frac{1}{3}\delta^i_j\nabla^2\right)(\Phi-\Psi)~.
  \label{eq:PiEinstein}
\end{equation}

\subsection{Two-fluid source tensor $S_{ij}$}\label{source2fluids}

We are now equipped to derive an expression for the source tensor $S_{ij}$ in
Eq.~(\ref{eq:tensorEOM}) for a generic two-fluid system.  In general,
we have seen that $S^{i}{}_j$ includes both a contribution from the geometric
term $Q^{i}{}_j$ defined in Eq.~(\ref{eq:QijOriginal}) and a contribution
from the second-order stress-energy tensor $T^{(2)i}{}_j$.
In order to evaluate the latter contribution, we begin by noting that 
$T^{(2)i}{}_j$ can be decomposed as 
\begin{eqnarray} \label{eq:T2ijsum}
  T^{(2)i}{}_j &=&\sum_{A}\Big[
    (\rho_A^{(0)}+P_A^{(0)})\,v_A^{(1)i}v^{(1)}_{A\,j} \nonumber \\
    & &~~+ P_A^{(0)}\Pi_A^{(2)i}{}_j
    +P_A^{(1)}\Pi_A^{(1)i}{}_j +P_A^{(2)}\delta^i_j \Big]~.~~~~~~~
  \label{eq:TijSecOrdMultiFluid}
\end{eqnarray}
When the TT projection is applied to this tensor, the $P_A^{(0)}\Pi_A^{(2)i}{}_j$ 
and $P_A^{(2)}\delta^i_j$ terms vanish.  What remains is a sum over terms
proportional to $v_A^{(1)i}v^{(1)}_{A\,j}$ and a sum over terms of the form 
$P_A^{(1)}\Pi_A^{(1)i}{}_j$.  We shall refer to the contributions to $S_{ij}$ 
which arise from these these two sums as $S_{ij}^{(v)}$ and $S_{ij}^{(\Pi)}$, 
respectively.

We begin by considering the overall contribution 
$S_{ij}^{(v)} = -\kappa^2a^2 \sum_A W_A v_{A i}v_{A j}$ 
from terms in $T^{(2)i}{}_j$ which involve the individual fluid velocities $v_A$.  
It can be shown that this contribution can be further decomposed into a piece involving 
the overall bulk velocity $v$ and a piece involving the overall slip velocity 
$\Delta v$, which we define as~\cite{Kumar:2024hsi}
\begin{equation}
  \Delta v_i ~\equiv~ v_{M i} - v_{\gamma i}~.
\end{equation}
From this definition and the definition of the bulk velocity 
in Eq.~(\ref{eq:bulkVdef}), we obtain the identity
\begin{equation}
  \sum_A  W_A \,v_{Ai}\, v_{Aj} ~=~ W v_i v_j + 
     \frac{W_M W_\gamma}{W}\,\Delta v_i\,\Delta v_j~.
\end{equation}
Using this identity, we find that $S^{(v)}_{ij}$ 
may be written as
\begin{equation}
  S^{(v)}_{ij} ~=~ S^{(v,{\rm bulk})} + S^{(v,{\rm slip})}~, 
  \label{eq:SvelSplit}
\end{equation}
where 
\begin{eqnarray}
  S^{(v,{\rm bulk})} &\equiv& -\kappa^2 a^2\,W v_i v_j
  \nonumber \\
  S^{(v,{\rm slip})} &\equiv& -\kappa^2 a^2\,
    \frac{W_M W_\gamma}{W}\,\Delta v_i\,\Delta v_j~.
  \label{eq:SijSlipAndSijBulk}
\end{eqnarray}

In order to express these contributions to the overall source term in Fourier space, 
we begin by noting that the Fourier transform of the product $v_i v_j$ 
can be expressed as 
\begin{equation}
  [v_i v_j](\kk) ~=~ -\int\frac{d^3p}{(2\pi)^{3/2}}
    \frac{p_i q_j}{p^2 q^2}\,\theta_{\pp}\theta_{\bm{q}}~,
\end{equation}
where $\bm{q} = \bm{k}-\bm{p}$ and where we have used Eq.~(\ref{eq:vtheta}) 
in order to express each $v_{\kk}^i$ in terms of the corresponding $\theta_{\kk}$.
Thus, we find that the bulk-velocity term in Eq.~(\ref{eq:SijSlipAndSijBulk})
may be written as
\begin{equation}
S^{(v,\mathrm{bulk})}_{ij}(\kk) ~=~ \kappa^2 a^2 W
    \int\frac{d^3p}{(2\pi)^{3/2}}\frac{p_i q_j}{p^2 q^2}\,
    \theta_{\pp}\theta_{\bm{q}}~.
\label{the_expression}
\end{equation}
Recognizing that $S=-4e^{ij}S_{ij}$ and $e^{ij}p_iq_j=-e^{ij}p_ip_j$, we find that substituting 
Eq.~(\ref{the_expression}) into Eq.~(\ref{eq:SkDef}) and contracting 
the indices according to Eq.~(\ref{eq:transversalityIdentity}) yields
the corresponding expression in Fourier space: 
\begin{eqnarray}
&&  S^{(v,\mathrm{bulk})}(\kk,\eta) ~=~
      4\kappa^2 a^2\,W \nonumber \\
    &&~~~~~~~~~~~~\times \int\frac{d^3p}{(2\pi)^{3/2}}
  \frac{e^{ij}(\kk)p_i p_j}{p^2 q^2}\,
  \theta_{\pp}(\eta)\theta_{\bm{q}}(\eta)~.~~~~~~~~~~~~ 
  \label{eq:SvBulk}
\end{eqnarray}

The Fourier transform of the slip-velocity term in 
Eq.~(\ref{eq:SijSlipAndSijBulk}) can be evaluated in a similar manner.
In order to express the result in a more compact form, we note that
the slip velocity can be expressed as 
\begin{equation}\label{eq:dvdtheta}
  \Delta v_i(\pp) ~=~ -i\frac{p_i}{p^2}\Delta\theta_{\pp}~,
\end{equation}
where $\Delta\theta_{\pp}$ is the Fourier transform of the slip-velocity
divergence $\Delta \theta \equiv \partial^j \Delta v_j = \theta_M-\theta_\gamma$.
Indeed, substituting the expression for $S^{(v,{\rm slip})}$ into Eq.~(\ref{eq:SkDef}) 
and contracting the indices according to Eq.~(\ref{eq:transversalityIdentity}), 
we find that the corresponding expression in Fourier space is
\begin{eqnarray} 
 && S^{(v,\mathrm{slip})}(\kk,\eta) ~=~
      4\kappa^2 a^2\frac{W_M W_\gamma}{W} \nonumber \\
    && ~~~~~\times \int\frac{d^3p}{(2\pi)^{3/2}}
      \frac{e^{ij}(\kk)p_i p_j}{p^2 q^2}\,
      \Delta\theta_{\pp}(\eta)\Delta\theta_{\bm{q}}(\eta)~.~~~~~~~
  \label{eq:SvSlip}
\end{eqnarray} 

We now consider the overall contribution 
$S_{ij}^{(\Pi)} = -\kappa a^2\sum_A P_A^{(1)}\Pi^{(1)}_{A ij}$ to $S_{ij}$ from 
terms in $T^{(2)i}{}_j$ which involve the anisotropic stress.  As with 
$S_{ij}^{(v)}$, we find that it is revealing to decompose this overall contribution 
into a sum of two pieces.  In particular, we note that if the quantity  
$P^{(1)}\Pi_{A ij}^{(1)}P_A^{(0)}/P^{(0)}$ is added to and subtracted from each 
term in the sum over $A$, our expression for $S_{ij}^{(\Pi)}$ may be rearranged 
to take the form
\begin{equation}
  S_{ij}^{(\Pi)} ~=~ -\kappa^2a^2
      \frac{P^{(1)}}{P^{(0)}}\sum_A P_A^{(0)}\Pi_{A ij}^{(1)} 
      -\kappa^2a^2 \Delta^{(\Pi)}_{ij}~,~~
  \label{eq:PiSplit}
\end{equation}
where we have defined
\begin{equation}
  \Delta^{(\Pi)}_{ij} ~\equiv~
    \sum_A\left(P_A^{(1)} 
    -\frac{P^{(1)}}{P^{(0)}}P_A^{(0)}\right)\Pi_{A ij}^{(1)}~.
  \label{eq:PiMixDef}
\end{equation}
The first term in Eq.~(\ref{eq:PiSplit}) can be simplified through the 
use of Eq.~(\ref{eq:PiTotalDef}), which implies that
\begin{eqnarray}
  \frac{P^{(1)}}{P^{(0)}}\sum_A P_A^{(0)}\Pi_{A ij}^{(1)} ~=~
     P^{(1)}\Pi_{ij}^{(1)}~.
\end{eqnarray}
Thus, using Eq.~(\ref{eq:PiEinstein}) in order to express 
the factor of $\Pi_{A ij}^{(1)}$ in this expression in terms of 
the potentials $\Phi$ and $\Psi$, we find that $S_{ij}^{(\Pi)}$ can be
decomposed as 
\begin{equation}
  S_{ij}^{(\Pi)} ~=~ S_{ij}^{(\Pi,{\rm fixed})} + S_{ij}^{(\Pi,{\rm mix})}~,    
\end{equation}
where we have defined
\begin{eqnarray}
  S_{ij}^{(\Pi,{\rm fixed})} &\equiv& \frac{P^{(1)}}{P^{(0)}}
    \left(\partial_i\partial_j-\frac13\delta_{ij}\nabla^2\right)(\Phi-\Psi)  
    \nonumber \\
  S_{ij}^{(\Pi,{\rm mix})} &\equiv& - \kappa^2 a^2\,\Delta^{(\Pi)}_{ij}~.
  \label{eq:SijPiFixAndSijPiMix}
\end{eqnarray}
Furthermore, we note that the factors of $P^{(1)}/P^{(0)}$ which appear within
these expressions can also be expressed in terms of $\Phi$, $\Psi$, 
and $P_{\rm nad}^{(1)}$.  In particular, by rearranging Eq.~(\ref{eq:Pnad})
and eliminating $\rho^{(1)}$ through use of Eq.~(\ref{eq:rho1}), we find 
\begin{equation}
  P^{(1)}~=~
    -\frac{2c_s^2}{\kappa^2 a^2}\Big[3\Hc(\Hc\Phi-\Psi')+\nabla^2\Psi\Big]
    +P_{\rm nad}^{(1)}~.
  \label{eq:P1expanded}
\end{equation}
Moreover, since the Friedmann equation implies 
\begin{equation}
  P^{(0)} ~=~ \expt{w}\rho^{(0)} ~=~ \frac{3\expt{w}\mathcal{H}^2}{\kappa^2a^2}~,  
\end{equation}
we find 
\begin{eqnarray}
  \frac{P^{(1)}}{P^{(0)}} &=&
    -\frac{2c_s^2}{3\expt{w}\Hc^2}\Big[3\Hc(\Hc\Phi-\Psi')+\nabla^2\Psi\Big]
    \nonumber \\ 
    && ~~~~+\frac{\kappa^2 a^2}{3\expt{w}\Hc^2}\,P_{\rm nad}^{(1)}~.
  \label{eq:P1overP0}
\end{eqnarray}

In order to express the results in Eqs.~(\ref{eq:SijSlipAndSijBulk}) in 
Fourier space, we begin by noting that the term in 
$S^{(\Pi,\mathrm{fixed})}_{ij}$ proportional to $\delta_{ij}$ vanishes when 
contracted with $e^{ij}$.  Simplifying the remaining term through use of 
the relation
\begin{equation}
  \left[\partial_i\partial_j(\Phi-\Psi)\right](\bm{q}) ~=~ 
    -q_i q_j(\Phi-\Psi)_{\bm{q}}~,
\end{equation}
we find 
\begin{eqnarray}
  S^{(\Pi,\mathrm{fixed})}(\kk,\eta) &=&
    4\int\frac{d^3p}{(2\pi)^{3/2}}\,e^{ij}(\kk)\,p_i p_j\,
    \frac{P^{(1)}_{\pp}(\eta)}{P^{(0)}}~~~~~~~\nn\\
    &~&\quad\quad\quad\quad\times (\Phi-\Psi)_{\bm{q}}(\eta)~,
\end{eqnarray}
where $P^{(1)}_{\pp}$ denotes the Fourier transform of $P^{(1)}$.
Expressing Eq.~(\ref{eq:P1overP0}) in Fourier space and defining
\begin{equation}
  \mathcal{B}_{\pp}(\eta) ~\equiv~ 
    3\Hc\big(\Hc\Phi_{\pp}-\Psi'_{\pp}\big)-p^2\Psi_{\pp}~,
\end{equation}
we find that $S^{(\Pi,\mathrm{fixed})}(\kk,\eta)$
can be written in the form
\begin{eqnarray}
  && S^{(\Pi,\mathrm{fixed})}(\kk,\eta) ~=~
      4\int\frac{d^3p}{(2\pi)^{3/2}}\,e^{ij}(\kk)p_i p_j
     ~~~~~~~~ \nonumber \\
    && ~~~~~~~~ \times\Bigg[
      \frac{2c_s^2}{3\expt{w}\Hc^2}\,\mathcal{B}_{\pp}(\eta)
      +\frac{\kappa^2 a^2}{3\expt{w}\Hc^2}\,
        P^{(1)}_{\rm nad,\pp}(\eta)
    \Bigg]~~~~~ \nonumber \\
    && ~~~~~~~~  \times (\Phi-\Psi)_{\bm{q}}(\eta)~.
  \label{eq:SPiFixedFinal}
\end{eqnarray} 
By the same token, the expression for $S_{ij}^{(\Pi,{\rm mix})}$ in Fourier space
is simply 
\begin{equation}
  S^{(\Pi,\mathrm{mix})}(\kk,\eta) ~=~
    4\kappa^2 a^2\,e^{ij}(\kk)\,\Delta^{(\Pi)}_{ij}(\kk,\eta)~,
  \label{eq:SPimix}
\end{equation}
where $\Delta^{(\Pi)}_{ij}(\kk,\eta)$ denotes the Fourier transform of 
$\Delta^{(\Pi)}_{ij}$

In summary, the source tensor for a generic two-fluid cosmological system
can be expressed as
\begin{eqnarray}
  S_{ij} &=& Q_{ij} + S_{ij}^{(v,{\rm bulk})} + S_{ij}^{(v,{\rm slip})}
    \nonumber \\
    & &~~~~~~~~ + S_{ij}^{(\Pi,{\rm fixed})} + S_{ij}^{(\Pi,{\rm mix})}~,
  \label{eq:SijFullReal}
\end{eqnarray}
where expressions for the individual contributions can be found in 
Eqs.~(\ref{eq:QijOriginal}), (\ref{eq:SijSlipAndSijBulk}),
and~(\ref{eq:SijPiFixAndSijPiMix}).  The corresponding expression in 
Fourier space is given by
\begin{eqnarray}
  S(\kk,\eta) &=&
      S^{(\mathrm{geom})}(\kk,\eta)
      + S^{(v,\mathrm{bulk})}(\kk,\eta) + S^{(\mathrm{slip})}(\kk,\eta) 
      \nonumber \\
    &&~ +S^{(\Pi,\mathrm{fixed})}(\kk,\eta)
      +S^{(\Pi,\mathrm{mix})}(\kk,\eta)~,~~~~~~
  \label{eq:SfullFinal}
\end{eqnarray}
where $S^{(v,\mathrm{bulk})}(\kk,\eta)$, $S^{(\mathrm{slip})}(\kk,\eta)$,
$S^{(\Pi,\mathrm{fixed})}(\kk,\eta)$, and $S^{(\Pi,\mathrm{mix})}(\kk,\eta)$ are
given in Eqs.~(\ref{eq:SvBulk}), (\ref{eq:SvSlip}), (\ref{eq:SPiFixedFinal}), 
and~(\ref{eq:SPimix}), respectively.  The geometric contribution 
$S^{(\mathrm{geom})}(\kk,\eta)$, which is obtained by contracting $Q_{ij}$ 
with $e^{ij}$ and taking the Fourier transform, can be written in the form
\begin{equation}\label{eq:SgeomKernel}
S^{(\mathrm{geom})}(\kk,\eta) ~=~
  4\int\frac{d^3p}{(2\pi)^{3/2}}\,e^{ij}(\kk)\,p_i p_j\,
  \mathcal{K}_{\mathrm{geom}}(\pp,\bm{q};\eta)\, ,
\end{equation}
where the expression for ${\cal K}_{\rm geom}$ follows from Eq.~\eqref{eq:QijOriginal}
and is not shown here for the  sake of brevity.

\subsection{Simplified geometric and velocity kernels for the \texorpdfstring{$\Phi=\Psi$}
\, special case\label{subsec:simplified_kernels}}

The expression in Eq.~(\ref{eq:SfullFinal}) is general and
when inserted into Eq.~(\ref{eq:modeEq}) yields a complete equation of motion 
for the scalar-induced tensor perturbation $h_{\kk}$ in the two-fluid 
case --- one which incorporates the effects of bulk flow, slip velocity, 
non-adiabatic pressure, and anisotropic stress. 
However, for simplicity in what follows, we shall follow 
Ref.~\cite{Baumann:2007zm} and neglect the contribution from terms associated 
with the anisotropic stress, as their contribution is typically subdominant. 
Thus, we shall henceforth approximate $\Pi_{ij} \approx 0$, whereupon we have 
\begin{equation}
  S(\bm{k},\eta) \,\approx\, S^{(\mathrm{geom})}(\bm{k},\eta)
    + S^{(v,\mathrm{bulk})}(\bm{k},\eta) 
    + S^{(\mathrm{slip})}(\bm{k},\eta)\,.
\label{eq:SdecompPhiEqPsi}
\end{equation} 

Moreover, not only do $S^{(\Pi,\mathrm{fixed})}(\kk,\eta)$ and 
$S^{(\Pi,\mathrm{mix})}(\kk,\eta)$ vanish when $\Pi_{ij} = 0$, but
the non-vanishing contributions to $S(\bm{k},\eta)$ further simplify as well 
due to the fact that $\Phi$ and $\Psi$ become equal in the 
special case in which $\Pi_{ij}$ vanishes.  First, we note that the geometric 
quadratic source term simplifies in this case to 
\begin{equation}
  Q_{ij} ~=~ 4\Phi\,\partial_i\partial_j\Phi +
    2\,\partial_i\Phi\,\partial_j\Phi~.
\label{eq:QijPhiEqPsi}
\end{equation}
Thus, taking the Fourier transform of this expression, contracting with the 
$e^{ij}$, and repeatedly using the transversality identity 
in Eq.~(\ref{eq:transversalityIdentity}), we obtain
\begin{equation}
  S^{(\mathrm{geom})}(\bm{k},\eta) ~=~
    8\int\frac{\mathrm{d}^3 p}{(2\pi)^{3/2}}\,
    e^{ij}(\bm{k})\,p_i p_j\,
    \Phi_{\bm{p}}(\eta)\,
    \Phi_{\bm{k}-\bm{p}}(\eta)~.~
  \label{eq:SgeomPhiEqPsi}
\end{equation}

The bulk- and slip-velocity contributions in Eqs.~(\ref{eq:SvBulk}) 
and~(\ref{eq:SvSlip}) also simplify considerably in the special case in which 
$\Pi_{ij}$ vanishes.  Indeed, in this case we find
\begin{eqnarray}
  S^{(v,\mathrm{bulk})}(\bm{k},\eta) &=&
      4\,\kappa^2 a^2 W \!
      \int \! \frac{\mathrm{d}^3 p}{(2\pi)^{3/2}}\,
      e^{ij}(\bm{k})\,p_i p_j \nonumber \\
    && ~~~~~~~~~~~~~\times \frac{\theta_{\bm{p}}(\eta)\,
      \theta_{\bm{k}-\bm{p}}(\eta)}{p^2\,|\bm{k}-\bm{p}|^2}
      \nonumber \\
  S^{(v,\mathrm{slip})}(\bm{k},\eta) &=&
      4\,\kappa^2 a^2 \frac{W_M W_\gamma}{W} \!
      \int \! \frac{\mathrm{d}^3 p}{(2\pi)^{3/2}}\,
      e^{ij}(\bm{k})\,p_i p_j ~~\nonumber  \\
      && ~~~~~~~~~~~~~\times
    \frac{\Delta\theta_{\bm{p}}(\eta)\,
      \Delta\theta_{\bm{k}-\bm{p}}(\eta)}{p^2\,|\bm{k}-\bm{p}|^2}\,.
      \nonumber \\
  \label{eq:SbulkAndSslipPhiEqPsi}
\end{eqnarray}
We will use these results when calculating the power spectrum and GW energy density for our two-fluid cosmology in Sect.~\ref{powerspectrumenergydens}.

Before proceeding, we point out that it is useful to think of general GWs 
from two slightly different perspectives. The first is that tensor modes, being 
sourced from putative events in the early universe, serve as probes of such events. 
The second is that tensor modes, once produced, evolve freely, with their amplitude 
redshifting according to the background equation of state; the resulting spectrum 
can thus retain memory of the cosmological epoch during which the mode entered the 
horizon. In this sense, GWs serve as witnesses to the history of cosmic expansion 
between the source event and the present day. The two perspectives therefore 
emphasize the complementary roles of GWs as \textit{source probes} and as 
\textit{historical witnesses}. For reviews, we refer to 
Refs.~\cite{Batell:2024dsi, Caldwell:2022qsj, Allahverdi:2020bys, Kane:2015jia}. 

For SIGWs, the two roles --- source probes and historical 
witness --- are not cleanly separable: the primordial scalar perturbations provide the initial 
seeds for these GWs, but the induced tensor modes are sourced by quadratic combinations of these 
scalar perturbations as the universe evolves.  The intervening cosmological history is therefore 
not merely a passive background through which GWs that have already been produced then propagate --- 
it also determines the active second-order source itself. 

This distinction is central to our analysis. As we have seen, the tensor source in 
the case of SIGWs is built from first-order scalar perturbations, fluid velocities, 
entropy perturbations, and anisotropic stresses. From Eq.~\eqref{singlefluidfirstorder1} 
it is clear that in the usual single-fluid case, the first-order Einstein constraints 
allow the relevant density and velocity perturbations to be eliminated in favor of the 
scalar gravitational potentials.  The induced tensor source can therefore be written 
entirely in terms of those potentials and their derivatives, as can be seen in 
Eq.~\eqref{singlefluidsignal}.  In a multi-fluid universe, by contrast, gravity 
constrains only total combinations of the perturbations.  Individual component 
velocities, relative velocities, entropy perturbations, and anisotropic stresses 
can remain independent physical degrees of freedom.  As a result, the SIGW spectrum 
depends not only on the averaged equation of state of the universe, but also on the 
internal perturbative structure of the cosmic medium, \ie, on its possible 
multi-fluid nature. These facts are evident from  Eq.~\eqref{eq:SfullFinal}.


\section{Tensor power spectrum and gravitational-wave energy density in multi-fluid cosmologies}\label{powerspectrumenergydens}


In this section, we first determine the tensor power spectrum and then use the result to
calculate the GW energy density generated at second order by scalar perturbations.  
The calculation proceeds in several steps.  First, in 
Sect.~\ref{subsec:SIGW_kernel_consistent}, we introduce a time dependent transfer kernel
in order to streamline our calculation of the unequal-time correlator for the source in 
terms of the primordial scalar power spectrum.  Next, in 
Sect.~\ref{subsec:SIGW_green_consistent}, we insert the source correlator into the 
Green's-function solution for the tensor equation, thereby obtaining our 
``master equations'' for the tensor power spectrum.  Finally, in 
Sect.~\ref{subsec:SIGW_Omega_consistent}, we calculate the corresponding GW energy density.

\subsection{Kernel representation of the tensor source\label{subsec:SIGW_kernel_consistent}}

We have already calculated the source tensor in previous sections. For the calculation 
of the tensor power spectrum using the source tensor, it is convenient to introduce a 
time-dependent transfer kernel. Our goal in this subsection is to derive this kernel function.

A tensor perturbation with comoving momentum $\bm{k}$ is sourced by pairs of scalar 
perturbations with momenta $\bm{p}$ and $\bm{q}=\bm{k}-\bm{p}$.  It is therefore 
standard to write the projected source as
\begin{equation}
  S(\bm{k},\eta) ~=~ \int \frac{d^3p}{(2\pi)^{3/2}}\,
    e^{ij}(\bm{k})\,p_i p_j\, \mathcal{F}(\bm{k},\bm{p},\eta)\,
    \psi_{\bm{p}}\psi_{\bm{q}}~.
\label{eq:source_full_kernel_consistent}
\end{equation}
Here $\psi_{\bm{k}}$ denotes the primordial scalar perturbation, while the
transfer kernel $\mathcal{F}(\bm{k},\bm{p},\eta)$ 
contains information about the time-evolution of $S(\bm{k},\eta)$ and includes 
the appropriate model-dependent normalization factors.  In a single-fluid cosmology, 
$\mathcal{F}(\bm{k},\bm{p},\eta)$ is determined by the scalar transfer function 
and its derivatives.  In a multi-fluid cosmology, as we have seen, it also 
receives contributions from the bulk velocity, from relative or slip velocities, 
from non-adiabatic pressure perturbations, and from anisotropic stresses.

In order to proceed further, it is useful to define the dimensionless ratios
\begin{equation}
  u~\equiv~ \frac{p}{k}\, , \quad
~~~ v~\equiv~ \frac{q}{k}\, ,
  \label{eq:define_uandv}
\end{equation}
and the dimensionless time variable
\begin{equation}
  x~\equiv~ k\eta~.
  \label{eq:uvx_def_consistent}
\end{equation}
For a statistically isotropic background, the transfer kernel 
$\mathcal{F}(\bm{k},\bm{p},\eta)$ depends on $\eta$ and on the vector quantities 
$\bm{p}$ and $\bm{q}$ only through the scalar combinations $u$, $v$, and the 
dimensionless time variable $x$.  Thus, in Eq.~(\ref{eq:source_full_kernel_consistent}), 
we may make the replacement $\mathcal{F}(\bm{k},\bm{p},\eta) ~\rightarrow~ f(u,v,x)$, 
where $f(u,v,x)$ is a function of only these scalar variables --- a function which we shall refer to as the 
reduced source kernel.  All dependence on the cosmological 
model, including the scalar evolution and possible multi-fluid effects, is encoded in 
$f(u,v,x)$.  

Moreover, in light of the manner in which we have decomposed 
$S(\bm{k},\eta)$ into a sum of three individual contributions $S_X(\bm{k},\eta)$ in 
Eq.~\eqref{eq:SdecompPhiEqPsi}, with $X\in \{\mathrm{geom},\mathrm{bulk},\mathrm{slip}\}$,
we may likewise define a corresponding reduced source kernel $f_X(u,v,x)$ for 
each of these three individual contributions via the relation
\begin{equation}
  S_X(\bm{k},\eta) ~=~ \int
    \frac{\mathrm{d}^3 p}{(2\pi)^{3/2}}\,
    e^{ij}(\bm{k})\,f_X(u,v,x)\,\,p_i \,p_j\, \psi_{\bm{p}}\,
    \psi_{\bm{q}}~.
\label{eq:sourceKernelDef}
\end{equation}

In order to evaluate the different $f_X(u,v,x)$ functions, we begin by noting that whenever the 
transfer kernel depends on $\eta$ only through $u$, $v$, and $x$, we may write
\begin{equation}
  \Phi_{\bm{p}}(\eta) ~=~ \psi_{\bm{p}}\,\Phi(ux)\, , \qquad
  \Phi_{\bm{q}}(\eta) ~=~
    \psi_{\bm{q}}\,\Phi(vx)\,,
  \label{eq:PhiTransferDef}
\end{equation}
where, as before, we have defined $\bm{q}= \bm{k}-\bm{p}$.  Accordingly, 
we may define dimensionless velocity-divergence transfer functions 
$\Theta(y)$ and $\Delta\Theta(y)$ of the dimensionless variable $y$ via
\begin{eqnarray}
  \theta_{\bm{p}}(\eta) &=& p\,\psi_{\bm{p}}\,\Theta(p\eta)
    ~=~ ku\,\psi_{\bm{p}}\,\Theta(ux)\, \nonumber \\
  \theta_{\bm{q}}(\eta) &=& q\,\psi_{\bm{q}}\,\Theta(q\eta)
    ~=~ kv\,\psi_{\bm{q}}\,\Theta(vx)\,,
  \label{eq:ThetaTransferDef}
\end{eqnarray}
together with
\begin{eqnarray}
  \Delta\theta_{\bm{p}}(\eta) &=&
    ku\,\psi_{\bm{p}}\,\Delta\Theta(ux) \nonumber \\
  \Delta\theta_{\bm{q}}(\eta) &=&
    kv\,\psi_{\bm{q}}\,\Delta\Theta(vx)~.
  \label{eq:DeltaThetaTransferDef}
\end{eqnarray}
Equivalently, we may also define corresponding transfer functions 
$\Theta_M(y)$ and $\Theta_\gamma(y)$ for the individual components via
\begin{eqnarray}
  \Theta_M(y) &\,\equiv\,& \Theta(y) 
    +\left(1-\frac{W_M}{W}\right)\Delta\Theta(y) \nonumber \\
  \Theta_\gamma(y) &\,\equiv\,& \Theta(y) 
    -\left(1-\frac{W_\gamma}{W}\right)\Delta\Theta(y)~.
  \label{eq:DeltaMandgammaTransferDef}
\end{eqnarray}

Inserting Eqs.~\eqref{eq:PhiTransferDef} through  \eqref{eq:DeltaThetaTransferDef} 
into Eqs.~(\ref{eq:SgeomPhiEqPsi}) and~(\ref{eq:SbulkAndSslipPhiEqPsi}), we obtain
\beqn
  f_{\mathrm{geom}}(u,v,x) &=& 8\,\Phi(ux)\,\Phi(vx)~\nonumber\\
  f_{\mathrm{bulk}}(u,v,x) &=&
    4\,\frac{\kappa^2 a^2 W}{k^2}\,
    \frac{\Theta(ux)\,\Theta(vx)}{uv}~\nonumber\\
 f_{\mathrm{slip}}(u,v,x) &=& 
    4\,\frac{\kappa^2 a^2}{k^2}\,
    \frac{W_M W_\gamma}{W}\,
    \frac{\Delta\Theta(ux)\,\Delta\Theta(vx)}{uv}~.\nonumber\\
\label{eq:triple}
\eeqn
The sum of these individual contributions, then, is the overall reduced source 
kernel $f(u,v,x)$.

The background coefficients appearing within the final two equations of 
Eq.~(\ref{eq:triple}) may be written in a more transparent form.  Using the 
Friedmann equation $\mathcal{H}^2=\kappa^2 a^2\rho/3$ and the relations
\begin{equation}
  W = (1+\expt{w})\rho~, \qquad 
  W_A ~=~(1+w_A)\rho\,\Omega_A~,
  \label{eq:WandFriedmann}
\end{equation}
which follow directly from the relations in Eq.~(\ref{eq:W_and_WA_defs}),
we find that
\begin{eqnarray}
  \frac{\kappa^2 a^2 W}{k^2} &=&
    3\,(1+\expt{w})\frac{\mathcal{H}^2}{k^2}~. \nonumber \\
  \frac{\kappa^2 a^2}{k^2}\frac{W_M W_\gamma}{W} &=&
    \frac{3\mathcal{H}^2}{k^2}\,
    \frac{(1+w_M)(1+w_\gamma)\Omega_M\Omega_\gamma}
    {1+\langle w\rangle}~.\nonumber \\
 \label{eq:bulkCoeffRewrite}
\end{eqnarray}
It therefore follows that
\begin{eqnarray}
  f_{\mathrm{geom}}(u,v,x) &=& 8\,\Phi(ux)\,\Phi(vx)~\nonumber\\
  f_{\mathrm{bulk}}(u,v,x) &=&
    12\,(1+\expt{w})\frac{\mathcal{H}^2}{k^2}\,
    \frac{\Theta(ux)\,\Theta(vx)}{uv}~\nonumber\\
 f_{\mathrm{slip}}(u,v,x) &=& 
    \frac{12\mathcal{H}^2}{k^2}\,
    \frac{(1+w_M)(1+w_\gamma)\Omega_M\Omega_\gamma}
    {1+\langle w\rangle} \nonumber \\ & & ~~~~\times
    \frac{\Delta\Theta(ux)\,\Delta\Theta(vx)}{uv}~.
\label{eq:triple_simplified}
\end{eqnarray}

The expressions in Eq.~\eqref{eq:triple_simplified} make explicit the structure of the 
simplified source in the case in which $\Phi=\Psi$.  In particular, we observe that 
$f_{\rm geom}(u,v,x)$ is bilinear in the gravitational potential, while $f_{\rm bulk}(u,v,x)$ 
and $f_{\rm slip}(u,v,x)$ are each bilinear in the corresponding velocity-divergence 
transfer function.

\subsection{Green's-function solution and source correlator\label{subsec:SIGW_green_consistent}}

Equipped with our expressions for $f_{\mathrm{geom}}(u,v,x)$, $f_{\mathrm{bulk}}(u,v,x)$, and 
$f_{\mathrm{slip}}(u,v,x)$, we are now ready to proceed to the next steps in evaluating 
the tensor power spectrum, namely obtaining the relevant Green's function and writing down the 
source correlator.

We solve Eq.~\eqref{eq:modeEq} by first removing the Hubble-friction term.  Defining
\begin{equation}
v_{\bm{k}}\, \equiv\, a\,h_{\bm{k}}\,,
\label{eq:v_redefinition_consistent}
\end{equation}
we find that 
the tensor equation becomes
\begin{equation}
  v_{\bm{k}}''+ \left(k^2-\frac{a''}{a}\right)v_{\bm{k}}
    ~=~ a\,S(\bm{k},\eta)~.
  \label{eq:v_equation_consistent}
\end{equation}
The corresponding retarded Green's function $g_k(\eta;\tilde{\eta})$ satisfies
\begin{equation}
  g_k''(\eta;\tilde{\eta}) +
    \left(k^2-\frac{a''}{a}\right)g_k(\eta;\tilde{\eta}) ~=~
    \delta(\eta-\tilde{\eta})~.
  \label{eq:green_equation_consistent}
\end{equation}
The sourced tensor mode is then
\begin{equation}
  h_{\bm{k}}(\eta) ~=~ \frac{1}{a(\eta)}
    \int_{\eta_{\rm init}}^{\eta} d\tilde{\eta}\,
    g_k(\eta;\tilde{\eta})\,
    a(\tilde{\eta})\, S(\bm{k},\tilde{\eta}_{\rm init})~.
  \label{eq:h_solution_consistent}
\end{equation}
The lower limit $\eta_{\rm init}$ denotes the initial time at which the sourced 
evolution starts. 

The tensor power spectrum is defined by
\begin{equation}
  \big\langle h_{\bm{k}}(\eta)\,h_{\bm{K}}(\eta)\big\rangle ~=~ 
    \frac{2\pi^2}{k^3}\, \delta^{(3)}(\bm{k}+\bm{K})\, P_h(k,\eta)~.
  \label{eq:Ph_definition_consistent}
\end{equation}
According to this convention, $P_h$ represents the contribution to the power 
spectrum associated with one of the two independent tensor-mode polarizations.
The total gravitational-wave energy density is simply the sum of these two 
individual contributions.

Substituting Eq.~\eqref{eq:h_solution_consistent} into Eq.~\eqref{eq:Ph_definition_consistent} 
yields
\beqn 
&& \!\!\big\langle h_{\bm{k}}(\eta) h_{\bm{K}}(\eta)\big\rangle ~=~
     \int_{\eta_{\rm init}}^{\eta}  d\eta_1
    \int_{\eta_{\rm init}}^{\eta}  d\eta_2\, \frac{a(\eta_1)a(\eta_2)}{a^2(\eta)} \nonumber\\
    &&~~~~~~~~\times \, g_k(\eta;\eta_1)\, g_K(\eta;\eta_2)\,
    \big\langle
    S(\bm{k},\eta_1)\, S(\bm{K},\eta_2) \big\rangle~.~~~~~~~~~
\label{eq:hh_from_SS_consistent}
\eeqn
Assuming that the primordial scalar perturbations are statistically homogeneous,
isotropic, and Gaussian, we have
\begin{equation}
  \big\langle \psi_{\bm{k}}\psi_{\bm{K}}\big\rangle ~=~
    \frac{2\pi^2}{k^3}\, \delta^{(3)}(\bm{k}+\bm{K})\,
    \mathcal{P}_{\psi}(k)~,
  \label{eq:psi_ps_consistent}
\end{equation}
where $\mathcal{P}_{\psi}(k)$ is the dimensionless primordial scalar power spectrum. 

Applying Wick's theorem to the four-point function that follows from the expression in 
Eq.~(\ref{eq:source_full_kernel_consistent}), we find that the unequal-time source 
correlator can be written as
\begin{equation}
\big\langle
  S(\bm{k},\eta_1)\, S(\bm{K},\eta_2) \big\rangle ~=~
    \frac{2\pi^2}{k^3}\, \delta^{(3)}(\bm{k}+\bm{K})\,
    \mathcal{I}(k;\eta_1,\eta_2)~,
  \label{eq:SS_compact_consistent}
\end{equation}
where we have defined
\begin{equation}
\begin{aligned}
  \mathcal{I}(k;\eta_1,\eta_2) &~\equiv~
    \frac{k^4}{4}\int_0^\infty dv \int_{|1-v|}^{1+v} du\,
    \mathcal{P}_{\psi}(ku)\, \mathcal{P}_{\psi}(kv) \\[2pt]
  &\quad\times\,\Lambda^2(u,v)\,
    f(u,v,x_1)\, f(u,v,x_2)~.
  \end{aligned}
  \label{eq:I_kernel_consistent}
\end{equation}
Here we have written $x_j = k\eta_j$ and defined the quantity
\begin{equation}
  \Lambda(u,v) ~\equiv~
    \frac{4v^2-\big(1+v^2-u^2\big)^2}{4uv}~.
  \label{eq:Lambda_def_consistent}
\end{equation}
We note that the limits of integration for $u$ and $v$ in 
Eq.~\eqref{eq:I_kernel_consistent} follow from the ranges   
\begin{equation}
  0<v<\infty~, \qquad |1-v|\le u\le 1+v~,
\label{eq:triangle_domain_consistent}
\end{equation}
as implied by the definitions in Eq.~\eqref{eq:define_uandv}.
We also note that the factor of $k^4/4$ appearing in this same equation
follows from the two Gaussian Wick contractions, from the
azimuthal integral, and from the fact that $e_{ij}^\lambda$ 
is normalized such that $e^\lambda_{ij}e^{ij}_\lambda=1$,
where $\lambda=\{+,\times\}$ indicates the two polarizations.
More details are provided in Appendix~\ref{app:numerical_approximation}.

Eq.~\eqref{eq:I_kernel_consistent} separates the unequal-time source correlator 
into three ingredients: the primordial scalar spectrum $\mathcal{P}_{\psi}$, the geometric 
transverse-traceless projection $\Lambda^2(u,v)$, and the reduced source function $f(u,v,x)$
which encapsulates information about the dynamical evolution of the source.
Using Eqs.~\eqref{eq:hh_from_SS_consistent} through \eqref{eq:I_kernel_consistent} and comparing 
with the definition in Eq.~\eqref{eq:Ph_definition_consistent}, we obtain the induced 
tensor power spectrum
\begin{equation}
\begin{aligned}
  P_h(k,\eta) &~=~ \int_0^\infty dv \int_{|1-v|}^{1+v} du\,
    \mathcal{P}_{\psi}(ku)\, \mathcal{P}_{\psi}(kv) \\[2pt]
  &~~~~~~~\times~
    \Lambda^2(u,v)\, \mathcal{T}^2(u,v,k,\eta)~,
  \end{aligned}
  \label{eq:Ph_master_consistent}
\end{equation}
where the time-integrated transfer function is given by
\begin{equation}
  \mathcal{T}(u,v,k,\eta) ~\equiv~
    \frac{k^2}{2a(\eta)} \int_{\eta_{\rm init}}^{\eta} d\tilde{\eta}~
    a(\tilde{\eta})\, g_k(\eta;\tilde{\eta})\,
    f(u,v,k\tilde{\eta})~.
  \label{eq:T_transfer_consistent}
\end{equation}
Eq.~\eqref{eq:Ph_master_consistent} then gives the spectrum for one polarization. We assume 
the two polarizations have equal power and include a factor of two for the total tensor 
power spectrum in the subsequent equations.

Eqs.~\eqref{eq:Ph_master_consistent} and \eqref{eq:T_transfer_consistent} are the master 
formulae for SIGWs in this notation.  The primordial statistics enter through 
$\mathcal{P}_{\psi}$;  the angular dependence of the transverse-traceless projection is 
contained in $\Lambda^2(u,v)$;  and the cosmological evolution is encoded in the reduced 
source kernel $f(u,v,x)$ together with the tensor Green's function.

\subsection{Gravitational-wave energy density\label{subsec:SIGW_Omega_consistent}}

We now turn to the calculation of the gravitational-wave energy density.
It is convenient to focus on the gravitational-wave energy density per logarithmic 
interval in comoving wavenumber $k$, \ie, 
\begin{equation}
  \Omega_{\rm GW}(k,\eta) ~\equiv~
    \frac{1}{\rho_{\rm tot}(\eta)}
    \frac{d\rho_{\rm GW}}{d\ln k}~.
  \label{eq:OmegaGW_definition_consistent}
\end{equation}
For subhorizon modes, this is related to the oscillation-averaged tensor power 
spectrum by
\begin{equation}
  \Omega_{\rm GW}(k,\eta) ~=~
    \frac{1}{24} \left(\frac{k}{\mathcal{H}(\eta)}\right)^2
    \overline{P_h(k,\eta)}~,
  \label{eq:OmegaGW_from_Ph_consistent}
\end{equation}
where the overline denotes an average over the rapid tensor oscillations discussed
in Appendix~\ref{app:numerical_approximation}.  
In Eq.~\eqref{eq:OmegaGW_from_Ph_consistent}, $P_h$ should be understood as the 
total tensor power summed over both polarizations.  If $P_h$ in 
Eq.~\eqref{eq:Ph_definition_consistent} is calculated for a single polarization, 
the polarization sum must be included before applying 
Eq.~\eqref{eq:OmegaGW_from_Ph_consistent}.

To compare the result with detector sensitivities, the comoving wavenumber is converted 
into the present-day physical frequency through
\begin{equation}
  f~=~\frac{k}{2\pi a_0}~.
  \label{eq:f_of_k_consistent}
\end{equation}
For $a_0=1$, this becomes 
\begin{equation}
  f ~\approx ~ 1.55\times 10^{-15}
    \left(\frac{k}{\mathrm{Mpc}^{-1}}\right)\,\mathrm{Hz}~
\label{eq:f_of_k_numeric_consistent}
\end{equation}
in SI units.
Thus, once $P_h(k,\eta)$ is known, the corresponding stochastic gravitational-wave 
background can be expressed directly as $\Omega_{\rm GW}(f)$.

In summary, we have shown that the spectrum of scalar-induced GWs which arises 
within the context of multi-fluid cosmologies can be evaluated in a manner 
analogous to the established procedure~\cite{Kohri:2018awv,Domenech:2019quo} 
through which it is typically evaluated for single-fluid cosmologies, but with
appropriate modifications to account for intrinsically multi-fluid effects, such
as the slip-velocity contribution $S_{ij}^{(v,\mathrm{slip})}$ to the source tensor
for tensor perturbations.  Given a primordial scalar-perturbation spectrum 
$\mathcal{P}_{\psi}(k)$, we may determine the corresponding reduced source kernel
$f(u,v,x)$.  Given this source kernel, we may then calculate the corresponding 
tensor power spectrum $P_h(k,\eta)$ and, in turn, $\Omega_{\rm GW}(f)$.
Indeed, we have seen that the cosmological history affects the prediction
for $\Omega_{\rm GW}(f)$ only through $f(u,v,x)$ and through the manner in which 
the background expansion history influences the Green's function 
in Eq.~(\ref{eq:green_equation_consistent}).


\section{The case of cosmological stasis}\label{stasissec}


In this section we will apply the formalism developed in Sects.~\ref{multifluus} 
and~\ref{powerspectrumenergydens}  to determine the GW spectrum which 
results from an epoch of cosmological matter/radiation stasis.  Indeed, such a stasis epoch is a prime 
example of a fully multi-fluid cosmological epoch.  We will first explicitly calculate the 
gravitational potential and the divergence of the velocity perturbations of 
the two constituent fluids (in this case, matter and radiation).  Then, in 
Sect.~\ref{results}, 
 following the methods 
developed in 
Sect.~\ref{powerspectrumenergydens},
we will use 
the sources derived in Sect.~\ref{multifluus} --- with stasis input values that we will derive  here ---   
in order to numerically evaluate the corresponding stasis
tensor power spectrum.

Normally, in cosmologies with multiple fluids, the energy densities of fluids with 
different equation-of-state parameters dilute at different rates as the universe expands.
This causes the composition of the mixture (and thus the effective equation-of-state 
parameter of the universe) to vary with time, causing the universe quickly to become 
dominated by the energy component with the smallest equation-of-state parameter.
As a result, any potential feature of the GW power spectrum that might signal the 
existence of multiple fluids is limited to a short transient period of the expansion 
history --- one which spans only a small range in the GW frequency.

There is, however, one important exception to this expectation --- one which arises 
in cosmologies experiencing {\it cosmological stasis}\/~\cite{Dienes:2021woi,Dienes:2023ziv}.
During periods of cosmological stasis, the abundances of different cosmological energy 
components (\eg, matter, radiation, vacuum energy, \etc)\  with different equations of 
state each nevertheless remain constant over an extended time interval  --- all despite 
the effects of Hubble expansion.  As a result, the effective equation-of-state parameter 
$\langle w \rangle$ for a universe exhibiting cosmological stasis also remains constant.
This stasis phenomenon has been shown to arise in a variety of different 
cosmologies~\cite{Dienes:2021woi,Dienes:2022zgd,Dienes:2023ziv,Dienes:2024wnu,Barrow:1991dn,
Halverson:2024oir,Barber:2024vui,Barber:2024izt,Huang:2025odd,Dienes:2025tox,Long:2025wjw}, 
each of which is associated with a different possible theory of physics beyond the Standard Model.  
Thus cosmological stasis represents a wholly new theoretical possibility --- indeed a wholly 
new kind of cosmological {\it epoch}\/  --- which may have occurred in the early universe.

In each of these realizations of stasis, the different energy components present during 
the stasis epoch have different equations of state.  The effect of cosmological 
expansion on the energy density associated with each of these energy components is therefore
different.  However, while the universe is in stasis, these differences are precisely compensated by 
the effects of additional processes which transfer energy density {\it between}\/ different 
energy components.  As a result, the energy densities of all such energy components scale 
in the same manner with $a$ during stasis and their abundances remain fixed.  
Moreover, this balance between the effect of expansion on each energy component 
and the effect of these other processes during stasis is not the result of a fine-tuning.   
Rather, it is the consequence of a {\it global dynamical attractor}\/ to which the 
equations which govern the cosmological system give rise.  As a result, throughout 
a broad range of the model-parameter space associated with each of these realizations of 
stasis, the universe necessarily evolves toward (and often eventually enters) a stasis epoch 
regardless of the cosmological initial conditions.  Such realizations of stasis also typically 
come equipped with their own mechanisms through which the stasis epoch naturally ends.   

With a constant abundance for each energy component and a constant effective 
equation-of-state parameter for the universe as a whole, and with both of these features 
persisting across an extended time interval, the stasis phenomenon offers an ideal setup 
for studying gravitational waves in multi-fluid cosmologies.  In addition, stasis not 
only modifies the cosmological expansion history at the background level,
but also affects the cosmic evolution of corresponding perturbations. 
It is therefore important to examine the implications of cosmological stasis for
tensor perturbations as well.

\subsection{Matter/radiation stasis from a tower of states}

We begin our analysis by quickly reviewing one kind of cosmological stasis:  
a matter/radiation stasis emerging from a tower of decaying states. While 
there exist many other kinds of  cosmological stasis, each with its own origins 
and potential signatures, we shall limit our considerations of stasis-induced 
GWs to this special case.  Indeed, as demonstrated in Ref.~\cite{Dienes:2025tox}, 
matter/radiation stasis epochs arising from a tower of decaying states can lead 
to a significant enhancement in the growth of density perturbations.
Such epochs are therefore likely to introduce new features into the corresponding 
GW spectrum as well.

Within stasis scenarios of this sort, the matter component of the universe is 
dominated by a tower consisting of $N$ matter states $\phi_\ell$, where 
$\ell = 0, 1,\dots, N-1$.  We assume that these states exhibit a mass spectrum 
of the form
\begin{equation}
  m_\ell ~=~ m_0 + (\Delta m) \ell^\delta ~, 
  \label{eq:ScalingRelsm}
\end{equation}
where $m_0 > 0$, $\Delta m > 0$, and $\delta > 0$ are general free parameters.
We shall also assume that the corresponding decay widths $\Gamma_\ell$ and initial 
abundances $\Omega_\ell(t_I)$ for these matter states at an early fiducial time 
$t_I$ are given by
\begin{eqnarray}
  \Gamma_\ell &=& \Gamma_0 \left(\frac{m_\ell}{m_0}\right)^\gamma \nonumber \\
  \Omega_\ell(t_I) &=& \Omega_0(t_I) \left(\frac{m_\ell}{m_0}\right)^\alpha~,
  \label{eq:ScalingRelsOmegaGamma}
\end{eqnarray}
where $\Gamma_0$ and $\Omega_0$ denote the decay width and abundance of the 
lightest tower mode, respectively, and where $\lbrace \alpha,\delta,\gamma\rbrace $ 
are general scaling exponents.  We shall further assume that these scaling exponents 
lie within the range
\begin{equation}
  0~ <~ \alpha + \frac{1}{\delta} ~\leq~ \frac{\gamma}{2}~.
  \label{eq:AlphaGammaCondit}
\end{equation}
Indeed, as shown in Ref.~\cite{Dienes:2021woi}, an epoch of cosmic stasis emerges for any scaling exponents which satisfy 
this condition.

Since we assume that each $\phi_\ell$ already behaves like massive matter at $t = t_I$, 
the corresponding equation-of-state parameters $w_\ell$ therefore satisfy $w_\ell \approx 0$ 
throughout the time of interest.  The total matter abundance in the universe is therefore given by 
$\Omega_M=\sum_\ell \Omega_\ell$.  In addition, we shall also assume for simplicity 
that all $\phi_\ell$ states decay primarily into products that are ultra-relativistic and
remain so until after the stasis epoch ends.  The decay of $\phi_\ell$ therefore sources the 
radiation component of the universe.

As discussed in Refs.~\cite{Dienes:2021woi,Dienes:2023ziv}, towers of states which exhibit 
all of these general properties arise naturally in many commonly studied theories of physics 
beyond the Standard Model.  Moreover, allowing the parameters 
$\lbrace \alpha,\gamma,\delta, m_0, \Delta m, \Gamma_0,\Omega_0(t_I)\rbrace$ to vary freely 
within the above constraints allows us to survey huge classes of BSM-derived cosmologies at once.
Indeed, within the regime wherein the condition in Eq.~(\ref{eq:AlphaGammaCondit}) is satisfied, 
the sequential decays of the states in such a tower drive the universe into an epoch of 
matter/radiation stasis during which the total matter and radiation abundances take the 
fixed values
\begin{equation}
  \barOmega_M = \frac{2\gamma - 4(\alpha+ 1/\delta)}{2\gamma - (\alpha + 1/\delta)}\,,
    ~~~~~~~\barOmega_\gamma = 1 - \barOmega_M\,.
\label{eq:fixedpointabundances}
\end{equation}
The effective equation-of-state parameter for the universe during stasis is then constant 
and given by $\expt{w}  =\wbar= \barOmega_\gamma/3$.

\subsection{Equations of motion for density and velocity perturbations}\label{theta_phi_stasis}

Given the background stasis cosmology described above, we shall now calculate the 
missing pieces within the source term in Eq.~\eqref{eq:SdecompPhiEqPsi}.  These are the 
gravitational potential $\Phi$ and the divergence of velocity perturbation $\theta$, both 
of which are necessary for calculating the tensor perturbations during the stasis epoch.

At first order in the perturbations, the $\mu = i$, $\nu = j$ components and the 
$\mu=\nu=0$ components of the Einstein equation respectively yield the relations 
\begin{eqnarray}
  4\pi Ga^2 \sum_A \delta_{A}  \rho^{(0)}_A c_{s,A}^2 &=& \Phi'' 
    + 3\mathcal{H}\Phi'
    + \left(2\frac{a''}{a} - \mathcal{H}^2\right)\Phi~~ \nonumber \\
  4\pi Ga^2 \sum_A \delta_A\rho^{(0)}_A &=& -3\mathcal{H}\Phi'
    - 3\mathcal{H}^2\Phi +\nabla^2\Phi ~,~~~
  \label{eq:G00andGijEqsPsiSub}
\end{eqnarray}
where $\delta_A \equiv (\rho_A -\overline{\rho}_A)/\overline{\rho}_A$ and
where we remind the reader that a prime indicates a derivative with respect 
to the conformal time $\eta$.  We note that these expressions differ from the 
corresponding expressions in Ref.~\cite{Dienes:2025tox} due to differences in 
notation and sign conventions.  Indeed, according to the conventions
established in Eq.~(\ref{eq:metric}), $\Phi$ denotes the Newtonian potential,
$\Psi$ denotes the spatial curvature, and the sign conventions for these 
quantities are such that $\Phi = \Psi$ in the absence of anisotropic stress.
By contrast, in Ref.~\cite{Dienes:2025tox}, $\Phi$ denotes the spatial
curvature, $\Phi$ denotes the Newtonian potential, and the condition which
obtains in the absence of anisotropic stress is $\Psi = -\Phi$.

The two equations in Eq.~(\ref{eq:G00andGijEqsPsiSub}) can be combined 
to yield a single second-order differential equation for $\Phi$.  During a period 
wherein the effective equation-of-state parameter $\langle w\rangle$ for the 
universe is constant, the Bardeen potential satisfies the differential 
equation~\cite{Dienes:2025tox}
\begin{eqnarray}
  && a^2\frac{d^2\Phi_k}{da^2} + \frac{(7+3\langle w\rangle)}{2}a \frac{d\Phi_k}{da}
     + \frac{\langle w\rangle k^2}{a^2H^2} \Phi_k \nonumber \\ 
   && ~~~~~~~~~~~~~~ 
   ~=~  -\frac{3}{2} \sum_{A} \Omega_{A} \delta_{kA} 
     \left(\langle w\rangle - c^2_{s,A}\right)\, ,~~~~~~
  \label{eq:dPrimePhiEqnArbwSimpFTDimVars}
\end{eqnarray}
where $\delta_{kA}$ is the $k$th Fourier mode of $\delta_A$, where $\expt{w}$ was 
defined in Eq.~(\ref{eq:rhoNoughtPNoughtBracketw}), and where $H \equiv a^{-1}da/dt$ 
is the usual Hubble parameter.  Note that in switching variables from the conformal 
time $\eta$ to the scale factor $a$, we have assumed that $a(t_I) = 1$ in order to 
avoid notational clutter.

In cosmologies in which a single perfect fluid with 
$w_{\rm PF}\approx \expt{w}$ dominates the universe, the sum over cosmological components 
on the right side of Eq.~\eqref{eq:dPrimePhiEqnArbwSimpFTDimVars} effectively includes only
a single term.  As a result, since $\expt{w}\approx c_{s, \rm PF}^2$ in the absence of 
significant non-adiabatic perturbations, the right side of this equation vanishes.
By contrast, during a stasis epoch, multiple components with different 
$w_A=c_{s,A}^2$ contribute non-negligibly to the total energy density of the universe.
Since $\expt{w}$ is in general different from any of the individual $c_{s,A}^2$, and since 
these cosmological components in general have different $\delta_{kA}$ at any given
time, the right side of Eq.~(\ref{eq:dPrimePhiEqnArbwSimpFTDimVars}) does not in general 
vanish during such an epoch.  This implies that the evolution of $\Phi_k$ does not decouple 
from the evolution of the $\delta_{kA}$.  Indeed, as discussed in Ref.~\cite{Dienes:2025tox},
this is the principal underlying reason why the density contrast for a decoupled matter 
component behaves differently as a function of $a$ during a stasis epoch once it enters 
the horizon than it does during an epoch of perfect-fluid domination with the same $\expt{w}$.

In order to proceed further, we note that a first-order equation of motion for $\Phi_k$ 
may also be obtained from the second line of Eq.~(\ref{eq:G00andGijEqsPsiSub}), \ie,
\begin{eqnarray}
a\frac{d\Phi_k}{da} &=& -\left(\frac{k^2}{3a^2H^2}+1\right)\Phi_k\nn\\
& &~~~-\frac{1}{2} \bigg(\Omega_\gamma\delta_{k\gamma}
  + \sum_{\ell=0}^{N-1}\Omega_\ell\delta_{k\ell}\bigg)~,
  \label{eq:PertSuitePhi} 
\end{eqnarray}
where $\delta_{k\ell}$ and $\delta_{k\gamma}$ denote $k$th Fourier modes of the density 
perturbations in the decaying matter field $\phi_\ell$ and in radiation, respectively.  
(Note in particular that $\delta_{k\ell}$ is {\it not}\/ a Kronecker delta between two tower 
levels $k$ and $\ell$.)  The manner in which $\delta_{k\ell}$ and $\delta_{k\gamma}$ evolve 
is governed by the set of differential equations~\cite{Dienes:2025tox}
\begin{eqnarray}
  a\frac{d\delta_{k\ell}}{da} &=& 3a\frac{d\Phi_k}{da} - \frac{\theta_{k\ell}}{aH}
    - \frac{\Gamma_\ell}{H} \Phi_k
   \nonumber \\
  a\frac{d\theta_{k\ell}}{da} &=& 
    -\theta_{k\ell} +\frac{k^2}{aH}\Phi_k 
    \nonumber \\
  a\frac{d\delta_{k\gamma}}{da} &=& 4a\frac{d\Phi_k}{da} 
    - \frac{4}{3}\frac{\theta_{k\gamma}}{aH}
    \nonumber \\ & &
    ~~~+\sum_{\ell=0}^{N-1}\frac{\Gamma_\ell}{H}
    \frac{\rho^{(0)}_\ell}{\rho^{(0)}_\gamma} 
    \Big( \delta_{k\ell} -\delta_{k\gamma} + \Phi_k \Big) 
    \nonumber \\
  a\frac{d\theta_{k\gamma}}{da} &=& 
    \frac{1}{4}\frac{k^2}{aH}\delta_{k\gamma} +\frac{k^2}{aH}\Phi_k 
    \nonumber \\ & &
    ~~~+ \sum_{\ell=0}^{N-1}\frac{\Gamma_\ell}{H}
    \frac{\rho_\ell^{(0)}}{\rho_\gamma^{(0)}}
    \left(\frac{3}{4}\theta_{k\ell} - \theta_{k\gamma} \right) ~.~ 
  \label{eq:PertSuiteAll} 
\end{eqnarray} 

\subsection{Calculating $\Phi_k$}\label{subsec:Phi}

In order to develop an understanding of how $\Phi_k$ evolves during the stasis epoch, 
we begin by noting that the $\phi_\ell$ which contribute significantly to $\Omega_M$ at 
each moment during this epoch are those with decay width $\Gamma_\ell \lesssim H$, 
whose energy densities have yet not been impacted significantly by particle decay.  
For all such $\phi_\ell$, the term proportional to 
$\Gamma_\ell$ on the right side of the differential equation for $\delta_{k\ell}$ in 
Eq.~(\ref{eq:PertSuiteAll}) is small in comparison to the other terms on 
the right side of this equation and can to a reasonable approximation be neglected.  
The differential equations for these $\phi_\ell$ are therefore essentially independent 
of $\ell$.  Likewise, the equation of motion in Eq.~(\ref{eq:PertSuiteAll}) for 
$\theta_{k\ell}$ contains no terms which depend meaningfully on $\ell$.  
It therefore follows that that the equations of motion for $\delta_{k\ell}$ and 
$\theta_{k\ell}$ for all $\phi_\ell$ with $\Gamma_\ell$ within this regime are 
essentially identical.  For these $\phi_\ell$, the first two lines in 
Eq.~\eqref{eq:PertSuiteAll} can be combined in the instantaneous-decay 
approximation --- \ie, the approximation that all particles decay precisely at 
$t=\Gamma_\ell^{-1}$ --- to yield a second-order differential equation for 
$\delta_{k\ell}$:
\begin{eqnarray}
  && \frac{d^2\delta_{k\ell}}{da^2} + \frac{3}{2a}\big(1-\wbar\big)\frac{d\delta_{k\ell}}{da}  
      ~\approx~ 3\frac{d^2\Phi_k}{da^2} \nonumber \\ 
  && ~~~~~~~~~~~~~~  +\frac{9}{2a}\big(1-\wbar\big)\frac{d\Phi_k}{da} 
    - \frac{k^2}{a^4H^2}\Phi_k~,~~~~
  \label{eq:FulldeltaellsEqApprox}
\end{eqnarray}
where we have set $\expt{w} = \wbar$.  In addition, we also note that the contributions to 
$\Omega_M$ from $\phi_\ell$ with decay widths within the opposite regime --- the regime 
in which $\Gamma_\ell \gtrsim H$ --- are highly suppressed.  
Thus, at any given moment, the $\phi_\ell$ which contribute meaningfully 
to the sums over $\ell$ in both Eq.~(\ref{eq:dPrimePhiEqnArbwSimpFTDimVars}) and 
Eq.~(\ref{eq:PertSuitePhi}) are those for which $\Gamma_\ell \lesssim H$,
whose $\delta_{k\ell}$ are essentially independent of $\ell$.  This implies that during
stasis, the value of each of these sums will not be significantly impacted if we approximate
$\delta_{k\ell}$ as being truly independent of $\ell$ and pull this quantity outside the 
sum.  

We first apply this approximation to Eq.~(\ref{eq:dPrimePhiEqnArbwSimpFTDimVars}).
Since $c_{s\gamma}^2 \approx w_\gamma = 1/3$ and $c_{s\ell}^2 \approx w_\ell = 0$, 
we find that in general, during a matter/radiation stasis,
this equation reduces to
\begin{eqnarray}
  && a^2\frac{d^2\Phi_k}{da^2} + \frac{(7+3\wbar)}{2}a\frac{d\Phi_k}{da}
     + \frac{\wbar k^2}{a^2H^2} \Phi_k \nonumber \\ 
  && ~~~~  ~=~ -\frac{3}{2} \left[ \barOmega_\gamma\delta_{k\gamma}
  \left(\wbar - \frac{1}{3}\right)
    + \wbar\sum_{\ell = 0}^{N-1} \Omega_\ell \delta_{k\ell} 
     \right] \,.~~~~~~~~
\end{eqnarray}
Pulling $\delta_{k\ell}$ outside the sum and noting that
\begin{eqnarray}
  \sum_{\ell = 0}^{N-1}\Omega_\ell ~=~ \barOmega_M ~=~ 1 - 3\wbar    
\end{eqnarray}
during stasis, we find that 
\begin{eqnarray}
  && a^2\frac{d^2\Phi_k}{da^2} + \frac{(7+3\wbar)}{2}a\frac{d\Phi_k}{da}
     + \frac{\wbar k^2}{a^2H^2} \Phi_k \nonumber \\ 
  && ~~~~~~~~~~~~~~ ~=~ \frac{3\wbar}{2}
  (1-3\wbar) \left(\delta_{k\gamma} - \delta_{k\ell} \right) \,.~~~~
  \label{eq:PhiDoublePrimeEqPullOutdkl}
\end{eqnarray}
Likewise, applying this same approximation to Eq.~(\ref{eq:PertSuitePhi}), we 
find that
\begin{equation}
  a\frac{d\Phi_k}{da} ~=~ -\left(\frac{k^2}{3a^2H^2} + 1\right)\Phi_k 
    - \frac{3\wbar}{2}\delta_{k\gamma} - \frac{1-3\wbar}{2}\delta_{k\ell}\,.
\label{eq:PhiPrimeEqPullOutdkl}
\end{equation}

Solving Eq.~(\ref{eq:PhiPrimeEqPullOutdkl}) for $\delta_{k\gamma}$ and substituting this 
result into Eq.~(\ref{eq:PhiDoublePrimeEqPullOutdkl}), we obtain a single equation which 
relates $\delta_{k\ell}$ to $\Phi_k$ and its derivatives:
\begin{eqnarray}
  & & a^2\frac{d^2\Phi_k}{da^2} + \frac{9-3\wbar}{2}a\Phi_k' 
    + \frac{k^2}{3a^2H^2}\Phi_k + (1-3\wbar)\Phi_k ~~~\nn\\
  & &~~~~~~~~~~~~~~~~~~~~~~~~~~ ~=~ 
    -\frac{1-3\wbar}{2}\delta_{k\ell}~.
  \label{eq:PhiEqnEliminatedkgamma}
\end{eqnarray}
This equation can in turn be combined with Eq.~(\ref{eq:FulldeltaellsEqApprox})
in order to obtain a fourth-order differential equation for $\Phi_k$ alone.  
After some algebra, we find that this equation may be expressed as
\begin{eqnarray}
  0 &=& a^4\frac{d^4\Phi_k}{da^4} + (10-3\wbar)a^3\frac{d^3\Phi_k}{da^3} \nonumber \\
    && + \left[\frac{1}{3}\frac{k^2}{a^2H^2} 
      + \frac{3}{4}(3\wbar^2 -30\wbar + 31)\right]a^2\frac{d^2\Phi_k}{da^2} \nonumber \\
    && + \left[\frac{1}{6}(9\wbar+7)\frac{k^2}{a^2H^2} 
      + \frac{3}{2}(9\wbar^2 -16\wbar + 7)\right]a\frac{d\Phi_k}{da}
      \nonumber \\
    && + \,\frac{1}{2}(3\wbar^2+7\wbar)\frac{k^2}{a^2H^2} \Phi_k~.
  \label{eq:FourthOrderPhiEq}
\end{eqnarray}
As in the standard cosmology, $\Phi_k$ is nearly constant prior to 
horizon entry.  However, Eq.~\eqref{eq:FourthOrderPhiEq} implies that once the 
mode is well within the horizon, the solution for $\Phi$ includes both a 
power-law term and an oscillatory term.  In particular, we find that
\begin{equation}
  \Phi_k ~\sim~ C_p^{(\Phi)} a^s + a^{-2} \left[C_{+}^{(\Phi)}
    e^{i \varphi_k(a)}+C_{-}^{(\Phi)}e^{-i \varphi_k(a)}\right]\,,
  \label{eq:Phi_power_law}
\end{equation}
where $C_p^{(\Phi)}$ and $C_{\pm}^{(\Phi)}$ are constants that specify the magnitudes 
of the power-law and oscillatory terms, where $\varphi_k(a)=ka^r/(\sqrt{3}H_I r)$ is a 
phase determined in part by $H_I \equiv H(t_I)$, and where we have defined
\begin{equation}
r \equiv \frac{1+3\wbar}{2}\,,~~~~~
  s ~\equiv~\frac{\sqrt{9\wbar^2 - 78\wbar + 25} 
  - (9\wbar+5)}{4}~.
\end{equation}
Note that since $s>-2$ for $0<\wbar<1/3$, the power-law part always dominates at late times.
Note that this behavior is completely different from the behavior which the gravitational potential 
exhibits in the standard cosmology, wherein $\Phi_k$ exhibits underdamped oscillations 
around $\Phi_k = 0$ after horizon entry.

Choosing a fiducial time $t_*$ within the stasis epoch, we find that since 
$H=H_* a^{-3(\wbar+1)/2}$ during stasis, Eq.~(\ref{eq:FourthOrderPhiEq}) above is 
invariant under the transformation
\begin{eqnarray}
  a&~\rightarrow~& \xi^{-2/(1+3\wbar)} a\nn\\
  k&~\rightarrow~& \xi k~,\label{eq:ak_transform}
\end{eqnarray}
as both the terms involving $k/(aH)$ and these in the form of $a^n d^n\Phi_k/da^n$ remain 
unchanged.  Indeed, it follows that derivatives of $\Phi_k$ exhibit the property
\begin{equation}
  X^n \frac
      {d^n\Phi_{\widehat k}(\widehat a)}
      {d\,{\widehat a}^n}
              \biggr|_{\widehat a =X a} ~=~ 
  \frac{d^n\Phi_k(a)}{da^n}~,
  \label{eq:invariance}
\end{equation}
where $X\equiv \xi^{-2/(1+3 \overline{w})}$ and $\widehat k \equiv Xk$.  
Thus, once $\Phi_k$ is obtained for a particular $k$-mode that enters the horizon during stasis, 
we can obtain an expression for any other mode $\Phi_{k'}$ with a wavenumber $k'$ within the
relevant range. 
Indeed, since $\eta\sim a^{(1+3\wbar)/2}$ during stasis, the quantity $k\eta$
is also invariant under the transformation in Eq.~\eqref{eq:ak_transform}.
This property is extremely useful as it allows us to write
\begin{equation}
  \Phi_k(a) ~=~ \Phi_k(a_*)\frac{\Phi_k(a)}{\Phi_k(a_*)} 
    ~=~ \psi_k^{(s)} \,\widehat{\Phi}(k\eta)~,
\end{equation}
where the scalar perturbation for a stasis cosmology at $t_*$ is defined as 
$\psi_k^{(s)}\equiv \Phi_k(a_*)$ 
and where $\widehat{\Phi}(k\eta)$ is the universal transfer function that encodes the 
subsequent evolution of $\Phi_k(a)$ for all relevant $k$-modes.  
Since the primordial curvature perturbation 
for modes which have not yet entered the horizon is conserved in this case, $\psi_k^{(s)}$ 
is related to the corresponding scalar perturbation $\psi_k^{(r)}$ for the standard 
cosmology by
\begin{equation}
  \psi_k^{(s)} ~=~ \frac{9(1+\wbar)}{2(5+3\wbar)} \,\psi_k^{(r)}~.    
\end{equation}
Note that $\widehat{\Phi}(k,\eta) = \widehat{\Phi}(k\eta)$ depends on 
the product $k\eta$ only. 

\begin{figure}[t!]
    \begin{center}
    \includegraphics[width=0.47\textwidth, keepaspectratio]{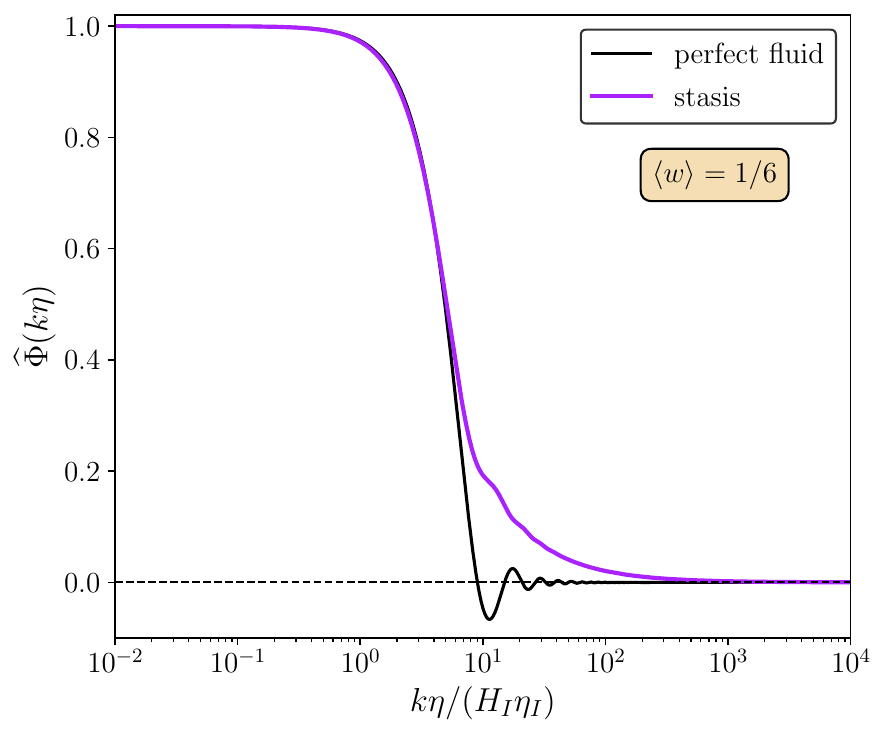}
        \caption{
        The transfer function $\widehat{\Phi}(k\eta)$ for the gravitational
        potential, shown as a function of $k\eta$ during an epoch of 
        matter/radiation stasis with an effective equation-of-state parameter $\wbar=1/6$
        (purple curve) and during an epoch of perfect-fluid domination with the 
        same effective equation-of-state parameter (black curve).  These results 
        highlight the differences that can arise at the perturbation level, even 
        between scenarios with the same background cosmology. 
    \label{fig:Phi_hat}}
    \end{center}
\end{figure}

An example of the evolution of $\widehat{\Phi}(k\eta)$ is shown in Fig.~\ref{fig:Phi_hat}. 
The purple curve shows the case of a stasis epoch with the background equation-of-state 
parameter $\wbar=1/6$, whereas the black curve shows the case of a perfect-fluid-dominated 
epoch wherein $w_{\rm PF} = \wbar$.  As is shown in the figure, $\widehat{\Phi}(k\eta)$ evolves 
similarly both when the $k$-mode is outside the horizon and shortly after horizon entry, \ie, 
when $k\eta/(H_I\eta_I) \lesssim \mathcal{O}(1)$.  However, the behavior in the two cases becomes
markedly different once $k\eta/(H_I\eta_I)$ is significantly larger than unity and the mode is 
therefore well within the horizon.  During an epoch of perfect-fluid domination, $\widehat{\Phi}$ 
oscillates around zero with a decreasing amplitude.  By contrast, during a stasis epoch, the
behavior of $\widehat{\Phi}(k\eta)$ is dominated by the power-law term in 
Eq.~\eqref{eq:Phi_power_law} at late times.  Oscillations rapidly damp away and the
transfer function asymptotes to zero far more gradually. 

\subsection{Calculating $\theta_{k\ell}$ and $\theta_{k\gamma}$} \label{velostasis}

As we have discussed above, we cannot in general rewrite the $v^iv_j$ term from 
the stress-energy tensor as combinations of the derivatives of the gravitational potential
when analyzing a cosmological epoch involving multiple fluids.  Instead, we
obtain the velocity perturbations $v^i$ directly by first obtaining the 
divergence of the velocity perturbation $\theta_{A}=\partial_i v_{A}^i$ from
Eq.~\eqref{eq:PertSuiteAll} and then using Eq.~\eqref{eq:vtheta} in order to 
obtain $v_{kA}$.
In doing this, we shall work in the instantaneous-decay approximation, as we
did in deriving the differential equation for $\Phi_k$ in Eq.~\eqref{eq:FourthOrderPhiEq}, 
and we shall make use of the full numerical solution for $\Phi_k$ that we obtain from 
this equation in what follows.  For convenience, 
we shall also establish our conventions for the scale factor such that $a_I=a(t_I)=1$ 
at some initial time $t_I$ at which the universe is already in stasis.

We begin by considering the second equation in Eq.~\eqref{eq:PertSuiteAll}, \ie,
\begin{equation}
  a\frac{d\theta_{k\ell}}{da} ~=~
    -\theta_{k\ell} +\frac{k^2}{aH}\Phi_k\,.
    \label{eq:dthetakl}
\end{equation}
Integrating this equation, we obtain
\begin{equation}
  a\theta_{k\ell} - a_I\theta_{k\ell}(t_I) ~=~ 
  \int_{a_I}^{a} da~\frac{k^2}{aH} \Phi_k~. 
  \label{eq:atheta_int}
\end{equation}

The manner in which $\theta_{k\ell}$ evolves with $a$ can be determined 
from this relation once a numerical solution for $\Phi_k$ is obtained from 
Eq.~\eqref{eq:FourthOrderPhiEq}.  However, it is also illuminating 
to consider the manner in which $\theta_{k\ell}$ behaves within particular regimes.  
At early times, well before the mode enters the horizon, the comoving Hubble radius 
scales as $(aH)^{-1}\sim a^{r}$ with $r\equiv (1+3\wbar)/2$, while $\Phi_k$ is 
nearly constant.  It therefore follows that within this early-time regime, we have
\beq
  \theta_{k\ell}~\sim~ -a^{r}
  \label{eq:thetak0_early}
\eeq
for all $\ell$.  By contrast, at later times, well after horizon entry, $\Phi_k$ is 
well approximated by the expression in Eq.~\eqref{eq:Phi_power_law}, which
includes both a power-law and and oscillatory term.
The oscillatory term in $\Phi_k$ in turn induces an oscillatory term in
$\theta_{k\ell}$.  However, this term includes an overall 
scaling factor of $a^{-2}$, just as does the corresponding term in $\Phi_k$
and therefore decays away rapidly with $a$.  The late-time behavior of 
$\theta_{k\ell}$ is therefore dominated by the power-law term in 
Eq.~\eqref{eq:Phi_power_law}, which scales as $\Phi_k\sim -a^s$.  We
therefore find that 
\begin{equation}
  \theta_{k\ell} ~\sim~ -a^{r+s}~ 
  \label{eq:thetak0_late}
\end{equation}
for all $\ell$ at late times.

Interestingly, we note that $r + s =0 $ for $\wbar = 1/6$, \ie, for the value of
$\wbar$ which corresponds to a stasis epoch with equal abundances of matter and 
radiation.  In other words, $\abs{\theta_{k\ell}}$ becomes approximately constant for 
all $\ell $at late times.  By contrast, for $0\leq \wbar < 1/6$, we find that $r+s > 0$ and 
thus $\abs{\theta_{k\ell}}$ grows with $a$ well after horizon entry, whereas for 
$1/6< \wbar \leq 1/3$ we find that $r+ s < 0$ and thus $\abs{\theta_{k\ell}}$ decreases.

Since the expression on the right side of Eq.~\eqref{eq:atheta_int} is independent 
of $\ell$, it follows that $\theta_{k\ell}$ depends on $\ell$ only through its 
dependence on its initial value $\theta_{k\ell}(t_I)$.  Moreover, $\theta_{k\ell}(t_I)$ 
also turns out not to be particularly sensitive to the value of $\ell$.  Indeed, 
to $\mathcal{O}(k^2)$, we find that
\begin{equation}
  \theta_{k\ell}(t_I) ~=~ \frac{2k^2}{3(1+\wbar)} 
    \frac{\Phi_k(t_I)}{H(t_I)}\,\label{eq:thetak0_ini}
\end{equation}
for all $\ell$.  Furthermore, we also note that within the regime in which the 
expression on the right side of Eq.~\eqref{eq:atheta_int} dominates over 
$\theta_{k\ell}(t_I)$, the manner in which $\theta_{k\ell}$ evolves is not particularly 
sensitive to the value of $\theta_{k\ell}(t_I)$.  Thus, within this regime, $\theta_{k\ell}$ 
would be to a good approximation independent of $\ell$ at times $t \gg t_I$ even if 
the spectrum of initial values for this velocity divergence {\it had}\/ exhibited a 
non-trivial scaling behavior with $\ell$.  It therefore follows that to a good 
approximation, we may treat $\theta_{k\ell}$ as a universal, $\ell$-independent 
function of $a$ for any relevant value of $k$ and 
replace $\theta_{k\ell}$ with $\theta_{k0}$, a quantity which we may then pull out of 
any sum over $\ell$.  As a result, the $N$ differential equations for each value 
$\theta_{k\ell}$ all collapse to a single equation.  Together with the similar 
treatment on $\delta_{k\ell }$ in Sect.~\ref{subsec:Phi}, the entire differential-equation 
system can be then rendered independent of the total number of states $N$ in the tower.

We now turn to $\theta_{k\gamma}$.  Within the regime in which $\theta_{k\ell}$ is to 
a good approximation independent of $\ell$, the fourth equation in 
Eq.~\eqref{eq:PertSuiteAll} reduces to
\begin{equation}
a\frac{d\theta_{k\gamma}}{da} ~=~ 
    \frac{1}{4}\frac{k^2}{aH}\delta_{k\gamma} +\frac{k^2}{aH}\Phi_k 
    + \left(\frac{3}{4}\theta_{k0} - \theta_{k\gamma} \right)(1-3\wbar)\,,
  \label{eq:theta_kg}
\end{equation}
where we have used the fact that~\cite{Dienes:2021woi} 
\begin{equation}
  \sum_\ell \Gamma_\ell \Omega_\ell ~=~ H(\overline{\Omega}_M-\overline{\Omega}_M^2)
  \label{eq:StasisGammaOmegaRel}
\end{equation}
during stasis in order to evaluate the sum.

Similar simplifications may be made within this regime to the equations of motion
for $\Phi_k$ and $\delta_{k\gamma}$ during stasis.  Approximating 
$\delta_{k\ell}\approx \delta_{k0}$ for all $\phi_\ell$ with $\Gamma_\ell^{-1} > t$ 
as in Sect~\ref{subsec:Phi} and once again making use of Eq.~\eqref{eq:StasisGammaOmegaRel}, 
we find that Eq.~\eqref{eq:PertSuitePhi} and the third equation in 
Eq.~\eqref{eq:PertSuiteAll} respectively reduce to
\begin{eqnarray}
  a\frac{d\Phi_k}{da} &=& -\left(1 +\frac{k^2}{3a^2H^2}\right)\Phi_k 
    - \frac{3\wbar}{2}\delta_{k\gamma} - \frac{1-3\wbar}{2}\delta_{k0} \nonumber \\
  a\frac{d\delta_{k\gamma}}{da} &=&  4a\frac{d\Phi_k}{da} 
    - \frac{4}{3}\frac{\theta_{k\gamma}}{aH}
    +(1-3\wbar)\Big( \delta_{k0} -\delta_{k\gamma} + \Phi_k \Big)~.\nonumber \\ 
  \label{eq:deltakg_approx_pre_final}
\end{eqnarray}
We may combine these two equations in order to eliminate $\delta_{k0}$, thereby obtaining  
\begin{eqnarray}
  a\frac{d\delta_{k\gamma}}{da} &=& 
-\left(1+3\wbar+\frac{2k^2}{3a^2H^2}\right)\Phi_k + 2a\frac{d\Phi_k}{da} 
      \nonumber \\ 
  & & -\frac{4\theta_{k\gamma}}{3aH}-\delta_{k\gamma}~.
  \label{eq:deltakg_approx_final}
\end{eqnarray}
Since the functional forms for $\Phi_k(a)$ and $H(a)$ are known during stasis, this differential
equation for $\delta_{k\gamma}$, along with the differential equations for $\theta_{k0}$ and 
$\theta_{k\gamma}$ in Eqs.~\eqref{eq:PertSuiteAll} and~\eqref{eq:theta_kg}, respectively,
together 
constitute a closed system which we may then solve in order to obtain explicit functional forms 
for these three quantities for all relevant $k$ --- functional forms which do not explicitly 
depend on the value of $N$.

We have already determined the manner in which $\theta_{k0}$ scales with $a$ at times
both well before and well after horizon entry.  In order to do the same for 
$\theta_{k\gamma}$, we begin by noting that well before horizon entry, when
$k \ll aH$, the gravitational potential $\Phi_k(t)$ remains effectively constant and 
equal to its initial value.
Moreover, since $\theta_{k\gamma}\sim kv_{k\gamma}$, it follows from dimensional analysis 
that the term associated with $\theta_{k\gamma}$ in Eq.~\eqref{eq:deltakg_approx_final} 
is at least as small as $\mathcal{O}(k/aH)$ and can be neglected while a mode is 
well outside the horizon.  Thus, within this regime, Eq.~\eqref{eq:deltakg_approx_final} 
reduces to
\begin{equation}
  a\frac{d\delta_{k\gamma}}{da} ~\approx~ 
    -\left(1+3\wbar\right)\Phi_k(t_I) - \delta_{k\gamma}~.
  \label{eq:deltakg_approx_final_earlyt}
\end{equation}
It therefore follows that well before horizon entry, we have the general solution to this 
first-order linear differential equation: 
\begin{equation}
  \delta_{k\gamma}(t) ~\approx~ - (1+3\wbar)\Phi_{k}(t_I) + \frac{ C_{k\gamma}}{a}~,
\end{equation}
where $C_{k\gamma}$ is an undetermined constant.  Since the term inversely 
proportional to $a$ diverges at early times, it follows that $C_{k\gamma} = 0$.  Thus,  
we find that $\delta_{k\gamma}$ is approximately constant at times well before horizon entry
and is  given by
\begin{equation}
  \delta_{k\gamma}(t_I) ~\approx~ - (1+3\wbar)\Phi_{k}(t_I) ~.
  \label{eq:the_early_constant_dkg}
\end{equation}

Given the result in Eq.~\eqref{eq:the_early_constant_dkg}, we are now equipped to assess 
how $\theta_{k\gamma}$ scales with $a$ at times well before horizon entry.  We begin by 
noting that Eqs.~\eqref{eq:thetak0_early} and~\eqref{eq:thetak0_ini}, taken together, 
suggest that at such times we have 
\begin{equation}
  \theta_{k0} ~\approx~ \frac{2k^2}{3(1+\wbar)aH}\Phi_k(t_I)~.
\end{equation}
With this result and the result in Eq.~\eqref{eq:the_early_constant_dkg}, we find 
that the evolution equation for $\theta_{k\gamma}$ in Eq.~\eqref{eq:theta_kg} reduces
at times well before horizon entry to 
\begin{equation}
  a\frac{d\theta_{k\gamma}}{da} + (1-3\wbar)\theta_{k\gamma} ~\approx~ D(a)~,
  \label{eq:thetakg_diff_early}
\end{equation}
where the inhomogeneous term $D(a)$ is given by
\begin{eqnarray}
D(a) &=& \frac{1}{4}\frac{k^2}{aH}\delta_{k\gamma}(t_I) +\frac{k^2}{aH}\Phi_k(t_I) 
      + \frac{3}{4} (1-3\wbar)\theta_{k0}\nn\\
&=&  \frac{5-6\wbar-3\wbar^2}{4(1+\wbar)} \frac{k^2}{a_IH_I}\Phi_k(t_I)\,a^{r}~.
\end{eqnarray}
The solution to this equation is
\begin{equation}
     \theta_{k\gamma} ~=~ C_1^{(\theta)} a^{2r-2} + C_2^{(\theta)} a^{r}~. 
    \label{eq:thetakg_diff_early_sol}
\end{equation}
The first term in this expression is the general solution to the corresponding 
homogeneous equation; the second is a particular solution that scales with $a$ in 
the same manner as $D(a)$ scales with $a$.  Since $2r-2= -(1-3\wbar) <0$ for 
$\wbar<1/3$, the first term decreases with $a$ and thus diverges at early times. 
Discarding this term by setting $C_1=0$, we find that the initial condition for 
$\theta_{k\gamma}$ is given by 
\begin{equation}
  \theta_{k\gamma}(t_I) ~=~ \frac{5-6\wbar-3\wbar^2}{6(1-\wbar^2)} 
    \frac{k^2}{a_IH_I}\,\Phi_k(t_I)\,.
\end{equation}
We also find that $\theta_{k\gamma}\sim a^{r}$ at early times well before horizon 
entry --- a scaling behavior which is similar to that experienced by $\theta_{k0}$.

We now turn to consider the behavior of $\theta_{k\gamma}$ in the opposite regime, 
wherein $k \gg aH$ and the mode is well within the horizon.  Since the comoving 
Hubble horizon $(aH)^{-1}$ grows with time within this regime, the contributions 
proportional to $(aH)^{-1}$ which appear in the inhomogeneous terms in both individual 
equations in Eq.~\eqref{eq:deltakg_approx_pre_final} increasingly come to dominate 
the behavior of $\theta_{k\gamma}$ and $\delta_{k\gamma}$.  In the approximation that 
the other, subleading terms can be neglected, we find that this coupled system of 
equations admits a particular solution wherein $\theta_{k\gamma}\sim a^{r+s}$ and 
$\delta_{k\gamma}\sim a^{s}$ both scale with $a$ according to power laws 
with different exponents.  Moreover, within this same approximation, we 
also note that the terms which couple these two equations naturally give rise to 
coupled oscillations of $\theta_{k\gamma}$ and $\delta_{k\gamma}$ --- oscillations 
which are ultimately driven by the oscillatory terms appearing the solution for 
$\Phi_k$ in Eq.~\eqref{eq:Phi_power_law}.

Since we have already derived the functional forms for $\Phi_k$ and $\theta_{k0}$ 
within this late-time regime, we can solve for $\delta_{k\gamma}$ and 
$\theta_{k\gamma}$ by applying a WKB-like approximation in which we posit that the
oscillatory components of $\theta_{k\gamma}$ and $\delta_{k\gamma}$ take the forms
\begin{eqnarray}
  \theta_{k\gamma}^{\rm osc} & = & A^{(\theta)}(a)
    \left[C^{(\theta)}_{d+}(a)e^{i\varphi_k(a)} 
    + C^{(\theta)}_{d-}(a)e^{-i\varphi_k(a)}\right] \nonumber \\
  \delta_{k\gamma}^{\rm osc} & = & A^{(\delta)}(a)
    \left[C^{(\delta)}_{d+}(a)e^{i\varphi_k(a)}
    + C^{(\delta)}_{d-}(a)e^{-i\varphi_k(a)}\right]\,, \nonumber \\
\end{eqnarray}
where the $C^{(X)}_{d\pm}$ are constant coefficients, where $\varphi_k(a)$ is the 
same phase function which appears in Eq.~(\ref{eq:Phi_power_law}), and where the 
$A^{(X)}(a)$ are functions which respectively encapsulate the dependence of 
$\theta_{k\gamma}^{\rm osc}$ and $\delta_{k\gamma}^{\rm osc}$ on $a$.  After some 
algebra, we find that the general solutions for these perturbations are well 
approximated within this regime by
\begin{eqnarray}
  \theta_{k\gamma} &=& C_p^{(\theta)} a^{r+s}+a^{r-3/2}
      \left[C_{h+}^{(\theta)}e^{i\varphi_k(a)} 
      +C_{h-}^{(\theta)} e^{-i\varphi_k(a)} \right]\nn\\
    && +a^{2r-2}\left[C_{d+}^{(\theta)}e^{i\varphi_k(a)}
      +C_{d-}^{(\theta)}e^{-i\varphi_k(a)}\right]\,\nn\\
  \delta_{k\gamma}&=&C_p^{(\delta)} a^{s}+a^{r-3/2}
      \left[C_{h+}^{(\delta)}e^{i\varphi_k(a)} 
      +C_{h-}^{(\delta)} e^{-i\varphi_k(a)} \right]\nn\\
    && +a^{2r-2}\left[C_{d+}^{(\delta)}e^{i\varphi_k(a)}
      +C_{d-}^{(\delta)}e^{-i\varphi_k(a)}  \right]~.
\label{eq:theta_delta_subhorizon}
\end{eqnarray}
The quantities $C_p^{(X)}$ appearing in these expressions are the coefficients of 
terms associated with the aforementioned power-law solutions for 
$\theta_{k\gamma}$ and $\delta_{k\gamma}$, while the $C_{h\pm}^{(X)}$ are the 
coefficients of terms associated with solutions of the corresponding homogeneous 
equations for $\theta_{k\gamma}$ and $\delta_{k\gamma}$.  Depending on the values of 
$r$ and $s$, which are ultimately determined by $\wbar$, different terms can dominate 
the behavior of these quantities in different stasis scenarios and at different stages 
during the late-time regime.

For convenience in later calculations, we use the fact that $d/da = (a^2H)^{-1}d/d\eta$ in order 
to rewrite the coupled system of evolution equations for $\theta_{k0}$, $\theta_{k\gamma}$, and
$\delta_{k\gamma}$ in Eqs.~\eqref{eq:PertSuiteAll}, \eqref{eq:theta_kg}, 
and~\eqref{eq:deltakg_approx_final}, respectively,
in terms of the product $k\eta$:
\begin{eqnarray} 
&&\frac{d}{d(k\eta)}\left(\frac{\theta_{k0}}{k}\right) = 
    -\frac{2}{(1+3\wbar)k\eta}\frac{\theta_{k0}}{k} +\Phi_k 
    \nonumber \\
&&\frac{d}{d(k\eta)}\left(\frac{\theta_{k\gamma}}{k}\right) = 
    \frac{1}{4}\delta_{k\gamma} +\Phi_k \nn\\
&&~~~~~~~~~~~~~~~~~~~~~~ 
  +\frac{2(1-3\wbar)}{(1+3\wbar)k\eta}
  \left(\frac{3}{4}\frac{\theta_{k0}}{k} - \frac{\theta_{k\gamma}}{k} \right) \nn\\
&&\frac{d\delta_{k\gamma}}{d(k\eta)}
= -\frac{2}{k\eta}\left(1+\frac{1+3\wbar}{6}k^2\eta^2\right)\Phi_k + 2\frac{d\Phi_k}{d(k\eta)}~~~~ \nn\\
&&~~~~~~~~~~~~~~~~~~~~~~ 
  -\frac{4}{3}\frac{\theta_{k\gamma}}{k} -\frac{2}{(1+3\wbar)k\eta}\delta_{k\gamma}~.
\label{theta_delta_coupled}
\end{eqnarray}
In these expressions, we have aligned the values of $\eta$ and $H$ at the initial time $t_I$ 
by making the identification that $\eta_I=2/[H_I(1+3\wbar)]=3(1+\wbar)t_I/(1+3\wbar)$, such 
that $aH=2/[\eta(1+3\wbar)]$.  Since we know that $\Phi_k$ has the same functional 
dependence on $k\eta$ for all $k$, it follows from Eq.~(\ref{theta_delta_coupled}) that 
each of the quantities $\theta_{k0}/k$, $\theta_{k\gamma}/k$, and $\delta_{k\gamma}$ likewise 
has the same functional dependence on the product $k\eta$, regardless of $k$.

\begin{figure}[t!]
    \begin{center}
\includegraphics[width=0.47\textwidth, keepaspectratio]{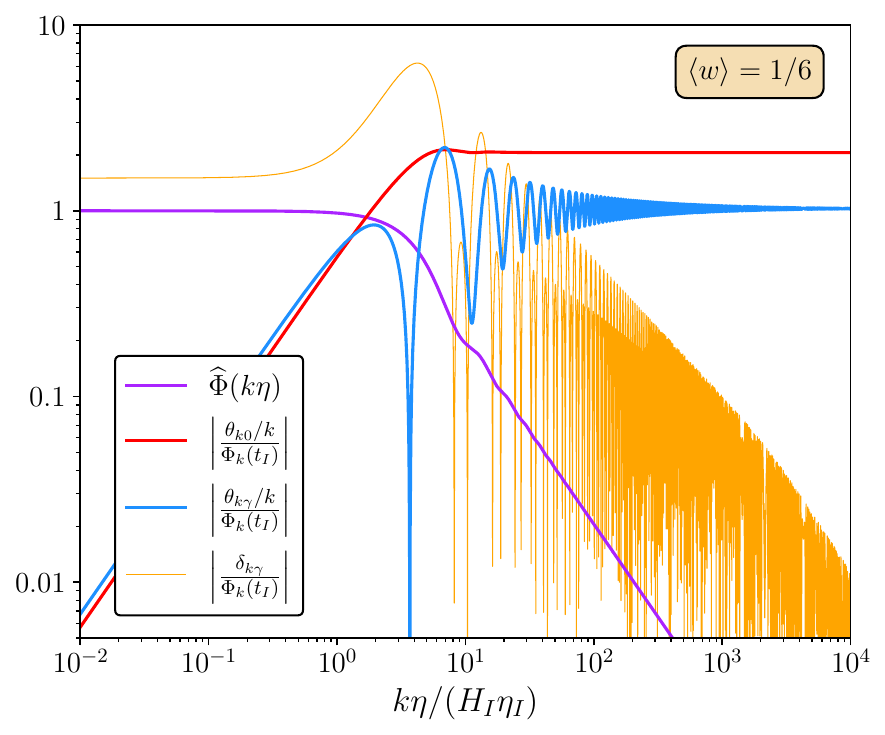}
\caption{The quantities $\widehat{\Phi}$, $\theta_{k0}/k$, $\theta_{k\gamma}/k$, 
  and $\delta_{k\gamma}$, normalized to the value $\Phi_k(t_I)$ of the 
  gravitational potential at some initial time $t_I$ during stasis and plotted 
  as functions of $k\eta/(H_I\eta_I)$.  The curves depend on $k$ only through the 
  product $k\eta$, and their shapes as functions of $k\eta$ therefore do not 
  depend on $k$.
\label{fig:thetas}}
\end{center}
\end{figure}

In Fig.~\ref{fig:thetas}, we show the evolution of $\widehat{\Phi}$, $\theta_{k0}/k$, 
$\theta_{k\gamma}/k$, and $\delta_{k\gamma}$ as functions $k\eta/(H_I\eta_I)$.
The magenta curve showing $\widehat{\Phi}$ is the same as the magenta curve which appears 
in Fig.~\ref{fig:Phi_hat}, but plotted on a logarithmic scale to show the behavior of this 
transfer function after 
 horizon entry --- \ie, a behavior consisting of damped oscillations within a power-law
envelope function which falls as $a^s$.  As we can see from the red, blue, and 
orange curves, if $k\eta/(H_I\eta_I) \lesssim \mathcal{O}(1)$, 
the absolute values of both $\theta_{k0}$ and $\theta_{k\gamma}$
experience a power-law growth wherein they scale as $\eta\sim a^r$,
whereas $\delta_{k\gamma}$ is approximately a constant.
As $k\eta/(H_I\eta_I)$ gets larger after the modes enter the horizon, 
$\theta_{k0}/k$, $\theta_{k\gamma}/k$, and $\delta_{k\gamma}$ also exhibit different behaviors.
For the particular choice of $\wbar=1/6$, we have $r+s=0>2r-2>r-3/2=s$.
As a result, $\theta_{k0}$ approaches a constant, as mentioned above. 
By contrast, it follows from Eq.~\eqref{eq:theta_delta_subhorizon} that
$\theta_{k\gamma}$ oscillates around a non-zero constant with a decreasing amplitude, 
whereas $\delta_{k\gamma}$ exhibits rapid oscillations around zero.

\FloatBarrier 
\section{Results\label{results}}


In the last few sections, we have assembled all the material required to finally 
calculate the tensor power spectrum and GW energy density in 
cosmologies with a single fluid, as well as those with two fluids, focusing on 
the case of cosmological stasis in particular.  The expression for $S(\kk,\eta)$ 
in the single-fluid case is given in Eq.~\eqref{singlefluidsignal}.
By contrast, the corresponding expression in the multiple-fluid case is given in 
Eq.~\eqref{eq:SfullFinal}.  After substituting for $\Phi$ 
and $\theta$ from Sect.~\ref{theta_phi_stasis} into Eq.~\eqref{eq:SfullFinal}, 
we have an expression $S(\kk,\eta)$ pertinent to the particular case in which the 
multi-fluid cosmology is a matter/radiation stasis cosmology.  
In each case, Eqs.~\eqref{eq:Ph_master_consistent} 
and~\eqref{eq:T_transfer_consistent} give the tensor power spectrum, while 
Eq.~\eqref{eq:OmegaGW_from_Ph_consistent} gives the GW energy density.

In this section, we will evaluate the total scalar-induced 
tensor power spectrum $P_h$ as well as the individual contributions to $P_h$ which arise
from different contributions to $S(\kk,\eta)$.  We will also calculate the resulting 
present-day GW abundance.  The results of this section will thereby allow us to determine which physical ingredients actually control the magnitude of the total SIGW
signal during stasis, and how these ingredients differ from the corresponding single-fluid 
(standard radiation) case.   The results we shall present here are based on a full numerical analysis,
the details of which are provided in Appendix~\ref{app:analytic_estimates}.
This Appendix also includes a discussion of the various 
approximations used in our numerical treatment, and of the various semi-analytical estimates we
have used in interpreting our results.

We begin by noting that during an epoch of matter/radiation stasis with 
an effective equation-of-state parameter $\langle w \rangle = \wbar$, 
the scale factor evolves as
\begin{equation}
  a(\eta)\propto \eta^\beta\,,
    \quad
   ~~~~ \beta \equiv\frac{2}{1+3\overline{w}}\, ,
  \label{eq:betaDef}
\end{equation}
and the conformal Hubble parameter is given by $\mathcal{H}=\beta/\eta$.
Thus, during such an epoch, we have $\mathcal{H}/k = \beta/x$ and the expressions in
Eq.~(\ref{eq:triple_simplified}) for the bulk- and slip-velocity contributions to 
the reduced source kernel simplify to
\begin{eqnarray}
  f_{\mathrm{bulk}}(u,v,x) &=&
    12\,(1+\overline{w})\,\frac{\beta^2}{x^2}\,
    \frac{\Theta(ux)\Theta(vx)}{uv} \nonumber \\
  f_{\mathrm{slip}}(u,v,x) &=& 
    12\,\frac{\beta^2}{x^2}\,
    \frac{(1+w_M)(1+w_\gamma)\barOmega_M\barOmega_\gamma}{1+\overline{w}} 
    \nonumber\\  &&~~~~~\times~
\frac{\Delta\Theta(ux)\Delta\Theta(vx)}{uv}~,
\label{eq:fbulkandslipFinalstasis}
\end{eqnarray}
where $\barOmega_M$, $\barOmega_\gamma$, and $\overline{w}$ are related in the 
manner specified in and directly below Eq.~\eqref{eq:fixedpointabundances}.
As an analysis benchmark in what follows, we shall consider a stasis epoch wherein
\begin{equation}
  \overline{w}=\frac16\,,
    \quad ~~~~\beta=\frac43\,,
\label{eq:results_benchmark_w}
\end{equation}
which corresponds the case in which $\barOmega_M = \barOmega_\gamma = 1/2$.

The results we obtain for a cosmological history involving a stasis epoch of this 
sort will be compared to those we obtain for a standard radiation-dominated history,
wherein the scale factor evolves as $a\sim \eta$ prior to the usual time of 
matter/radiation equality.  In deriving our results for both of these cosmological 
histories, we work in the special case in which $\Phi~=~\Psi$,
and the anisotropic-stress contributions in Eq.~\eqref{eq:SfullFinal} vanish.  
The source in the case of standard cosmology is therefore built from
the geometric and bulk-velocity terms, while the stasis source also contains the 
slip-velocity term associated with the relative motion of matter and radiation.  
The tensor spectrum is then obtained from Eqs.~\eqref{eq:Ph_master_consistent} 
and~\eqref{eq:T_transfer_consistent}, and the present-day energy density is calculated using
Eq.~\eqref{eq:OmegaGW_from_Ph_consistent}.

\begin{figure}[t]
    \centering
    \includegraphics[width=0.47\textwidth]{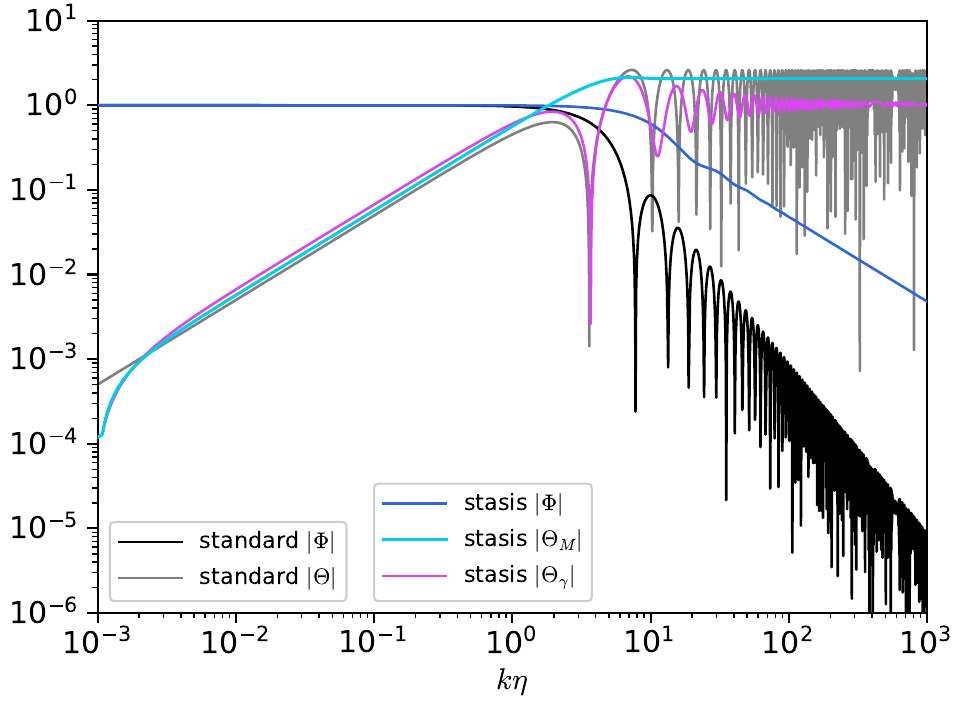}
    \caption{
    Transfer functions entering the reduced scalar-induced tensor source kernel,
    the absolute values of which are  plotted as functions of $k\eta$. 
    The gravitational-potential and velocity-divergence transfer functions are
    defined in Eqs.~\eqref{eq:PhiTransferDef} and
    \eqref{eq:DeltaMandgammaTransferDef}, respectively, and enter the geometric,
    bulk-velocity, and slip-velocity source kernels through
    Eq.~(\ref{eq:triple_simplified}).  The black and gray curves show the 
    gravitational-potential and velocity-divergence transfer functions that 
    would have arisen for the standard radiation-dominated case.
    In this single-fluid case, the velocity transfer is fixed by the momentum
    constraint.  By contrast, the blue, cyan, and magenta curves show the 
    gravitational-potential, matter-velocity-divergence, and 
    radiation-velocity-divergence transfer functions for the benchmark 
    matter/radiation  stasis epoch with $\overline{w}=1/6$.  During stasis, 
    the matter and radiation velocity divergences remain
    independent perturbative degrees of freedom and therefore generate both
    bulk-velocity and relative-slip contributions to the tensor source.
    Consequently, even when a stasis epoch and a perfect fluid epoch have the
    same effective equation-of-state parameter, their perturbation dynamics
    and scalar-induced tensor sources are physically distinct.
  \label{fig:transfer_functions}}
\end{figure}

In Fig.~\ref{fig:transfer_functions}, we plot the individual transfer 
functions that enter into the expression for the tensor source, as a function of $k\eta$.  
The black and gray curves show the results for a standard radiation-dominated universe, 
while the blue, cyan, and magenta curves show the results for a universe which undergoes 
an epoch of matter/radiation stasis with $\overline{w} = 1/6$.  
This figure already illustrates our central physical point.  In the case of 
standard radiation domination, the velocity transfer is fixed by the single-fluid
momentum constraint, so the source can be written entirely in terms of the usual 
scalar transfer function.  By contrast, in the case of an epoch of matter/radiation 
stasis with an effective equation-of-state parameter $\overline{w}$, matter and 
radiation are both dynamically important whenever $0<\overline{w}<1/3$.  Their 
relative background abundances are fixed by Eq.~(\ref{eq:fixedpointabundances}), but 
their perturbations and velocity divergences are not described by a single 
perfect-fluid degree of freedom. Thus, the stasis source cannot be captured by a
single perfect fluid with the same averaged equation of state.  Instead, the
bulk- and slip-velocity transfer functions must be evolved directly via the system
of equations in Eq.~(\ref{theta_delta_coupled}).

In our study, we assume that the spectrum of the primordial curvature perturbation 
takes the form of a log-normal distribution~\cite{Domenech:2021ztg}:
\begin{equation}
  \mathcal{P}_\psi(k) ~=~
    A_\psi \exp\!\left\{
      -\frac{\left[\log\!\left(k/k_p\right)\right]^2}{2\sigma^2}\right\}~,
\label{eq:app_gaussian_power}
\end{equation}
where $k_p$ denotes the peak value and where henceforth we take
$A_\psi=10^{-3}$ and $\sigma=0.25$.  We emphasize that we shall be comparing our 
results for a stasis cosmology to results obtained for the standard cosmology for 
this same input spectrum.

\begin{figure*} 
\centering
  \includegraphics[width=0.47\textwidth]{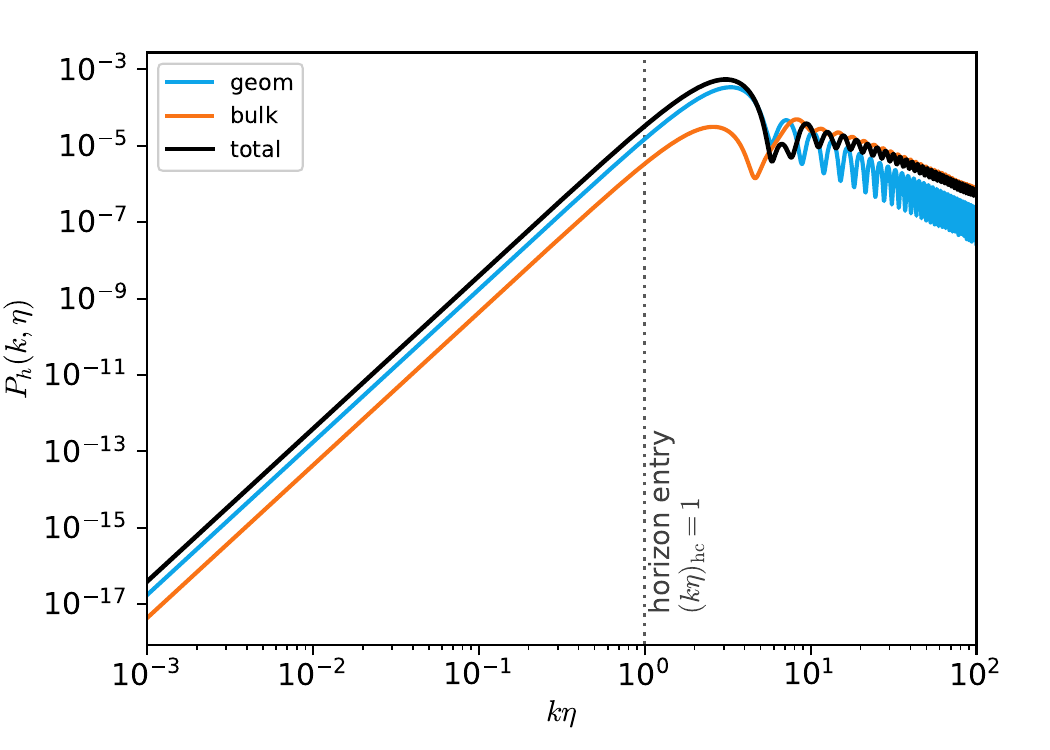}
  \includegraphics[width=0.47\textwidth]{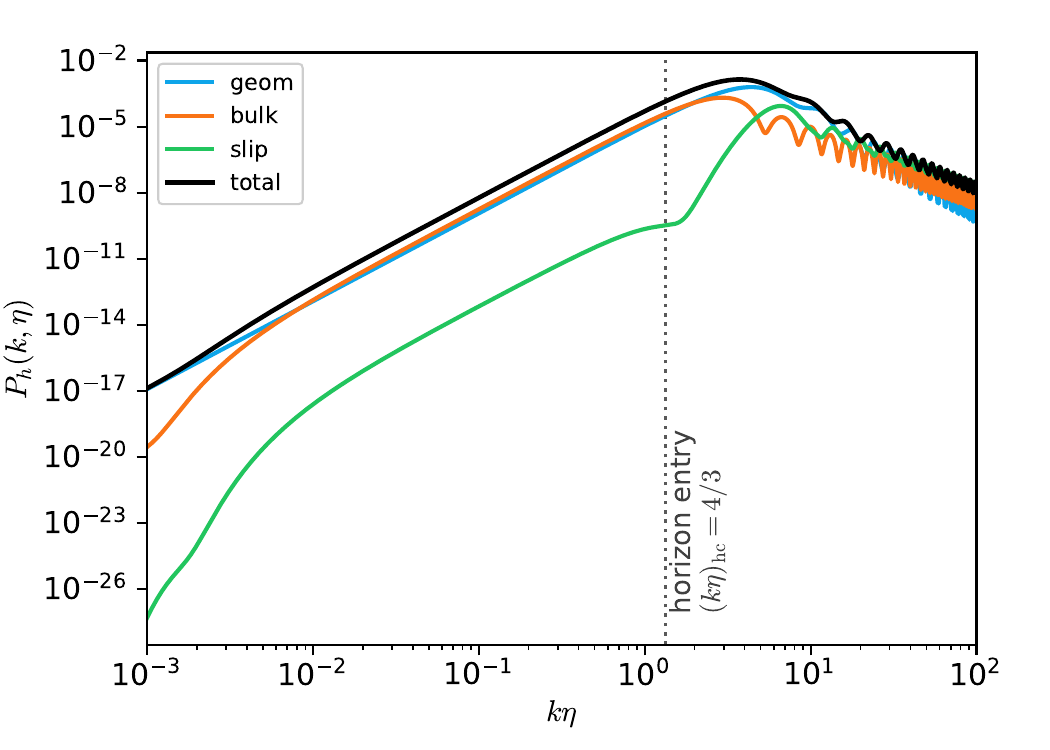}
\caption{Contributions to the scalar-induced tensor power spectrum $P_h(k,\eta)$, 
  plotted as functions of $k\eta$ and  evaluated via Eqs.~\eqref{eq:Ph_master_consistent} 
  and~\eqref{eq:T_transfer_consistent} using the log-normal primordial scalar
  spectrum in Eq.~\eqref{eq:app_gaussian_power} with $A_\psi=10^{-3}$ and $\sigma=0.25$.
  The vertical gray dotted line in each panel marks horizon entry, with $k=\mathcal{H}$.
  For a general background satisfying $a(\eta)\propto\eta^\beta$ with 
  $\mathcal{H}=\beta/\eta$, horizon entry occurs at $(k\eta)_{\rm hc}=\beta$. 
  \textit{Left panel:} The geometric (blue) and bulk velocity (orange) contributions 
  to $P_h(k,\eta)$ for a standard radiation-dominated universe, with $(k\eta)_{\rm hc}=1$.  
  The geometric contribution controls the principal production peak around horizon entry, 
  where the scalar source has its strongest overlap with the tensor Green function.
  \textit{Right panel:} The geometric (blue), bulk velocity (orange), and relative slip
  contributions (green) for a benchmark matter/radiation stasis epoch with $\wbar=1/6$ 
  and hence $(k\eta)_{\rm hc}=4/3$.  The slip contribution is sourced by the relative motion 
  of matter and radiation and therefore has no single-fluid analogue. For the benchmark 
  shown, the geometric contribution continues to control the principal production
  peak, while the slip contribution provides a direct signature of the multi-fluid nature 
  of stasis.  In both panels, the solid black curve is the coherent physical total,
  obtained by adding the source transfer amplitudes before squaring;  thus
  the relevant combinations are $\left|\mathcal{T}_{\rm geom}+\mathcal{T}_{\rm bulk}\right|^2$
  for radiation domination (left panel) and 
  $\left|\mathcal{T}_{\rm geom}+\mathcal{T}_{\rm bulk} +\mathcal{T}_{\rm slip}\right|^2$
  for stasis (right panel). The black curve in each case therefore includes all interference 
  and cross terms and is not, in general, equal to an incoherent sum of the different 
  component contributions.   
\label{fig:standard_vs_stasis_source}}
\bigskip
    \centering
\includegraphics[width=0.47\textwidth]{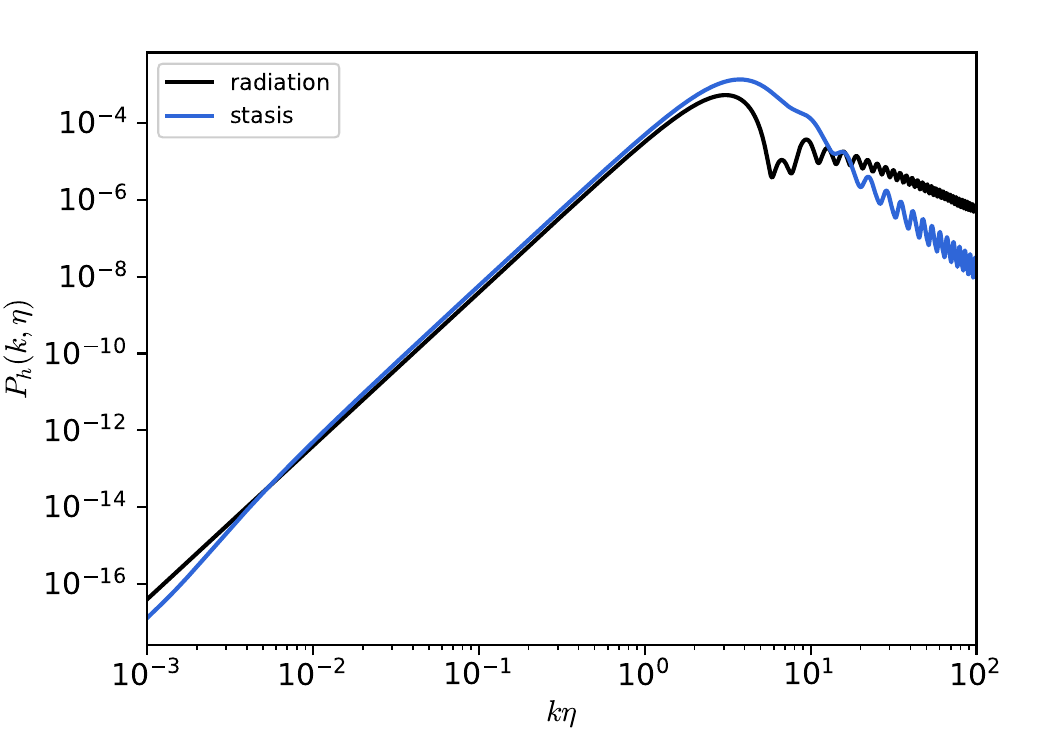}
\includegraphics[width=0.48\textwidth]{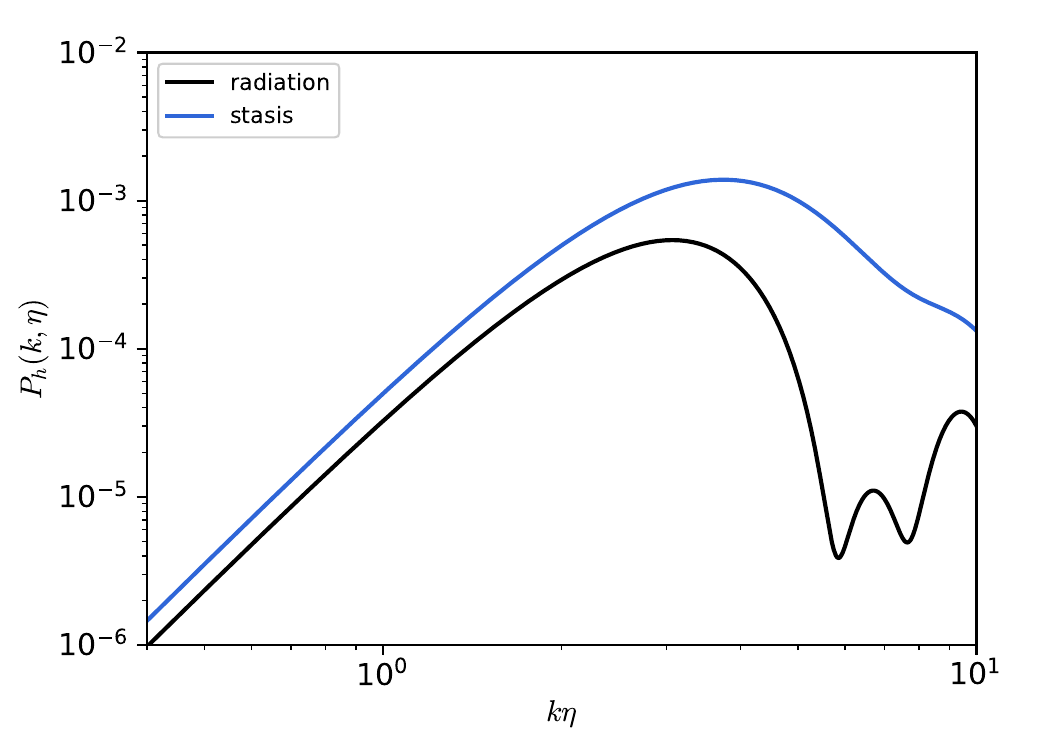}
\caption{
  {\it Left panel}: The integrated tensor power spectrum $P_h(k,\eta)$ for a 
  radiation-dominated universe (black curve) as well as for a universe experiencing 
  matter/radiation stasis (blue curve), plotted as functions of $k\eta$ for the same 
  parameter choices $A_{\psi} = 10^{-3}$ and $\sigma = 0.25$ as in 
  Fig.~\ref{fig:standard_vs_stasis_source}.  We see that the power is generated most 
  efficiently near horizon entry in both cases, where the scalar source overlaps strongly 
  with the tensor Green's function. At later times the source becomes inefficient and both 
  spectra approach the free-propagation envelopes discussed in Sect.~\ref{app:late_time_Ph}.
  {\it Right panel}: Magnification of the portion of the left panel near horizon entry, 
  where the difference between radiation domination and stasis is visible.  The difference 
  between the two curves in this region indicates that stasis impacts not only the 
  redshifting of tensor modes at late times but also the source term itself.
\label{fig:tensor_power_vs_keta}}
\end{figure*}

The source-component budgets in $P_h(k,\eta)$, calculated from Eq.~(\ref{eq:Ph_master_consistent}) 
and Eq.~\eqref{eq:T_transfer_consistent} using the expressions in
Eqs.~(\ref{eq:triple_simplified}), are displayed as functions of $k\eta$ in
Fig.~\ref{fig:standard_vs_stasis_source}.  As shown in the left panel of
Fig.~\ref{fig:standard_vs_stasis_source}, for the standard case of a radiation-dominated epoch, 
the geometric contribution dominates around the time of horizon entry when 
$x\gtrsim \mathcal{O}(1)$, while the bulk-velocity contribution remains smaller 
because it carries explicit powers of $\mathcal H/k$.  In a general matter/radiation-stasis 
epoch (displayed on the right panel), the bulk and slip terms are controlled by the 
inertial weights
\begin{equation}
  \frac{W_M}{W} ~=~ \frac{1-3\overline{w}}{1+\overline{w}}, \qquad 
  \frac{W_\gamma}{W} ~=~ \frac{4\overline{w}}{1+\overline{w}},
  \label{eq:results_inertial_weights_general}
\end{equation}
and by the relative velocity transfer $\Delta\theta = \theta_M-\theta_\gamma$ that 
appears within the third equation in Eq.~\eqref{eq:triple}. This slip contribution is therefore 
proportional to the relative motion between the two fluids and vanishes in any 
single-fluid description.  

In order to assess the relative strengths of the bulk and slip contributions 
to the source, one can compare the prefactors which appear in the corresponding 
contributions to reduced source kernel in Eq.~\eqref{eq:triple_simplified}.  
Explicitly, for the case of a matter/radiation stasis, the prefactor in the 
expression for the slip-velocity contribution includes the quantity
\begin{equation}
  \frac{(1+w_M)(1+w_\gamma)\barOmega_M\barOmega_\gamma}{1+\overline{w}} ~=~
    \frac{4\overline{w}(1-3\overline{w})}{1+\overline{w}}
  \label{eq:results_slip_prefactor_general}
\end{equation}
For our benchmark value $\overline{w}=1/6$ for the effective equation-of-state parameter, 
this prefactor becomes
\begin{equation}
  \frac{4\overline{w}(1-3\overline{w})}{1+\overline{w}}
    \bigg|_{\overline{w}=1/6} ~=~ \frac{2}{7}~.
  \label{eq:results_slip_prefactor_w16}
\end{equation}
In comparison, the prefactor in the expression for the bulk-velocity contribution
contains a factor of  $1 + \overline{w} = 7/6$.  For the benchmark shown, therefore, 
the bulk contribution typically dominates over the slip contribution, while the geometric 
contribution is larger than both. The slip piece is physically important because it encodes 
the multi-fluid character of the background.  Analytical estimates are given in 
Appendix~\ref{app:analytic_estimates}. 

In Fig.~\ref{fig:tensor_power_vs_keta}, 
the blue curve in the left panel shows the 
tensor power spectrum as a function of $x=k\eta$ for stasis, assuming that the horizon 
entry and subsequent evolution occur during stasis. The black curve shows the corresponding 
results for a  standard radiation-dominated epoch. We can see that the tensor power is 
generated most efficiently near horizon entry, where
$x\sim{\cal O}(1)$, because this is where the scalar source has the largest
overlap with the tensor Green's function. This is also the regime where the effects of 
a multi-fluid scalar source are most visible. The right panel of 
Fig.~\ref{fig:tensor_power_vs_keta} shows a zoomed-in version of the left panel, highlighting 
the difference between the standard and stasis source histories in the horizon-entry region.
At times well after horizon entry, the source becomes inefficient and the tensor mode propagates 
freely. The late-time envelope follows from the redshifting of a free tensor mode.
More detailed discussions are given in Appendix~\ref{app:late_time_Ph}. 

\begin{figure}[t]
    \centering
\includegraphics[width=0.47\textwidth]{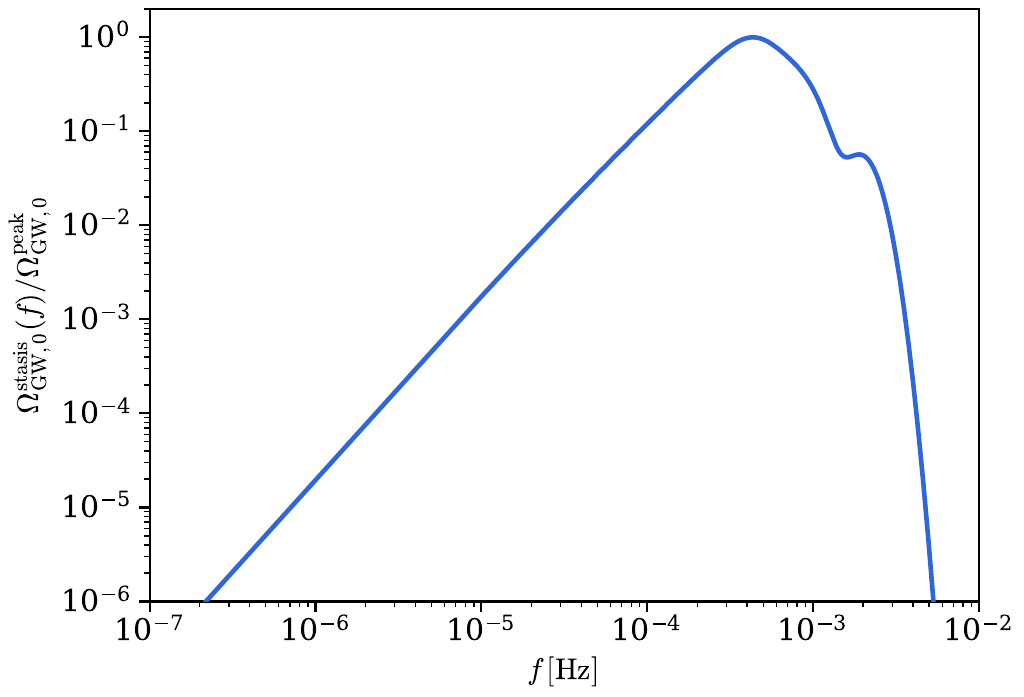}
\caption{The present-day GW spectrum $\Omega_{\mathrm{GW},0}^{\rm sta}(f)$, normalized to 
  its peak value $\Omega_{\mathrm{GW},0}^{\rm peak}=7.2839\times10^{-12}$ and plotted as 
  a function of the GW frequency $f$, for a universe which experiences an epoch of 
  matter/radiation stasis ending at temperature $T_{\rm end}=200\,\mathrm{GeV}$ and a 
  primordial curvature-perturbation spectrum of the log-normal from given in 
  Eq.~\eqref{eq:app_gaussian_power}.  The frequency at which the peak in the GW spectrum 
  occurs in this case is $f_{\rm peak}=4.3472\times10^{-4}\,\mathrm{Hz}$.  
\label{fig:omega_normalized}}
\end{figure}

In Fig.~\ref{fig:omega_normalized}, we show the present-day GW spectrum 
$\Omega_{{\rm GW},0}(f)$ induced by scalar perturbations which enter the horizon 
during the stasis epoch, normalized to its peak value $\Omega_{{\rm GW},0}^{\rm peak}$
across the full range of $f$.  For purposes of illustration, we take stasis to end at
the time when the total energy density of the universe is such that the temperature 
$T_{\rm end}$ to which the universe reheats immediately after the end of stasis is 
$T_{\rm end} = 200$~GeV.  The shape of this normalized spectrum for $f$ around the frequency 
$f_p$ at the peak is set by the scalar perturbation spectrum when it enters horizon.  
Once SIGWs are produced, their sub-horizon evolution is controlled by $\wbar$, which 
determines both the overall normalization of the spectrum due to redshifting and the 
functional dependence of $\Omega_{{\rm GW},0}(f)/\Omega_{{\rm GW},0}^{\rm peak}$ 
on $f$ in the low-frequency tail. See Eq.~\eqref{eq:app_Omega_freq_general} 
and~\eqref{eq:app_Omega_freq_w16} for details on the subhorizon evolution of tensor modes.

\begin{figure*}[t]
    \centering
\includegraphics[width=0.32\textwidth]{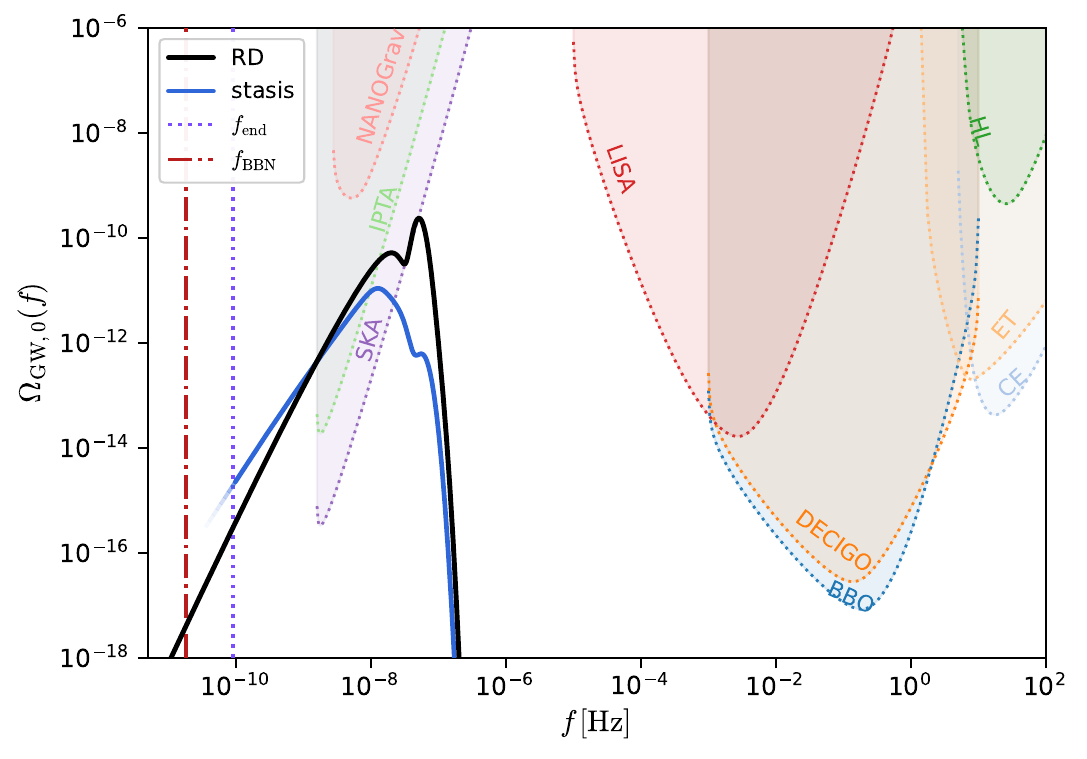}
\includegraphics[width=0.32\textwidth]{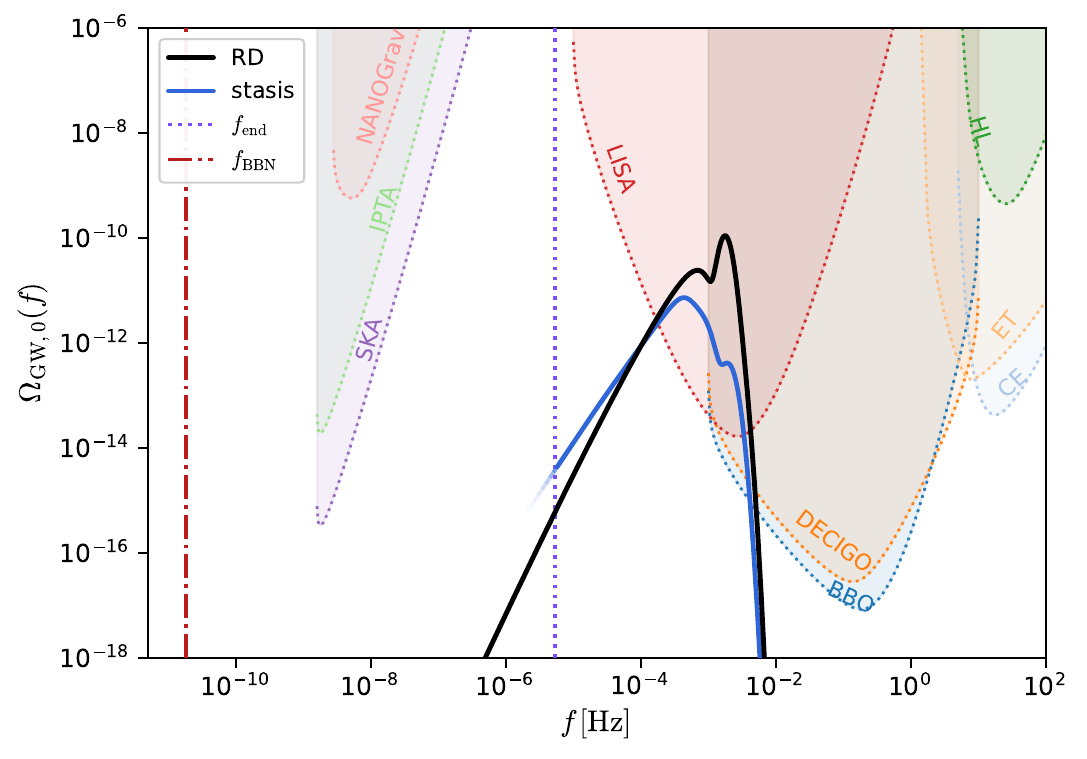}
\includegraphics[width=0.32\textwidth]{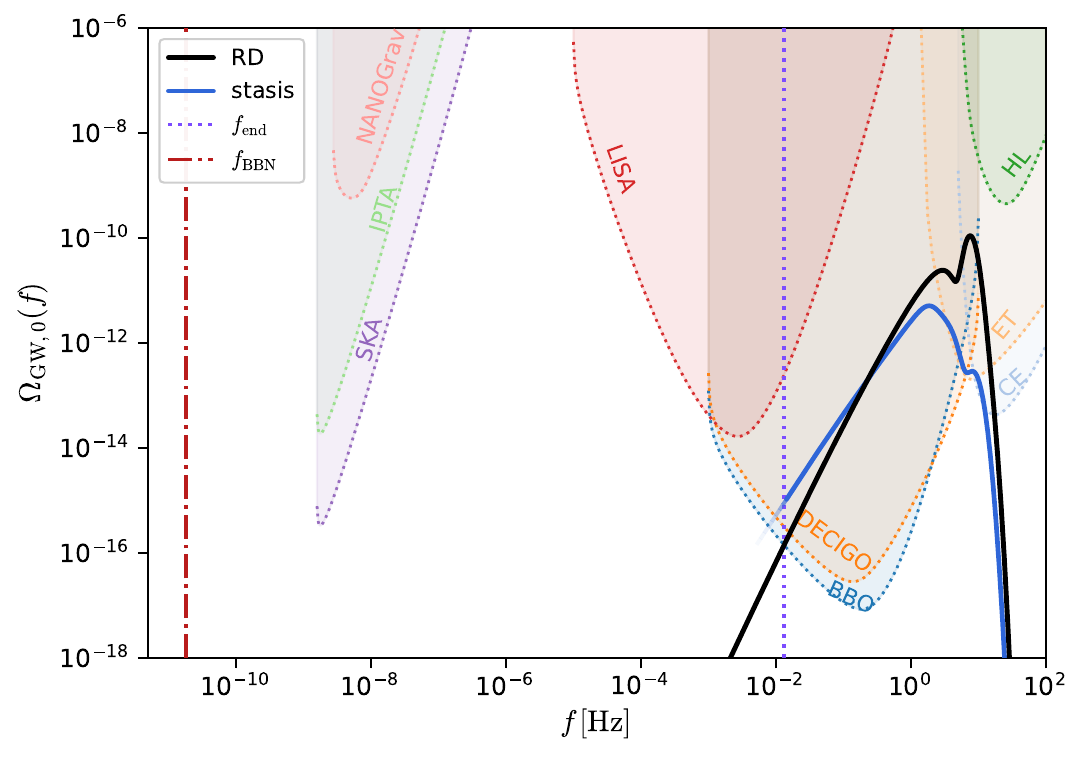}
\includegraphics[width=0.57\textwidth]{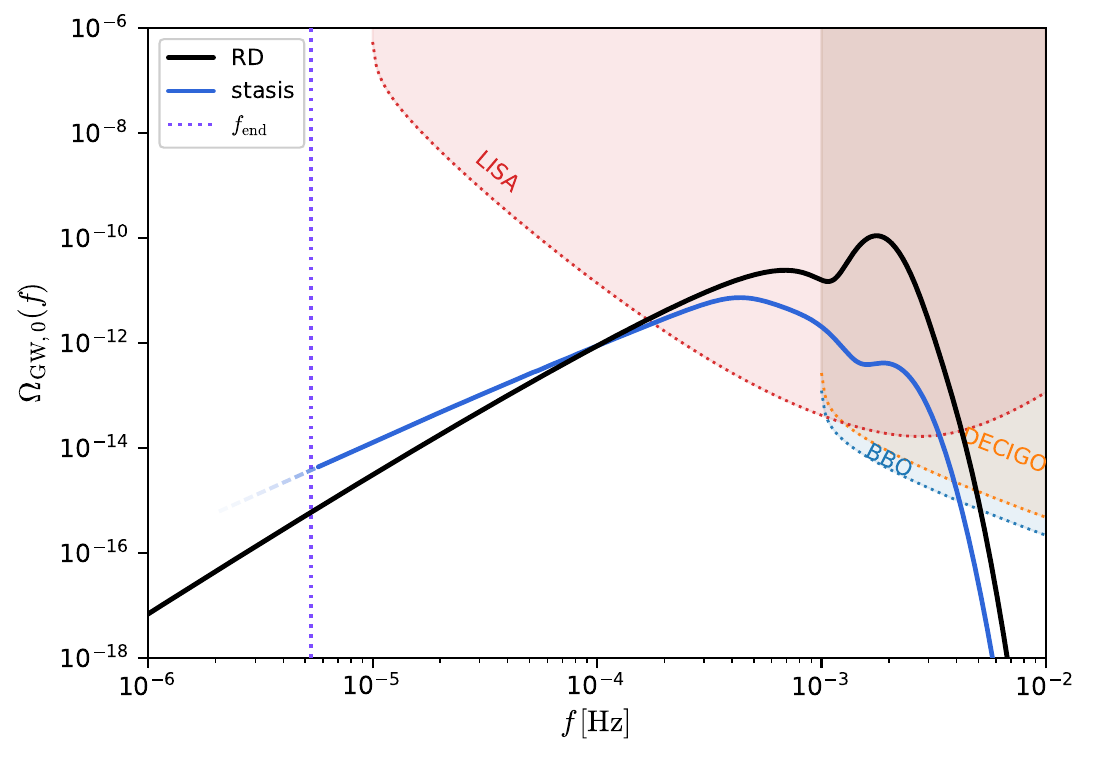}
\caption{Present-day scalar-induced gravitational-wave energy density
  $\Omega_{\mathrm{GW},0}(f)$ generated by the log-normal primordial
  scalar spectrum in Eq.~\eqref{eq:app_gaussian_power}, with
  $A_\psi=10^{-3}$ and $\sigma=0.25$. The solid black curves show the
  prediction for a standard radiation-dominated (RD) cosmology, while
  the solid blue curves show the prediction for the benchmark matter/radiation-stasis 
  cosmology with $\overline{w}=1/6$.  In each panel, the same primordial scalar 
  spectrum is assumed for the RD and stasis calculations.
  {\it Upper panels:} Results for stasis ending at $T_{\rm end}=5\,\mathrm{MeV}$ 
  (left panel), $200\,\mathrm{GeV}$ (middle panel), and $500\,\mathrm{TeV}$ 
  (right panel). The peak scale of the primordial scalar spectrum is chosen 
  separately for each of the three panels so that the resulting GW signal 
  lies in a frequency range relevant for the displayed experiments. The vertical 
  dotted purple line indicates the present-day frequency $f_{\rm end}$ corresponding 
  to the mode that enters the horizon at the end of stasis, while the dot-dashed 
  red line when shown marks $f_{\rm BBN}$, corresponding to $T=1\,\mathrm{MeV}$. 
  Modes with $f>f_{\rm end}$ enter the horizon during stasis and therefore receive 
  both the modified multi-fluid source contribution and the modified redshifting 
  associated with the stasis expansion history. Modes with $f<f_{\rm end}$ enter after
  stasis has ended and follow the standard RD evolution. The faint dashed blue segment 
  below $f_{\rm end}$ is included only as a visual continuation of the stasis spectral 
  slope toward lower frequencies.
  {\it Lower panel:} Magnified view of the $T_{\rm end}=200\,\mathrm{GeV}$ case shown 
  in the upper middle panel, centered on the signal peak and the frequency range probed 
  by planned space-based interferometers. This view highlights the modification
  of both the amplitude and shape of the spectrum produced by stasis relative to the RD 
  prediction. The experimental sensitivity curves are taken from Ref.~\cite{Schmitz:2020syl}.
\label{fig:omega_gw_gaussian}}
\end{figure*}

In Fig.~\ref{fig:omega_gw_gaussian}, we compare the results for 
$\Omega_{{\rm GW},0}(f)$ for a cosmology involving an epoch of matter/radiation stasis 
(solid blue curve in each panel) to the corresponding results for the standard cosmology
(solid black curve in each panel).  Once again, as in Fig.~\ref{fig:omega_normalized},
we assume a log-normal scalar-perturbation spectrum of the form given in 
Eq.~(\ref{eq:app_gaussian_power}), about the origin of which we remain agnostic.
We assume that the universe is already in stasis when the spectrum of primordial scalar 
perturbations enters the horizon.  In the different panels in the top row of the figure, we 
show the results obtained for several different choices of $k_p$ and $k_{\rm end}$.
In particular, in these three panels we have chosen different combinations of $k_p$ 
and $k_{\rm end}$ values which can be accessed by different future GW 
 detectors --- detectors whose sensitivities are also indicated within each panel.  
In the bottom panel, we show a zoomed-in version of middle panel of the top row.
In each case, $k_{\rm end}$ is calculated via the matching procedure described in
Appendix~\ref{app:finite_end_stasis}.  

In order to map values of $k$ onto physical frequencies $f$ at present time, we begin 
by noting that the GW frequency associated with $k_{\rm end}$ is given according to
Eq.~\eqref{eq:f_of_k_numeric_consistent} by
\begin{equation}
f_{\rm end}
~\equiv~
1.55\times 10^{-15}
\left(
\frac{k_{\rm end}}{{\rm Mpc}^{-1}}
\right)
{\rm Hz}~.
\label{eq:results_fend_def}
\end{equation}
Modes with $k<k_{\rm end}$ enter the horizon after stasis has ended and therefore
they evolve in the same manner as they would in the standard cosmology.  By contrast, 
modes with $k>k_{\rm end}$ enter the horizon during stasis, evolve in a modified manner 
during the stasis epoch, and then redshift in the standard manner after stasis ends.
In general, since the conformal time at which such a mode with wavenumber $k$ enters 
the horizon within the context of our stasis cosmology is $\eta=\beta/k$, 
it follows that the value of $x$ that this mode has at the end of the stasis epoch is
\begin{equation}
  x_{\rm end}(k) ~=~ k\eta_{\rm end}
    ~=~ \beta_s\frac{k}{k_{\rm end}} 
    ~=~ \frac{2}{1+3\overline{w}}\frac{k}{k_{\rm end}}\,.
\label{eq:results_xend_general}
\end{equation}
The present-day GW frequency for a mode with wavenumber $k$ is therefore simply 
\begin{equation}
  f ~=~ \frac{1+3\wbar}{2}
  \,f_{\rm end} 
  \, x_{\rm end}(k)~.
\end{equation}

As we can see, stasis deforms the shape of the GW spectrum relative to that obtained 
in the standard cosmology.  The shape of the GW spectrum in the stasis case can be 
understood as a consequence of two features which are particular to stasis.  
First, the source of SIGWs is determined by the nature of perturbations in the cosmic 
fluids, which inevitably includes slip-velocity contributions when multiple 
fluids coexist.  This feature of multi-fluid cosmologies is reflected in the corresponding 
contribution to $P_h(k,\eta)$ shown in Fig.~\ref{fig:standard_vs_stasis_source}.  The overall 
impact of this contribution on $P_h(k,\eta)$ is illustrated in 
Fig.~\ref{fig:tensor_power_vs_keta}.   

Second, cosmological stasis is characterized by an effective equation of state 
$\overline{w}\neq1/3$ and its expansion history different from that of a RD universe. 
The evolution of the GW energy density, which redshifts in the same way as other 
radiation components, is therefore modified relative to the standard RD case during 
the stasis epoch.  In our matter/radiation-stasis model, the sustained decay of matter 
states keeps the universe at $\overline{w}=1/6$, resulting in additional redshift 
suppression of the radiation abundance. This effect is already evident in 
Fig.~\ref{fig:tensor_power_vs_keta}, where the blue curves (corresponding to the 
stasis case) drop faster than the black curves (corresponding to the standard cosmology) 
in the large $k\eta$ region. As a result, the stasis curves in 
Fig.~\ref{fig:omega_gw_gaussian} exhibit a relatively low amplitude near $f_p$.
We note that modes that entered the horizon before the onset of stasis suffer dilution 
from entropy generation at the end of stasis, although in Fig.~\ref{fig:omega_gw_gaussian} 
we do not show such modes.


\section{Discussion\label{discussions}}


In this paper, we have studied the production and evolution of SIGWs during 
cosmological epochs involving multiple fluids.  During such epochs, the intermediate 
cosmological epoch is not merely a passive medium through which tensor perturbations 
generated during previous epochs propagate.  Rather, first-order perturbations in 
the densities of these fluids, along with their velocities, entropy perturbations, 
and anisotropic stresses act as active sources for tensor perturbations at second order.  
The spectrum of SIGWs that emerges is therefore sensitive not only to the manner in which 
the scale factor evolves during a multi-fluid epoch, but also to the internal perturbative 
structure of the medium.  In a genuine multi-fluid cosmology, gravity constrains only 
the total perturbation of each sort, whereas the relative velocities of the individual 
fluids and non-adiabatic degrees of freedom remain as independent physical sources.

We first derived the second-order tensor source for a general two-fluid cosmology.  
This derivation makes explicit the sense in which cosmologies with multiple fluids 
differ from those with a single fluid.  In a single-fluid universe, the first-order 
Einstein constraints allow the density perturbation and velocity perturbation to be 
eliminated in favor of the scalar gravitational potentials.  The induced tensor source 
can therefore be written entirely in terms of these potentials and their derivatives.  
In a two-fluid universe, by contrast, the tensor source contains additional physical data.  
After decomposing the total velocity into a bulk velocity and a relative, or ``slip,'' 
velocity, we found that the source naturally decomposes into geometric, bulk-velocity, 
slip-velocity, non-adiabatic pressure, and anisotropic-stress contributions.  This 
decomposition is general and can be performed irrespective of the details of the background 
cosmology.

We have also applied this general framework to the particular case of an epoch of cosmological 
stasis involving matter and radiation.  Stasis provides a particularly clean realization of a 
genuine multi-fluid epoch because the fractional abundances of cosmological energy components 
with different equations of state can remain approximately constant for an extended period.  
At the background level, such a universe can mimic a universe containing a single perfect fluid with 
a constant effective equation-of-state parameter.  At the perturbative level, however, it remains 
intrinsically multi-fluid, with perturbations growing differently than they do during an epoch 
of perfect-fluid domination with the same effective equation-of-state parameter.  For concreteness,  
we have focused specifically on the case of a matter/radiation stasis involving a tower of 
decaying matter states.  We have focused on the case wherein anisotropic stress is negligible 
and worked in the approximation that the transition from stasis to the subsequent 
radiation-dominated epoch is effectively instantaneous. 

We have found that even when a multi-fluid epoch and a single-fluid epoch share the same 
effective equation-of-state parameter, they need not give the same SIGW spectrum.  The reason 
is that the SIGW spectrum is sensitive to the manner in which the scalar source itself evolves, 
not only to the subsequent redshifting of the tensor modes which it sources.  Within the stasis 
epoch we have considered here, the scalar potential and the overall velocity divergence
evolve differently from the manner in which they evolve during a single-fluid epoch.
These differences enter directly into the source kernel for the induced tensor perturbations.
Moreover, the relative velocity divergence between the matter and radiation components,
which has no analogue within a single-fluid epoch, likewise enters the source kernel and 
yields an additional contribution to the SIGW spectrum.

Our numerical results illustrate this point for a benchmark matter/radiation stasis model with 
effective equation-of-state parameter $\wbar=1/6$ --- a value of $\wbar$ which corresponds to the 
case wherein the abundances of matter and radiation are equal during stasis and the multi-fluid 
nature of the stasis cosmology is most manifest.  These results show that the induced spectrum 
differs from the the corresponding result within the standard cosmology in two crucial ways.  
First, the active production of tensor modes is modified because the scalar and velocity transfer 
functions evolve according to the multi-fluid stasis dynamics.  Second, modes that enter the 
horizon before stasis ends redshift differently as a result of the modified cosmological 
expansion rate during the stasis epoch until this epoch ends and the universe transitions to 
an epoch of radiation domination.  

The broader lesson is that SIGWs can probe more than just the background expansion history of 
the early universe.  They can also probe the internal composition of the cosmic medium during 
the epoch within which these tensor modes are sourced.  This is especially important for cosmologies 
in which a constant effective equation-of-state parameter for the universe emerges from a 
nontrivial mixture of cosmological energy components.  In such cases, the background history 
alone is not sufficient to determine the SIGW signal.  The perturbative degrees of freedom 
carried by the individual components, and especially their relative velocities and entropy 
perturbations, can leave observable imprints. This makes SIGWs a particularly sensitive 
probe of multi-fluid cosmologies, and cosmological stasis provides a concrete example wherein 
the internal structure of the early universe can be imprinted directly on the GW spectrum.

Our work sits within a large body of existing literature.  For example,
the extent to which SIGWs can provide testimony regarding the presence of non-standard single-fluid
epochs within the cosmological timeline has been explored in 
Refs.~\cite{Chianese:2025mll, Domenech:2024wao, Pearce:2025ywc}.  
The extent to which GWs can provide testimony of a kination-dominated era in particular has been 
studied in Refs.~\cite{Gouttenoire:2021jhk, Gouttenoire:2021wzu}. 
Likewise, the extent to which GWs can provide testimony regarding the existence and properties of 
cosmic strings has been considered in Refs.~\cite{Cui:2018rwi, Ghoshal:2023sfa}.  The sensitivity of 
LISA and LIGO to the effective equation-of-state parameter of the universe at early times has been 
explored in Refs.~\cite{Bernal:2019lpc,Figueroa:2019paj,DEramo:2019tit}.  These and other ways in which GWs can serve as a probe 
of the early history of the universe are reviewed in Ref.~\cite{Caldwell:2022qsj}.

As we have seen in this paper, gravitational waves can also provide information 
regarding potential stasis epochs in the early universe.  For example, the authors of
Refs.~\cite{Dienes:2022zgd,Barenboim:2026zgj,Barenboim:2026txj} studied {\it primordial}\/ tensor 
modes generated before the stasis epoch. The subsequent evolution of these tensor modes is 
governed by the homogeneous tensor equation, and the imprint of a later stasis epoch 
is captured by the transfer function associated with a constant effective equation of state.
By contrast, the SIGWs studied in this paper are sourced {\it during the stasis epoch
itself}\/, arising directly from the stasis phenomenon.  As a result, the spectra associated 
with these gravity waves is sensitive not only to the background scale factor, but also to 
the multi-fluid composition of the universe during stasis.

There has also been prior work studying SIGWs in cosmologies containing multiple fluids.
For example, the authors of Ref.~\cite{Kumar:2024hsi} study SIGWs in a cosmology 
containing both matter and radiation components.  However, the physical setting and 
emphasis are different from those of our work.  In Ref.~\cite{Kumar:2024hsi} the non-standard 
cosmological era was taken to be an early matter-dominated epoch (EMDE) generated by 
a spectator/curvaton field which also provides a concrete inflationary origin for the 
enhanced small-scale curvature perturbations.   Within this framework, the induced 
GW spectrum was calculated by numerically tracking the onset and end of the EMDE, 
including the effect of the  relative velocity between the matter and radiation 
components during the transition.  By contrast, the present work is not primarily 
concerned with constructing the primordial scalar spectrum from a specific inflationary 
mechanism.  Instead, our goal has been to identify the general structure of the 
scalar-induced tensor source in a genuinely multi-fluid cosmology, and to show how 
this structure differs from that of a single perfect fluid with the same effective 
equation of state.  In particular, while the relative velocity contribution in 
Ref.~\cite{Kumar:2024hsi} is important during the finite transition intervals between 
radiation domination and matter domination, in this work we adopted cosmological 
stasis as a framework in which the multi-fluid character of the background cosmology 
persists over an extended epoch.  As a result, the slip velocity in our cosmological 
setting is not just a transient correction near the transition, but part of the active 
tensor source throughout the stasis era.  

Given our results, there are many opportunities for future research.

Within this paper, we have treated the stasis epoch in a very direct way as consisting 
of two fluids whose abundances somehow remain constant despite cosmological expansion.
However, stasis is more than that:  stasis also intrinsically involves a process (called 
a ``pump'' in the prior stasis literature) whereby energy density is transferred from 
one of the fluids to the other.   Indeed, it is this transfer of energy which 
counterbalances the effects of the cosmological expansion and thereby leads to constant 
abundances for each fluid.   The question that arises, then, is whether this additional 
transfer of energy density between the fluids involved might 
source additional gravitational waves beyond those we have already considered.  Indeed, 
in the case of the matter/radiation stasis which has been our focus within this paper, 
this transfer arises from the successive decays of the states $\phi_\ell$ within our 
initial tower.   

It is natural to expect that such decays might act as additional sources of 
gravitational waves.  After all, it is known that the Hawking evaporation of primordial 
black holes --- a process which intrinsically involves the decays of states near the 
event horizon --- gives rise to high-frequency gravitational waves~\cite{Aggarwal:2025noe}. 
However, it is not clear whether that conclusion carries over to the case of a decaying 
tower of states.   This is therefore an interesting question for future research.

In this paper we have focused on the case of a {\it matter/radiation}\/ stasis induced 
by a tower of decaying states.  However, there are many other kinds of stasis, some 
involving different fluids entirely and some involving different sorts of pump 
mechanisms.  Indeed, some of these pumps involve sequences of scalars experiencing 
overdamped/underdamped transitions~\cite{Dienes:2023ziv, Dienes:2024wnu}, while others 
result from particle {\it annihilation}\/ rather than decay~\cite{Barber:2024vui}.  
In each case, the underlying phenomenology of the stasis epoch can be different, and 
these differences might also be reflected in the resulting gravitational waves.  This too merits further exploration.

Stasis epochs, of course, are not eternal.   Indeed, there are times at which the universe is 
{\it entering into}\/ a stasis epoch, and  later times at which the universe is 
{\it exiting from}\/ such an epoch.  These transition periods are more complicated 
to analyze, since the universe during such transition periods will have 
{\it time-dependent}\/ overall equations of state.   However, these periods can be 
important for the overall cosmology of the early universe.   This too remains a 
possible topic for future research.

Finally, in this paper we have concentrated on the features of the gravitational waves 
that are sourced by a period of stasis.   However, there remains the complementary 
question of determining the extent to which we can learn about the stasis epoch itself 
by studying these gravitational waves --- \ie, these ``witnesses'' to the early universe.  
Learning how to ``interrogate'' these witnesses in order to extract this information 
will also be an interesting future area of investigation.


\begin{acknowledgments}

The research activities of KRD are supported in part 
by the U.S.\ Department of Energy under Grant DE-FG02-13ER41976 / DE-SC0009913 
and by the U.S.\ National Science Foundation through its employee IR/D program. 
The work of FaH is supported by the Homer L.~Dodge Postdoctoral Fellowship at the 
University of Oklahoma, while the work of LH  was supported by the STFC under UKRI 
grant ST/X000753/1 and by an Ernest Rutherford Fellowship from the UKRI Science 
and Technology Facilities Council (Grant ST/D005634/1).  The research activities of KS are supported 
in part by the U.S.\ National Science Foundation under Grant PHY-2412671, 
while those of BT are supported in part by the U.S.\ National Science Foundation 
under Grant PHY-2310622.  
TX acknowledges support from the HKUST Jockey Club Institute for Advanced Study. 
KRD, FaH, KS, BT, and TX would like to thank the Center for Theoretical Underground 
Physics and Related Areas (CETUP$^\ast$), the Institute for Underground Science 
at Sanford Underground Research Facility (SURF), and the South Dakota Science 
and Technology Authority for hospitality and financial support at different stages 
while this work was being completed.  KRD, KS, and BT would also like to thank the 
Aspen Center for Physics, which is supported by the National Science Foundation 
under Grant PHY-2210452, for hospitality during the course of this work, while KS and BT 
also thank the Aspen Center for financial support.  The opinions and conclusions 
expressed herein are those of the authors and do not represent any funding agencies. 

\end{acknowledgments}

\appendix


\section{Analytical estimates and numerical approximations\label{app:analytic_estimates}}


In this Appendix we summarize the analytical estimates used to interpret the
numerical calculation in the main text.  We avoid repeating the derivation of
the tensor source, the source correlator, and the tensor power spectrum, which
are already given in Sects.~\ref{subsec:simplified_kernels},
\ref{subsec:SIGW_kernel_consistent}, and~\ref{subsec:SIGW_green_consistent}.~ 
In particular, the numerical results are
based on Eqs.~(\ref{eq:Ph_master_consistent}) 
and~(\ref{eq:T_transfer_consistent}), while the GW abundance is
obtained from Eq.~(\ref{eq:OmegaGW_from_Ph_consistent}).

Throughout this Appendix we use the dimensionless time variable $x~=~ k\eta$ 
defined in Eq.~(\ref{eq:uvx_def_consistent}).  For any epoch with constant 
effective equation of state $w$, we have
\begin{equation}
  a(\eta)~\propto~ \eta^\beta, \qquad
  \beta\equiv\frac{2}{1+3w}
\label{eq:app_beta_def}
\end{equation}
as stated in Eq.~\eqref{eq:betaDef}.  During a radiation-dominated epoch, for
example, we have $\beta_{\rm rad}=1$.  During an epoch of matter/radiation stasis with
effective equation-of-state parameter $0 < \wbar < 1/3$, by contrast, we have
\begin{equation}
  \beta_s ~\equiv~ \frac{2}{1+3\wbar}
  \label{eq:app_beta_s_def}
\end{equation}
We note that the corresponding range of $\beta_s$ is $1<\beta_s<2$.  For our 
stasis benchmark, for which with $\wbar=1/6$, we have
\begin{equation}
  \beta_s\Big|_{\wbar=1/6} ~=~ \frac{4}{3}~.
  \label{eq:app_beta_s_w16}
\end{equation}

\subsection{Scalar transfer function: 
Single-fluid case\label{app:single_fluid_phi_estimate}}

In this section we derive an analytical estimate for the scalar transfer function
in the single-fluid case.  This transfer function, which we denote as $\Phi_{\rm PF}$,
is approximated by the usual adiabatic perfect-fluid result for constant $w>0$.  For
$c_s^2=w$, the regular solution can be written as~\cite{Redmond:2018xty}
\begin{equation}
  \Phi_{\rm PF}(x) ~=~
    \Phi_{{\rm PF},i}\,{\cal N}_\nu\, z^{-\nu}J_\nu(z)~,
  \label{eq:app_phi_pf}
\end{equation}
where $z\equiv c_x x$, where $J_\nu(z)$ is a Bessel function of the first kind, where
we have defined
\begin{equation}
  \mathcal{N}_\nu~\equiv~ 2^\nu\,\Gamma(\nu+1)\,,
    \qquad
    \nu~=~ \frac{5+3w}{2(1+3w)}~,
  \label{eq:app_phi_pf_norm}
\end{equation}
and where $\Phi_{{\rm PF},i}$ is a normalization factor chosen such that 
$\Phi_{\rm PF}\to\Phi_i$ as $x\to 0$.  Expanding this expression at early times,
when $x \ll 1$, gives
\begin{equation}
  \Phi_{\rm PF}(x) ~\approx~
    \Phi_{{\rm PF},i}\left[1- \frac{c_s^2x^2}{4(\nu+1)} +\mathcal{O}(x^4)\right]~.
\label{eq:app_phi_pf_smallx}
\end{equation}
By contrast, at late times, when $x\gg1$, the form of the envelope function within
which $\Phi_{\rm PF}$ oscillates can be obtained from the standard 
Bessel-function expansion for large argument, which yields
\begin{equation}
  |\Phi_{\rm PF}(x)| ~\propto~ x^{-\nu-\frac12} ~=~
    x^{-\frac{3(1+w)}{1+3w}}~.
  \label{eq:app_phi_pf_largex}
\end{equation}
For the special case of a radiation-dominated epoch, this expression reduces to
\begin{equation}
  |\Phi_{\rm rad}(x)|~\propto~ x^{-2}~.
  \label{eq:app_phi_rad_largex}
\end{equation}
The corresponding result for the case of a matter-dominated epoch, wherein $w=c_s^2=0$, 
must be derived separately.  In this case, the regular solution for $\Phi_{\rm PF}$
turns out to be
\begin{equation}
  \Phi_{\rm PF}(x)~\approx~ [{\rm constant}]~.
  \label{eq:app_phi_matter}
\end{equation}

\subsection{Scalar transfer function: Stasis case\label{app:stasis_phi_estimate}}

In this section we derive an analytical estimate for the scalar transfer function
in the stasis case, which we denote $\Phi_s$.  In this case, the gravitational 
potential is not governed by a single-fluid acoustic equation, but rather by
Eq.~\eqref{eq:FourthOrderPhiEq}.  In order to obtain an analytical estimate for
$\Phi_s$, we only need the leading subhorizon behavior of this potential.  In
the regime $k\gg aH$, the resulting envelope equation is
\begin{equation}
  \frac{1}{3} a^2\frac{d^2\Phi}{da^2}+ \frac{9\wbar+7}{6}a \frac{d\Phi}{da}
    + \frac{3\wbar^2+7\wbar}{2}\Phi ~\approx~0~.
  \label{eq:app_phi_stasis_reduced}
\end{equation}
The power-law ansatz $\Phi\propto a^\xi$ yields 
\begin{equation}
  2\xi^2+(9\wbar+5)\xi + 9\wbar^2+21\wbar ~=~0~.
  \label{eq:app_phi_stasis_xi_eq}
\end{equation}
The two roots of this equation are
\begin{equation}
  \xi_\pm(\wbar) ~=~
    \frac{-5-9\wbar \pm \sqrt{25-78\wbar+9\wbar^2}}{4}~.
  \label{eq:app_phi_stasis_xi_roots}
\end{equation}
For $0\le \wbar < 1/3$, the late-time envelope is controlled by the larger 
(\ie, less negative) root $\xi_+(\wbar)$.  Since
\begin{equation}
  a~\propto~ x^{\beta_s}~,
  ~~~~~~
  \beta_s~=~\frac{2}{1+3\wbar}~,
\end{equation}
the envelope function for the gravitational potential, and therefore for the 
transfer function $\Phi_s$, takes the form
\begin{equation}
  |\Phi_s(x)| ~\propto~ x^{\beta_s\xi_+(\wbar)}
    ~=~ x^{\frac{2\xi_+(\overline{w})}{1+3\overline{w}}}~.
  \label{eq:app_phi_stasis_largex_general}
\end{equation}
Equivalently,
\begin{equation}
  |\Phi_s(x)| ~\propto~
    x^{\frac{ -5-9\overline{w} +
    \sqrt{25-78\overline{w}+9\overline{w}^2}}
    {2(1+3\overline{w})}}~.
  \label{eq:app_phi_stasis_largex_explicit}
\end{equation}

Having obtained this general expression valid for all $0\le \wbar < 1/3$, we now 
specialize to our stasis benchmark, for which $\overline{w}=1/6$.  Since this 
choice of $\wbar$ yields $\beta_s = 4/3$ and $\xi_+ = 3/4$, it follows that 
\begin{equation}
  |\Phi_s(x)|~\propto~ x^{-1}~.
  \label{eq:app_phi_stasis_w16}
\end{equation}
Thus, we see that $|\Phi_s(x)|$ decays more slowly with $x$ for our stasis 
benchmark --- and indeed for a matter/radiation stasis with any other value of
$0 < \bar{w} < 1/3$ as well --- than does the corresponding envelope function 
$|\Phi_{\rm rad}(x)|$ in Eq.~\eqref{eq:app_phi_rad_largex} for the case of a 
radiation-dominated universe.  This difference is one of the reasons why the 
source term contributes significantly to the generation of tensor perturbations 
for a longer time interval during a stasis epoch than it does during a 
radiation-dominated epoch.

\subsection{Late-time tensor-power-spectrum envelope\label{app:late_time_Ph}}

At late times, when the source term on the right side of Eq.~\eqref{eq:SfullFinal} no 
longer contributes significantly, the tensor mode evolves according to the homogeneous 
limit of Eq.~\eqref{eq:v_equation_consistent}.  Moreover, since $k^2\gg a''/a$ for modes
well outside the horizon, this equation of motion for $v_{\kk}$ reduces for such modes 
to the simple-harmonic-oscillator equation.  Since this implies that $v_{\kk}$ oscillates 
with an approximately constant amplitude, it follows that tensor perturbation $h_{\kk}$ 
oscillates within a envelope whose amplitude decays as  
\begin{equation}
  |h_{\kk}|\,\propto\, a^{-1}\,.
  \label{eq:app_h_free_scaling}
\end{equation}
Since $a\propto x^\beta$, the tensor power spectrum scales at late times as
\begin{equation}
  P_h(k,x) ~\approx~ P_{h,*}(k) \left(\frac{x_*}{x}\right)^{2\beta}~,
  \label{eq:app_Ph_free_scaling}
\end{equation}
where $x_*$ denotes some $\mathcal{O}(1)$ fiducial value of $x$ near the point at which 
$S(\kk,\eta)$ effectively peaks or saturates and where $P_{h,*}(k) \equiv P_h(k,x_*)$ 
denotes the value of $P_h(k,x)$ at this fiducial value of $x$.  A useful choice for $x_*$ 
is the value $x_{\rm hc}$ of $x$ at horizon crossing~\cite{Baumann:2007zm,Domenech:2021ztg}.  Since 
$k=\mathcal H=\beta/\eta$ when a mode with wavenumber $k$ enters the horizon, this 
choice of $x_*$ is equivalent to taking
\begin{equation}
  x_* ~=~ \beta\,.
  \label{eq:app_xstar_hc}
\end{equation}

For the  radiation-domination case, $\beta_{\rm rad}=1$, and hence
\begin{equation}
  P_h^{\rm rad}(k,x) ~\approx~ P_{h,*}^{\rm rad}(k)
    \left(\frac{x_*}{x}\right)^2.
  \label{eq:app_Ph_rad}
\end{equation}
By contrast, for the case of matter/radiation stasis with a value of $\overline{w}$
within the range $0 \leq \wbar < 1/3$, we have $\beta_s=2/(1+3\overline{w})$.  We  
therefore have
\begin{equation}
  P_h^{s}(k,x) ~\approx~ P_{h,*}^{s}(k)
    \left(\frac{x_*}{x}\right)^{2\beta_s} =~
    P_{h,*}^{s}(k) \left(\frac{x_*}{x}\right)^{\frac{4}{1+3\overline{w}}}~.
  \label{eq:app_Ph_stasis_general}
\end{equation}
For our stasis benchmark, for which $\overline{w}=1/6$ and $\beta_s=4/3$, this general 
expression reduces to
\begin{equation}
  P_h^{s}(k,x) ~\approx~ P_{h,*}^{s}(k) \left(\frac{x_*}{x}\right)^{8/3}~.
  \label{eq:app_Ph_stasis_w16}
\end{equation}
This result implies that for this benchmark, $P_h^{s}(k,x)$ falls more rapidly with $x$ 
when $x \gg x_*$ than does $P_h^{\rm rad}(k,x)$ --- a conclusion borne out in the left
panel of Fig.~\ref{fig:tensor_power_vs_keta}.

\subsection{End of stasis\label{app:finite_end_stasis}}

Since we are working in the approximation in which the stasis epoch ends abruptly
at some conformal time $\eta = \eta_{\rm end}$, we may unambiguously associate 
a corresponding value of the scale factor $a_{\rm end} \equiv a(\eta_{\rm end})$ 
and a corresponding wavenumber $k_{\rm end}$ with the end of stasis.  In accord 
with standard conventions~\cite{Baumann:2007zm,Domenech:2021ztg}, we
define $k_{\rm end}$ such that
\begin{equation} 
  k_{\rm end}~\equiv~ \mathcal H_{\rm end}~.
  \label{eq:app_kend_def}
\end{equation}
During an epoch of matter/radiation stasis with equation-of-state parameter 
$0 \leq \overline{w} < 1/3$, we have $\mathcal H=\beta_s/\eta$ and 
$\beta_s=2/(1+3\overline{w})$.  It therefore follows that
\begin{equation}
  x_{\rm end}(k) ~=~ k\eta_{\rm end}
    ~=~ \beta_s\frac{k}{k_{\rm end}}
    ~=~ \frac{2}{1+3\overline{w}}\frac{k}{k_{\rm end}}~.
  \label{eq:app_xend_stasis_general}
\end{equation}
For our stasis benchmark, for which $\overline{w}=1/6$, this yields
\begin{equation}
  x_{\rm end}(k) ~=~ \frac{4}{3}\frac{k}{k_{\rm end}}~.
  \label{eq:app_xend_stasis_w16}
\end{equation}

For modes with $k<k_{\rm end}$, horizon entry occurs after stasis ends, and
the standard radiation result applies.  For modes with $k>k_{\rm end}$, 
by contrast, the tensor power at the end of a general stasis epoch is estimated as
\begin{eqnarray}
  P_{h,{\rm end}}^{s}(k) &\,=\,& P_{h,*}^{s}(k)
    \left( \frac{x_{*,s}}{\beta_s} \frac{k_{\rm end}}{k}
    \right)^{2\beta_s} \nonumber \\
   &\,=\,& P_{h,*}^{s}(k)
    \left[ \frac{(1+3\wbar)x_{*,s}}{2}\frac{k_{\rm end}}{k}
    \right]^{\frac{4}{1+3\overline{w}}}~.
  \label{eq:app_Ph_end_stasis_explicit}
\end{eqnarray}
For $\overline{w}=1/6$, this becomes
\begin{equation}
  P_{h,{\rm end}}^{s}(k) ~=~
    P_{h,*}^{s}(k)  \left( \frac{3x_{*,s}}{4}
    \frac{k_{\rm end}}{k} \right)^{8/3}~.
  \label{eq:app_Ph_end_stasis_w16}
\end{equation}

After stasis ends, the power in the sub-horizon tensor modes redshifts as $a^{-2}$.  
Taking $a_0=1$, we have
\begin{equation}
  P_{h,0}^{s}(k) ~=~
    P_{h,{\rm end}}^{s}(k)\,a_{\rm end}^2~.
  \label{eq:app_Ph0_stasis}
\end{equation}
The corresponding result for a radiation-dominated universe is
\begin{equation}
  P_{h,0}^{\rm rad}(k) ~=~
    P_{h,*}^{\rm rad}(k) \left( x_{*,{\rm rad}}
    \frac{k_{\rm end}}{k} \right)^2
    a_{\rm end}^2~.
  \label{eq:app_Ph0_rad}
\end{equation}
Thus, we find that the ratio of the tensor power obtained in the stasis case
for a mode of wavenumber $k$ to the tensor power obtained in the case of a 
standard radiation-dominated universe is
\begin{equation}
  \frac{P_{h,0}^{s}}{P_{h,0}^{\rm rad}} ~=~
    \frac{P_{h,*}^{s}}{P_{h,*}^{\rm rad}}
    \frac{(x_{*,s}/\beta_s)^{2\beta_s}}{x_{*,{\rm rad}}^2}
    \left(\frac{k_{\rm end}}{k} \right)^{2\beta_s-2}~.
  \label{eq:app_Ph0_ratio_general}
\end{equation}
Using $\beta_s=2/(1+3\overline{w})$, we find that 
the exponent is
\begin{equation}
  2\beta_s-2 ~=~
  \frac{2(1-3\overline{w})}{1+3\overline{w}}~.
  \label{eq:app_ratio_exponent_general}
\end{equation}
For the choice $x_* = x_{\rm hc}$ of our fiducial value of $x_*$, as discussed above 
Eq.~\eqref{eq:app_xstar_hc}, we have $x_{*,{\rm rad}}=1$ and $x_{*,s}=\beta_s$.  Hence,
we find that
\begin{equation}
  \frac{P_{h,0}^{s}}{P_{h,0}^{\rm rad}} ~\approx~
    \frac{P_{h,*}^{s}}{P_{h,*}^{\rm rad}}
    \left(\frac{k_{\rm end}}{k}
    \right)^{\frac{2(1-3\overline{w})}{1+3\overline{w}}}~.
  \label{eq:app_Ph0_ratio_hc_general}
\end{equation}
For our stasis benchmark, for which $\overline{w}=1/6$, this general expression reduces to
\begin{equation}
  \frac{P_{h,0}^{s}}{P_{h,0}^{\rm rad}} ~\approx~
    \frac{P_{h,*}^{s}}{P_{h,*}^{\rm rad}}
    \left(\frac{k_{\rm end}}{k}\right)^{2/3}~.
  \label{eq:app_Ph0_ratio_w16}
\end{equation}

\subsection{Present-day abundance\label{app:omega_today_estimate}}

The present-day abundance $\Omega_{\rm GW,0}(k)$ of GWs per logarithmic interval 
in $k$ for modes well within the horizon is in general given by 
Eq.~\eqref{eq:OmegaGW_from_Ph_consistent} with the conformal time $\eta$ set 
to its present-day value.  For the purpose of obtaining an analytic estimate 
for $\Omega_{\rm GW,0}(k)$ in any, it is convenient to 
write~\cite{Baumann:2007zm,Kohri:2018awv}
\begin{equation}
  \Omega_{\rm GW,0}(k) ~\approx~
    \frac{1}{24}\left(
    \frac{k}{\mathcal H_0}\right)^2
    P_{h,0}(k)~.
  \label{eq:app_Omega_today}
\end{equation}
The wavenumber $k_{\rm end}$ of the mode which enters the horizon at the 
moment stasis ends is related to the Hubble parameter $H(a_{\rm end})$ at this 
time, as defined below Eq.~(\ref{eq:dPrimePhiEqnArbwSimpFTDimVars}), by  
\begin{equation}
  k_{\rm end} ~=~ a_{\rm end}H(a_{\rm end})~.
  \label{eq:app_kend_aend_relation}
\end{equation}
If the stasis epoch ends well before the usual time of matter-radiation equality 
within the standard cosmology, then
\begin{equation}
  a_{\rm end} ~\approx~ 
    \frac{\mathcal H_0\sqrt{\Omega_r}}{k_{\rm end}}~,
  \label{eq:app_aend_rad}
\end{equation}
where $\Omega_r$ denotes the radiation abundance today, defined with respect to the 
present-day critical density and with $a_0 = 1$.
Using this approximation yields an estimate for $\Omega_{\rm GW,0}^{(w)}(k)$ within 
any modified cosmology in which an epoch with a constant effective equation-of-state
parameter $w$ is followed immediately by a transition directly to the usual period of 
radiation domination.  In particular, we find that
\begin{equation}
  \Omega_{\rm GW,0}^{(w)}(k)~\approx~
    \frac{\Omega_r}{24}P_{h,*}^{(w)}(k)
    \left(\frac{x_*}{\beta}\right)^{2\beta}
    \left(\frac{k}{k_{\rm end}}\right)^{2-2\beta}.~~
  \label{eq:app_Omega_general}
\end{equation}
For the case of a radiation-dominated universe, we have $\beta_{\rm rad}=1$, and for the 
choice $x_* = x_{\rm hc}$ of fiducial dimensionless time parameter, we therefore also 
have $x_*=1$.  Thus, we obtain
\begin{equation}
  \Omega_{\rm GW,0}^{\rm rad}(k) ~\approx~
    \frac{\Omega_r}{24} P_{h,*}^{\rm rad}(k)~.
\label{eq:app_Omega_rad}
\end{equation}

By contrast, for an epoch of matter/radiation stasis with $0 \leq \wbar < 1/3$,  
we have
\begin{equation}
  \Omega_{\rm GW,0}^{s}(k) ~\approx~
    \frac{\Omega_r}{24} P_{h,*}^{s}(k)
    \left(\frac{x_{*,s}}{\beta_s}\right)^{2\beta_s}
    \left(\frac{k}{k_{\rm end}}\right)^{2-2\beta_s}~.
  \label{eq:app_Omega_stasis_general}
\end{equation}
Noting that $\beta_s = 2/(1+3\overline{w})$ and again taking $x_* = x_{\rm hc}$,
we find that the present-day GW abundance in the stasis case is given by
\begin{equation}
  \Omega_{\rm GW,0}^{s}(k) ~\approx~
    \frac{\Omega_r}{24} P_{h,*}^{s}(k)
    \left(\frac{k}{k_{\rm end}}\right)^{
    -\frac{2(1-3\overline{w})}{1+3\overline{w}}}~.
  \label{eq:app_Omega_stasis_hc_general}
\end{equation}
For our stasis benchmark, for which $\overline{w}=1/6$, this general expression 
reduces to
\begin{equation}
  \Omega_{\rm GW,0}^{s}(k) ~\approx~
    \frac{\Omega_r}{24} P_{h,*}^{s}(k)
    \left(\frac{k}{k_{\rm end}}\right)^{-2/3}~.
  \label{eq:app_Omega_stasis_w16}
\end{equation}

In order to render the effect of stasis on the GW spectrum more manifest, we note that
the ratio of $\Omega_{\rm GW,0}^{s}(k)$ to the corresponding result in 
Eq.~\eqref{eq:app_Omega_rad} for the standard cosmology is
\begin{equation}
  \frac{\Omega_{\rm GW,0}^{s}}
    {\Omega_{\rm GW,0}^{\rm rad}} ~\approx~
    \frac{P_{h,*}^{s}}{P_{h,*}^{\rm rad}}
    \left(\frac{k_{\rm end}}{k}\right)^{
    \frac{2(1-3\overline{w})}{1+3\overline{w}}}~.
  \label{eq:app_Omega_ratio_general}
\end{equation}
For our benchmark value $\overline{w}=1/6$, this reduces to
\begin{equation}
  \frac{\Omega_{\rm GW,0}^{s}}
    {\Omega_{\rm GW,0}^{\rm rad}} ~\approx~
    \frac{P_{h,*}^{s}}{P_{h,*}^{\rm rad}}
    \left(\frac{k_{\rm end}}{k}\right)^{2/3}~.
  \label{eq:app_Omega_ratio_w16}
\end{equation}

It is also illuminating to consider how $\Omega_{\rm GW,0}^{s}(f)$ scales with the GW 
frequency $f$.  Using the conversion relation between $k$ and $f$ in 
Eq.~\eqref{eq:f_of_k_numeric_consistent}, we find that for all $f>f_{\rm end}$, the 
general scaling behavior of $\Omega_{\rm GW,0}^{s}(f)$ with $f$ for a scale-invariant 
tensor perturbation spectrum is
\begin{equation}
  \Omega_{\rm GW,0}^{s}(f) ~\propto~
    \Omega_{\rm GW,0}^{s}(f_{\rm end})
    \left(\frac{f}{f_{\rm end}}\right)^{
    -\frac{2(1-3\overline{w})}{1+3\overline{w}}} ~.
  \label{eq:app_Omega_freq_general}
\end{equation}
For our benchmark value $\overline{w}=1/6$, this reduces to
\begin{equation}
  \Omega_{\rm GW,0}^{s}(f) ~\propto~
    \Omega_{\rm GW,0}^{s}(f_{\rm end})
    \left(\frac{f}{f_{\rm end}}\right)^{-2/3}~.
  \label{eq:app_Omega_freq_w16}
\end{equation}
By contrast, modes with frequencies $f<f_{\rm end}$ enter the horizon after stasis 
ends and therefore follow the standard radiation-domination result.

\subsection{Comments on numerical implementation\label{app:numerical_approximation}}

Finally, we provide a few comments about the procedures used in producing the plots in  
Sect.~\ref{results}.~  

As discussed in the main text, we focus on the special 
case in which the anisotropic-stress terms in Eq.~\eqref{eq:SfullFinal} vanish.  
Moreover, as discussed in Sect.~\ref{subsec:simplified_kernels}, we have $\Phi=\Psi$ in this case.
Our results for the velocity-divergence transfer functions in
Eqs.~\eqref{eq:ThetaTransferDef} and~\eqref{eq:DeltaThetaTransferDef} for the 
stasis case are obtained by solving Eq.~\eqref{theta_delta_coupled}, rather than
by using the single-fluid momentum constraint.  We also evaluate the time integral
in Eq.~\eqref{eq:T_transfer_consistent} directly, rather than replacing 
$\mathcal T^2(u,v,k,\eta)$ by an integral over $f^2$.

In performing the numerical integration over the variables $u$ and $v$ in 
Eq.~(\ref{eq:I_kernel_consistent}), we find that in order to obtain an appropriate 
sampling within the triangular domain for these variables specified below this equation, 
it is convenient to perform a change of variables from $\{u,v\}$ to the 
alternative variables $\{s,t\}$~\cite{Kohri:2018awv}, which are implicitly defined through 
the inverse relations
\begin{equation}
  u~=~\frac{t+s+1}{2}\,,
    \quad
    v~=~\frac{t-s+1}{2}\,,
  \label{eq:app_uv_sampling}
\end{equation}
with $s\in[-1,1]$.  In evaluating the resulting integrals, we impose a finite upper cutoff 
on $t$ and increase the this cutoff until the results stabilize.  

While we focus primarily on a scalar spectrum of the form in Eq.~(\ref{eq:app_gaussian_power}) 
in this paper, we make the following observations concerning the results for $\Omega_{\rm GW,0}^{s}(f)$  
which would instead follow from a flat --- \ie, $k$-independent --- $\mathcal{P}_\psi(k)$ spectrum.
For such a spectrum, the present-day GW spectrum obtained within the standard cosmology is 
approximately flat in $f$ because $P_{h,*}^{\rm rad}$ is approximately scale-independent and
Eq.~\eqref{eq:app_Omega_rad} contains no residual power of $k/k_{\rm end}$.
On the other hand, for a cosmology involving a matter/radiation stasis with our benchmark 
effective equation-of-state parameter $\wbar = 1/6$, Eq.~\eqref{eq:app_Omega_stasis_w16} yields 
a GW spectrum with a high-frequency envelope which is approximately given --- up to the residual 
frequency dependence of $P_{h,*}^{s}$ --- by
\begin{equation}
  \Omega_{\rm GW,0}^{s}(f)~\propto ~f^{-2/3}
\label{eq:app_stasis_flat_slope}
\end{equation}
for modes with $f~>~f_{\rm end}$.  By contrast, for the Gaussian spectrum 
in Eq.~(\ref{eq:app_gaussian_power}), the shape of the GW spectrum is instead dominated 
by the convolution of ${\cal P}_\psi(ku){\cal P}_\psi(kv)$ in Eq.~\eqref{eq:Ph_master_consistent}.

When performing our GW energy-density calculations, it was necessary for us to smooth 
our numerical tensor power spectra.  This was done by averaging the rapidly oscillating 
tensor phase at a common late-time value
\begin{equation}
  x_{\langle \mathcal{T} \rangle} ~=~ (k\eta)_{\langle \mathcal{T} \rangle} =~10^{4}\,
\end{equation}
rather than by filtering neighboring frequency bins. For $a(\eta)\propto\eta^{\beta}$, 
the homogeneous tensor solutions are $b_{1}(x)=\sqrt{x}\,J_{\nu}(x)$ and 
$b_2(x) = \sqrt{x}\,Y_{\nu}(x)$ with $\nu=|\beta-1/2|$, where $J_{\nu}(x)$ and 
$Y_{\nu}(x)$ denote Bessel functions of the first and second kind, respectively.
The corresponding Green's-function coefficients are
\begin{equation}
  I_i^{\rm tot}(u,v) ~=~ 
    \sum_X \int_0^{x_{\langle \mathcal{T} \rangle}} dy\,
  y^\beta\, b_i( y)\, f_X(u,v,y)\,,
\end{equation}
where $X\in \{\mathrm{geom},\mathrm{bulk},\mathrm{slip}\}$ and where 
$y = k \tilde{\eta}$ is a dummy integration variable which ranges from 
$0 \leq y \leq x_{\langle \mathcal{T} \rangle}$. The components are
summed before squaring, so all interference terms are retained.

For $x_{\langle \mathcal{T} \rangle}\gg1$, averaging the asymptotic sine and cosine 
phases in evaluating the square of time-integrated transfer function in 
Eq.~(\ref{eq:T_transfer_consistent}) yields 
\begin{equation}
  \overline{|\mathcal{T}(u,v)|^2} ~=~
    \frac{1}{\pi}
    \left( \frac{\pi\kappa}{4x_{\langle \mathcal{T}\rangle}^\beta} \right)^2
  \Big[
    \big|I_1^{\rm tot}(u,v)\big|^2
      +\big|I_2^{\rm tot}(u,v)\big|^2
  \Big]~.
\end{equation}
Replacing $\mathcal{T}^2(u,v,k,\eta)$ by this 
$\overline{\mathcal{T}^2(u,v)}$ in
Eq.~\eqref{eq:Ph_master_consistent} yields a smoothed tensor power spectrum 
for $k\eta > x_{\langle \mathcal{T}\rangle}$.  For general values of $k$ and $\eta$, 
then, the total smoothed tensor power spectrum, summed over tensor-mode polarizations 
takes the form 
\begin{eqnarray}
  \overline{P_h(k,\eta)} &=& 
    2\int_0^\infty dv \int_{|1-v|}^{1+v}  du\,
    \mathcal P_\psi(ku)\,\mathcal P_\psi(kv) \Lambda^2(u,v)\nonumber\\
  &&~~\times\,
    \begin{cases}
    \mathcal{T}^2(u,v,k,\eta) &
      k\eta < x_{\langle \mathcal{T}\rangle} \\
    \overline{|\mathcal{T}(u,v)|^2} &
      k\eta > x_{\langle \mathcal{T}\rangle}~,
   \end{cases}
\end{eqnarray}
where the overall factor of two accounts for the polarization sum.  This 
$\overline{P_h(k,\eta)}$ is the quantity which appears in 
Eq.~\eqref{eq:OmegaGW_from_Ph_consistent}.

We emphasize that this smoothing procedure removes only the unobservable, rapid tensor 
oscillations while preserving the spectral envelope, the peak structure, and the 
interference between the individual reduced-source kernels $f_X(u,v,x)$ associated 
with the different $S_X(\kk,\eta)$.
 
\bibliography{ref}
\end{document}